\documentclass[aps,12pt,superscriptaddress,preprint,tightenlines,floats,%
               nofootinbib,hyperref]{revtex4}

\usepackage{graphicx}
\usepackage{color}
\usepackage{amsmath} 
\usepackage{dcolumn} 

\makeatletter
\@ifundefined{onlinecite}{}{} 
\@ifundefined{inlinecite}{\let\inlinecite=\onlinecite}{}
{\catcode`\%=12 \gdef\percent{
\makeatother

\let\oldcolor=\color
\renewcommand\color[1]{\oldcolor{black}}
\newcommand\ds{\displaystyle}
\newcommand\etal{{\it et al.\spacefactor1000}}
\newcommand\ibid{{\it ibid.\spacefactor1000}}

\newcommand\MSbar{\ensuremath{\text{$\overline{\rm MS}$}}}
\newcommand\lhs{\hbox{l.h.s.}}
\newcommand\rhs{\hbox{r.h.s.}}
\newcommand\ie{\hbox{\it i.e.\/}}
\newcommand\eg{\hbox{\it e.g.\/}}
\def\half{{\textstyle{\frac{1}{2}}}}
\def\quarter{{\textstyle{\frac{1}{4}}}}
\def\sixth{{\textstyle{\frac{1}{6}}}}
\def\CC{{\ensuremath{\cal C}}}
\def\CD{{\ensuremath{\cal D}}}
\def\CF{{\ensuremath{\cal F}}}
\def\CL{{\ensuremath{\cal L}}}
\def\CO{{\ensuremath{\cal O}}}
\def\CS{{\ensuremath{\cal S}}}
\def\CT{{\ensuremath{\cal T}}}
\def\CW{{\ensuremath{\cal W}}}
\def\gam#1{\ensuremath{\overline{(\gamma_{#1}\otimes I)} }}
\def\ixi#1{\ensuremath{\overline{(I\otimes\xi_{#1})} }}
\def\sfno#1#2{\ensuremath{\overline{(\gamma_{#1}\otimes\xi_{#2})}}}
\def\semitimes{\ensuremath{\mathrel>\joinrel\mathrel\triangleleft}}
\def\slash#1{\ensuremath{\mbox{$\not \!\! #1$}}}
\def\MeV{{\ensuremath{\mathop{\rm MeV}\nolimits}}}
\def\GeV{{\ensuremath{\mathop{\rm GeV}\nolimits}}}
\def\Tr{{\ensuremath{\mathop{\sf Tr}}}}
\def\Re{{\ensuremath{\mathop{\sf Re}}}}
\def\bar{\overline}
\def\hat{\widehat}
\def\tilde{\widetilde}
\def\gsim{{\mathrel{\raise2pt\hbox to 8pt{\raise -5pt\hbox{$\sim$}\hss{$>$}}}}}
\def\rsim{{\mathrel{\raise2pt\hbox to 8pt{\raise -5pt\hbox{$\sim$}\hss{$>$}}}}}
\def\lsim{{\mathrel{\raise2pt\hbox to 8pt{\raise -5pt\hbox{$\sim$}\hss{$<$}}}}}

\newcommand\overleft[1]{\ensuremath{\mathord{\mathop{#1}\limits^\leftarrow}}}
\newcommand\overright[1]{\ensuremath{\mathord{\mathop{#1}\limits^\rightarrow}}}
\newcommand\overleftright[1]{\ensuremath{\mathord{\mathop{#1}\limits^\leftrightarrow}}}
\newcommand\psibar{{\ensuremath{\mathord{\overline\psi}}}}
\newcommand\slashnext[1]{\mathpalette{\bgroup\let\style=}
                                     {\setbox0=\hbox{$\style #1$}%
                                      \setbox2=\hbox to\wd0{\hss$\style/$\hss}%
                                      \wd2=0pt\dp2=0pt\box2\box0\egroup}}
\newcommand\onelink{%
   {\mathchoice{\mathord{\rm\hbox{\the\textfont\fam 1-link}}}%
               {\mathord{\rm\hbox{\the\textfont\fam 1-link}}}%
               {\mathord{\rm\hbox{\the\scriptfont\fam 1-link}}}%
               {\mathord{\rm\hbox{\the\scriptscriptfont\fam 1-link}}}}}
\newcommand\Dslash{{\ensuremath{\mathord{\slashnext D}}}}
\newcommand\Wilson{{\ensuremath{\mathord{\cal W}}}}

\newcommand\lc[1]{\lowercase{#1}}

\ifx\pdfoutput\undefined\else
  \ifnum\pdfoutput=1\relax\usepackage{thumbpdf}\fi
\fi
\usepackage[hyperindex]{hyperref}

\begin{document}

\preprint{LA-UR-03-3849}
\pacs{????}
\title{Scaling behavior of discretization errors in renormalization and improvement constants}

\author{Tanmoy Bhattacharya}
\email{tanmoy@lanl.gov}\homepage{http://t8web.lanl.gov/t8/people/tanmoy/}
\author{Rajan Gupta}
\email{rajan@lanl.gov}\homepage{http://t8web.lanl.gov/t8/people/rajan/}
\affiliation{Theoretical Division, Los Alamos National Lab, Los Alamos,
         New Mexico 87545, USA}
\author{Weonjong Lee}
\email{wlee@snu.ac.kr}\homepage{http://lgt.snu.ac.kr/}
\affiliation{School of Physics, Seoul National University, Seoul, 151-747, Korea}
\author{Stephen R. Sharpe}
\email{sharpe@phys.washington.edu}
\affiliation{Physics Department, University of Washington,
         Seattle, Washington 98195, USA \vspace* {2cm}}

\date{\today}

\begin{abstract}

Non-perturbative results for improvement and renormalization constants
needed for on-shell and off-shell $O(a)$ improvement of bilinear
operators composed of Wilson fermions are presented. The calculations
have been done in the quenched approximation at $\beta=6.0$, $6.2$ and
$6.4$.  To quantify residual discretization errors we compare our data
with results from other non-perturbative calculations and with
one-loop perturbation theory.





\vskip 1 cm

\noindent PACS numbers: 11.15.Ha and 12.38.Gc

\end{abstract}

\maketitle

\color{black}
\section{Introduction}
\label{sec:intro}

In this paper we present our final results for the renormalization and
improvement constants for quark bilinear operators using Wilson's
gauge action and the $O(a)$ improved Dirac action first proposed by
Sheikholeslami and Wohlert~\cite{SW:IA:85}.  The calculations have been
done at three values of the gauge coupling, $\beta=6.0$, $6.2$, and
$6.4$ in the quenched approximation.\footnote{%
Preliminary results were presented in~\cite{LANL:Zfac:lat01} and are updated here.}
Our results represent a realization of Symanzik's improvement
program for systematically reducing discretization errors in lattice
simulations~\cite{Symanzik:IA:83A,Symanzik:IA:83B}.  Results for the
improvement of the Dirac action have been obtained previously by the
ALPHA collaboration and we have used these in our calculation. This
paper deals with the improvement of external bilinear operators, 
${\cal O}$ with $\CO$ being one of the five Lorentz structures $A, V, P, S, T$.

{\color{red}

The mixing with extra operators, both for on-shell and off-shell
improvement of the operators, and the introduction of mass dependence
in the renormalizaton constants has been discussed in detail in
Section II of Ref.~\inlinecite{LANL:Zfac:00}. To summarize that
discussion, and to remind the reader of the notation, the fully
improved and renormalized bilinear
operators at $O(a)$ are
\begin{eqnarray}
(A^R_{I})_{\mu}  & \equiv & Z_A^0( 1+{\tilde b}_A a {\tilde m}_{ij} ) 
                   \big( A_{\mu} + a c_A \partial_\mu P  - a \frac14 c'_A (
                   \bar{\psi}^{(i)} \gamma_\mu\gamma_5 \overrightarrow{\CW}\psi^{(j)} - 
		   \bar{\psi}^{(i)} \overleftarrow{\CW}\gamma_\mu\gamma_5 \psi^{(j)} )\big)
\nonumber \\
(V^R_{I})_{\mu}  & \equiv & Z_V^0( 1+{\tilde b}_V a {\tilde m}_{ij} ) 
                   \big( V_{\mu} + a c_V \partial_\nu T_{\mu\nu}   - a \frac14 c'_V (
                   \bar{\psi}^{(i)} \gamma_\mu \overrightarrow{\CW}\psi^{(j)} - 
		   \bar{\psi}^{(i)} \overleftarrow{\CW}\gamma_\mu \psi^{(j)} ) \big)
\nonumber \\
(T^R_{I})_{\mu\nu}& \equiv & Z_T^0( 1+{\tilde b}_T a {\tilde m}_{ij} ) 
                   \big( T_{\mu\nu} + a c_T ( \partial_\mu V_\nu - \partial_\nu V_\mu)
                   - a \frac14 c'_T (
                   \bar{\psi}^{(i)} i\sigma_{\mu\nu} \overrightarrow{\CW}\psi^{(j)} - 
		   \bar{\psi}^{(i)} \overleftarrow{\CW}i\sigma_{\mu\nu} \psi^{(j)} ) \big)
\nonumber \\
(P^R_{I})        & \equiv & Z_P^0( 1+{\tilde b}_P a {\tilde m}_{ij} ) 
                   \big( P - a \frac14 c'_P (
                   \bar{\psi}^{(i)} \gamma_5 \overrightarrow{\CW}\psi^{(j)} - 
		   \bar{\psi}^{(i)} \overleftarrow{\CW}\gamma_5 \psi^{(j)} ) \big)
\nonumber \\
(S^R_{I})        & \equiv & Z_S^0( 1+{\tilde b}_S a {\tilde m}_{ij} ) 
                   \big( S - a \frac14 c'_S (
                   \bar{\psi}^{(i)} \overrightarrow{\CW}\psi^{(j)} - 
		   \bar{\psi}^{(i)} \overleftarrow{\CW} \psi^{(j)} ) ) \big)
\nonumber \,,
\label{eq:impbilinears}
\end{eqnarray}
Here $(ij)$ (with $i\ne j$) specifies the flavor. 
The $Z_\CO^0$ are renormalization constants in the chiral
limit and ${\tilde m}_{ij}$ is the quark mass defined in
Eq.~(\ref{eq:cA}) using the axial Ward identity (AWI)\@.
$\overrightarrow{ \CW} \psi_j =
(\overrightarrow{\slash{D}}+m_j)\psi_j +O(a^2)$ is defined to be the
full $O(a)$ improved Dirac operator for quark flavor $j$ 
(See Appendix in Ref.~\inlinecite{LANL:Zfac:00}). This ensures that the equation 
of motion operators give rise
only to contact terms, and thus cannot change the overall
normalization $Z_\CO$.  The normalization is chosen 
such that, at tree level, $c'_\CO=1$ for all Dirac structures. 

We determine the improvement and renormalization constants using Ward
identities. When implementing these, we have a number of choices. Two
are of particular importance. First, we need to pick a discretization
of the total derivatives appearing in the improvement terms
proportional to $c_{A,V,T}$. Note that, because the derivatives are
external to the operators, rather than internal, this choice should
not impact the result for the $Z$'s or $c$'s, aside from corrections
of $O(a^2)$ which are not controlled. In fact, we will find that such
higher order corrections are largely kinematical, and can be removed
by the chiral extrapolations. Second, we need to choose the external
states. As far as we know, there are no standard choices, and so we
take either the state giving the best signal, or an average if there
are several giving similar accuracy. We then use the difference of the
results with those from other states as part of the estimate of the
uncertainty. Although this is somewhat {\em ad hoc}, it is a
well-defined \color{red}procedure as long as we make consistent choices for all
lattice spacings.
}
{\color{red} We stress that the coefficients \(\tilde b_\CO\)
differ from the \(b_\CO\) used by earlier authors.  These are 
related as 
\begin{equation}
b_\CO am_{ij}  = {\tilde b}_\CO a {\tilde m}_{ij} + O(a)
\label{eq:brel1}
\end{equation}
where $m_{ij} \equiv ( m_i + m_j)/2$ is the average bare quark
mass defined as $a m_i=1/2\kappa_i -
1/2\kappa_c$, $\kappa$ being the hopping parameter in the
Sheikholeslami-Wohlert action and $\kappa_c$ its value in the 
\color{red}chiral
limit. At the level of \(O(a)\) improvement, one has 
\begin{equation}
\tilde b_\CO = (Z_A^0 Z_S^0 / Z_P^0) b_\CO  \,.
\label{eq:brel}
\end{equation}
The analogous relation between $m$ and $\tilde m$ 
is given in Eq.~(\ref{eq:massVI}). 
}

{\color{red} In this paper we present results for those overall
normalization constants, $Z^0_{\cal O}$, that are scale independent
and the improvement constants $b_{\cal O}$, $c_{\cal O}$, and ${\tilde
c}_{\cal O}$. A detailed discussion of the methods has already been
presented in Refs.~\inlinecite{LANL:Zfac:00}
and~\inlinecite{LANL:Zfac:98}, and we do not repeat them here. The extension
of the method to full QCD has been presented in ~\inlinecite{LANL:Zfull:06}. }
Instead we concentrate on presenting the final results and new aspects
of the analyses. In particular, using three lattice spacings we are
able to significantly improve our understanding of residual
discretization and perturbative errors by comparing our results with
those obtained by the ALPHA collaboration using a non-perturbative
method based on the Schr\"odinger functional and with the predictions
of perturbation theory at one-loop order.

The remainder of this paper is organized as follows. In the next section we
describe the essential features of our simulations and the types of propagator
we use.
Section~\ref{sec:details} gives an overview of the methods we use to implement
Ward identities and a summary of the results. We then run through the results
from the different Ward
identities that are needed to calculate $c_A$ (Secs.~\ref{sec:cA} and 
\ref{sec:cAmom}
for zero and non-zero spatial momenta, respectively),
$Z^0_V$ and $b_V$ (Sec.~\ref{sec:ZV}), $c_V$ and $\tilde b_A-\tilde b_V$
(Sec.~\ref{sec:cV}), $Z^0_A$ (Sec.~\ref{sec:ZA}), 
$Z^0_P/Z^0_S$ and $\tilde b_S-\tilde b_P$ (Sec.~\ref{sec:ZPZS}), 
$\tilde b_P-\tilde b_A$ and $\tilde b_S$ (Sec.~\ref{sec:bPbA}), 
$c_T$ (Sec.~\ref{sec:cT}), and
the coefficients of the equation-of-motion operators (Sec.~\ref{sec:offshell}).
We compare our results with those of others in Sec.~\ref{sec:ALPHA}
and with one-loop perturbation theory in Sec.~\ref{sec:alphaanda}.
We close with brief conclusions in Sec.~\ref{sec:conc}.

\section{Details of Simulations}
\label{sec:parameters}
The parameters used in the simulations at the three values of $\beta$
are given in Table~\ref{tab:lattices}. The table also gives the labels
used to refer to the different simulations.  For the lattice scale $a$
we have taken the value determined in
Ref.~\inlinecite{Guagnelli:Scale:98} using $r_0$ as it does not rely
on the choice of the fermion action for a given $\beta$.
The values of the hopping parameter $\kappa$, along with the
corresponding results for the quark mass $a{\tilde m}$, determined
using the Axial Ward Identity (AWI), and $aM_\pi $ are given in
Table~\ref{tab:qmasses} .

\begin{table}
\begin{center}
\begin{tabular}{|l|c|c|c|c|c|c|c|}
\hline
\multicolumn{1}{|c|}{Label}&
\multicolumn{1} {c|}{$\beta$}&
\multicolumn{1} {c|}{$c_{SW}$}&
\multicolumn{1} {c|}{$a^{-1}$ (GeV)}&
\multicolumn{1} {c|}{Volume}  &
\multicolumn{1} {c|}{$L$ (fm)}&
\multicolumn{1} {c|}{Confs.}  &
\multicolumn{1} {c|}{$x_4$}  \\
\hline       	     
{\bf 60NPf}  & 6.0 & 1.769  & 2.12      & $16^3 \times 48$  & 1.5  & 125      & $4 - 18$  \\
{\bf 60NPb}  &     &        &           &                   &      & 112      & $27 - 44$ \\
\hline       	     
{\setbox0=\vbox{\hrule width 0pt\relax
 \vskip 5pt\hbox{\bf 62NP }}\dp0=0pt\relax
 \ht0=0pt\relax\box0}
& 6.2 & 1.614  & 2.91      & $24^3 \times 64$  & 1.65 & 70       & $6 - 25$  \\
             &     &        &           &                   &      & 70       & $39 - 58$ \\
\hline       	     
{\bf 64NP }  & 6.4 & 1.526  & 3.85      & $24^3 \times 64$    & 1.25 & 60       & $8 - 56$  \\
\hline    
\end{tabular}
\end{center}
\caption{Simulation parameters and statistics. $x_4$ denotes the time
interval over which the chiral rotation is performed in the AWI\@. The
initial Wuppertal source is placed at $t=0$.}
\label{tab:lattices}
\end{table}

Four major changes have been made in the analysis compared to
our previous work~\cite{LANL:Zfac:00}. 
First, the addition of the data set at $\beta=6.4$ to
those at $\beta=6.0$ and $6.2$ (the latter two being unchanged
from Ref.~\inlinecite{LANL:Zfac:00}) allows the
identification of higher order contributions in the chiral
extrapolations. As a result we now use quadratic or linear fits in the
chiral extrapolations for all three $\beta$ values as opposed to
the linear or constant fits used in \cite{LANL:Zfac:00}.  Second, the
improvement in the signal with increasing $\beta$ allows us to better
determine which values of $\kappa$ to keep in the fits. We are able to
use all seven values, $\kappa_1-\kappa_7$, at $\beta=6.2$ and
$6.4$ whereas $\beta=6.0$ data at $\kappa=\kappa_7$ are too noisy (no
clear plateaus in the ratios of correlators), and in some cases even
the data at $\kappa=\kappa_6$ are too noisy to include in the fits.

The third improvement is with respect to the discretization of the
derivatives in the operators. 
As in Refs.~\inlinecite{LANL:Zfac:98,LANL:Zfac:00}, we use
two discretization schemes in order to estimate the size
of $O(a^2)$ uncertainties.
Most of our central values come from the ``two-point scheme'' 
(which is changed from
Refs.~\inlinecite{LANL:Zfac:98,LANL:Zfac:00}).
This uses two-point discretization\footnote{%
$f(x+0.5a) \to [f(x+a)+f(x)]/2$, \\
$\partial_x f(x+0.5a) \to [f(x+a)-f(x)]/a$ and \\
$\partial_x^2 f(x+0.5a) \to [f(x+2a)-f(x+a)-f(x)+f(x-a)]/(2a^2)$.
}
throughout the calculation, i.e. both in the axial rotation of the
action, $\delta S$, and in the operators.  It improves upon the scheme
with the same name that we used in
refs.~\inlinecite{LANL:Zfac:98,LANL:Zfac:00}, in which we only used
two-point discretization in the calculation of $c_A$ and $\delta S$,
but all other operators were discretized using three-point
discretization.

We estimate discretization errors using a hybrid
scheme in which we use
three-point discretization\footnote{%
$\partial_x f(x) \to [f(x+a)-f(x-a)]/(2a)$, and \\
$\partial_x^2 f(x) \to [f(x+a)- 2 f(x) + f(x-a)]/a^2$.
}
in all the operators but retain the two-point discretization
in $\delta S$ (using the corresponding two-point values for
$c_A$ and $\widetilde m$).
 We refer to this as the ``three-point scheme''.
We did not use three-point derivatives in the
discretization of $\delta S$ in the present calculation for
reasons of computational cost.
We stress, however, that both schemes have 
errors starting at $O(a^2)$. By comparing them we
obtain information about about the size of these errors.
Further details on the two schemes are explained later.

Lastly, we have also added the calculation of $c_A$ using a ``four-point''
discretization of derivatives\footnote{%
$f(x+0.5a) \to (9[f(x+a)+f(x)] - [f(x+2a)+f(x-a)])/16$, \\
$\partial_x f(x+0.5a) \to (9([f(x+a)-f(x)]-[f(x+2a)-f(x-a)]/27)/8a$ and \\
$\partial_x^2 f(x+0.5a) \to [f(x+2a)-f(x+a)-f(x)+f(x-a)]/(2a^2)$.
}
which is improved to $O(a^3)$ at the classical level. 
This allows us to further study discretization errors.

The fourth improvement is in the definition of the central value $x$
obtained from the jackknife fits. We now include an $O(1/N)$
correction in the single elimination jackknife
procedure~\cite{Jackknife:02} and define
\begin{equation}
x = {\overline x} + N(x_0 - {\overline x})
\label{eq:jackknife}
\end{equation}
where ${\overline x} = \sum_N x_{jk}/N$ is the uncorrected (and
previously used) estimate, $N$ is the sample size, and $x_0$ is the
result of the fit to the full data sample. 

The reanalysis changes many of the results presented in
\cite{LANL:Zfac:00}. The most significant changes 
(with final results changing by more than $1\sigma$) arise
from the order and range of the fit used (for example, 
changing from linear to quadratic extrapolation). The other changes
in the analysis lead to smaller changes in the final results.
We comment below on the changes at appropriate
places. Because of these changes we present here estimates from all four
sets of simulations listed in Table~\ref{tab:lattices}, and these
revised estimates supersede previously published numbers.

To highlight the improvement in the signal in various ratios of
correlation functions with $\beta$, we include in
Figures~\ref{fig:cVfit}, \ref{fig:cVfit1}, \ref{fig:cVfit2},
\ref{fig:ZAcomp}, \ref{fig:ZPZS}, \ref{fig:cT2}, and \ref{fig:cT3}
previous data from {\bf 60NP} and {\bf 62NP} sets for comparison.
Whereas the signal is marginal at $\beta=6.0$, it improves rapidly,
and by $\beta=6.4$ reliable estimates for all constants can be
obtained with $O(100)$ independent configurations.

For each set of simulation parameters the quark propagators are
calculated using Wuppertal smearing~\cite{LANL:HMD2:91}.  The hopping
parameter in the 3-dimensional Klein-Gordon equation used to generate
the gauge-invariant smearing is set to $0.181$, which gives mean
squared smearing radii of $(r/a)^2 \approx 2.9$, $3.9$, and $5.4$ for
$\beta=6.0$, $6.2$, and $6.4$ respectively.

In Table~\ref{tab:lattices} we also show the time extent of the region
of chiral rotation in the three-point axial Ward identities.
The dependence of our results on this region was investigated
at $\beta=6.0$, as shown by the two different time intervals
listed under {\bf 60NPf} and {\bf 60NPb}. We
observed no significant difference in the two results, so for our
final results we average the two values weighted by their errors.  In
the {\bf 62NP} calculation, we used two separate rotation regions with
equal time extent and placed symmetrically about the source. This
allowed us to average the correlation functions to improve the
statistical sample. In the {\bf 64NP} data set we were able to further
improve the efficiency of the method by enlarging the region of
insertion to include the whole lattice except for a few time slices
placed symmetrically on either side of the source for the original
propagator at $t=0$. This construct allows us to average the signal
from forward and backward propagation with a single insertion region,
(time slices $8-56$), and reduces the computational time
significantly because only five inversions are required instead of the
eight needed in the {\bf 60NP} and {\bf 62NP} studies 
(where forward and
backward propagating correlators were calculated separately).

The five kinds of propagators we use in our calculation
at $\beta=6.4$ are as follows.  The 
initial quark propagator is calculated with a Wuppertal source on
time-slice $t = 0$ for all the lattices.  To make explicit the
construction of sources for propagators with insertions we label the
two ends of the time integration region by $(t_i,t_f)$, which for {\bf
64NP} data are $t_i=8$ and $t_f=56$ as listed in
Table~\ref{tab:lattices}. We define the insertion operator $\delta S$
using the two-point discretization of the derivatives, whereby the
discretizations for the three terms in $\delta S = 2 \tilde m P - \partial_4
A_4 - a c_A \partial^2_4 P$ are
\begin{eqnarray}
\int d^4x \partial_4 A_4 &\to& \int d^3x [A_4(t_f,x) - A_4(t_i,x)] \,, \nonumber \\
\int d^4x \partial^2_4 P &\to& \int d^3x 
[ P(t_f+1) - P(t_f-1) - P(t_i+1) + P(t_i-1) ]/2 \,, 
\label{eq:inversions}
\\
\int d^4x P &\to& \int d^3x [P(t_i)/2 + P(t_i+1) + 
\ldots + P(t_f-1) + P(t_f)/2] \,. \nonumber
\end{eqnarray}
Starting with the original Wuppertal source propagator we construct
the three quantities defined in Eq.~\ref{eq:inversions} and use these
as sources to create the propagators with insertions. 
The final, fifth, propagator is calculated by inserting $\gamma_5$ at
zero 3-momentum on time slice $t= 23$, $20$ and $24$ respectively for
the three $\beta$ values. This is needed to study the vector Ward
identity used to extract $Z_V$. 

The quark and
antiquark in the operators in $\delta S$, which have flavors
we call  ``1'' and ``2'' respectively, are always taken to be
degenerate, i.e. $m_1 = m_2$.  This choice is made for computational
simplicity.


\section{Overview of Methodology and Results}
\label{sec:details}

In this section we discuss technical details relevant to the implementation
of all the Ward identities, and give a summary of our results.

The Ward identities can be implemented on states having any spatial momentum,
and we collected data for ($0,\ 1,\ \sqrt2,\ \sqrt3,\ 2$) units of lattice momenta. 
In the extraction of
$c_A$ we find that the results from all momenta are consistent,
but only after errors proportional to
$(pa)^2$ are taken into account.  Because of these additional
discretization errors, and the larger statistical errors in
correlators with non-zero spatial momentum, we did not find that the
results at non-zero momenta added useful information. Thus, for the calculation
of all other renormalization and improvement coefficients we present results
only from correlators with zero spatial momentum.

We were unable to determine the covariance matrix to sufficient accuracy
to do fully correlated fits. Thus, when fitting the time dependence of correlators,
or ratios of correlators, we use
only the diagonal part of the covariance matrix. Similarly,
fits to the quark mass dependence (which are done
within the jackknife procedure) ignore correlations between the
results at different masses.
Also, in the analysis of the three-point axial
WI identities we do not propagate the errors associated with estimates
of $c_A$ and $c_V$ as we do not have a corresponding error estimate
on each jackknife sample. The fully self-consistent method would be to do a
simultaneous fit to all the unknown parameters, but we do not have enough
statistical power to do this. 
Because of these shortcomings, we can
make no quantitative statement about goodness of fit. Nevertheless,
assuming that the fits are good, the errors in the fit parameters,
which are obtained using the jackknife procedure, should be reliable.

\begin{figure}[tbp]  
\begin{center}
\includegraphics[width=0.7\hsize]{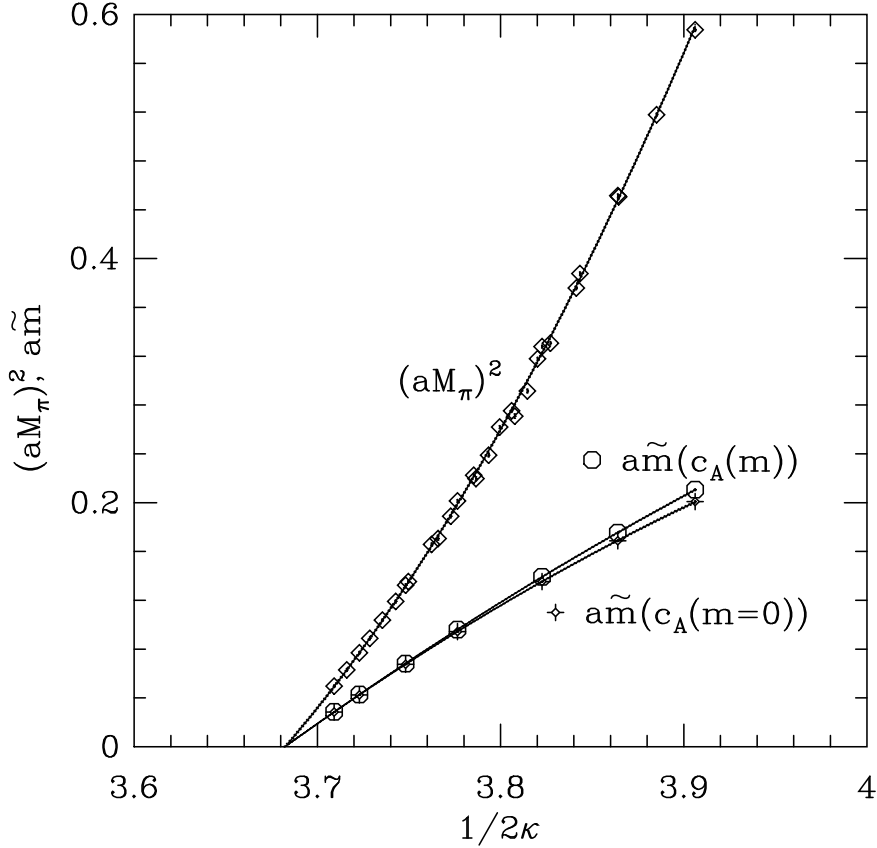}
\end{center}
\caption{Fits used to determine $\kappa_c$ by extrapolating {\bf 64NP} 
results for
${\tilde m}$ and $M_\pi^2$.  We show quadratic fits to the 
two-point version of ${\tilde m}$
for the two cases discussed in text (octagons label points with
$c_A({\tilde m})$ and pluses label points with chirally
extrapolated $c_A$), and a quadratic fit to $M_\pi^2$ (diamonds).}
\label{fig:kappac}
\end{figure}

%
\begin{table}
\begin{center}
\advance\tabcolsep by -1pt
\begin{tabular}{|c|c|c|c|c|c|c|c|c|c|}
\hline
\multicolumn{1}{|c|}{}&
\multicolumn{3} {c|}{{\bf 60NP}}&
\multicolumn{3} {c|}{{\bf 62NP}}&
\multicolumn{3} {c|}{{\bf 64NP}}\\
\multicolumn{1}{|c|}{Label}  &
\multicolumn{1} {c|}{$\kappa$}  &
\multicolumn{1} {c|}{$a{\tilde m}$}  &
\multicolumn{1} {c|}{$aM_\pi $}  &
\multicolumn{1} {c|}{$\kappa$}  &
\multicolumn{1} {c|}{$a{\tilde m}$}  &
\multicolumn{1} {c|}{$aM_\pi$}  &
\multicolumn{1} {c|}{$\kappa$}  &
\multicolumn{1} {c|}{$a{\tilde m}$}  &
\multicolumn{1} {c|}{$a M_\pi $}  \\
\hline    
%
%
%
%
$\kappa_1$ & $0.1300$ & $0.1442(10)$ & $0.711( 2)$ &  $0.1310$ & $0.1345(6)$  & $0.609(1)$ & $0.1280$ & $0.2106(6)$ & $0.766(1)$ \\
$\kappa_2$ & $0.1310$ & $0.1182(08)$ & $0.631( 2)$ &  $0.1321$ & $0.1053(4)$  & $0.522(1)$ & $0.1294$ & $0.1754(5)$ & $0.672(1)$ \\
$\kappa_3$ & $0.1320$ & $0.0913(06)$ & $0.544( 2)$ &  $0.1333$ & $0.0728(3)$  & $0.418(1)$ & $0.1308$ & $0.1391(4)$ & $0.573(1)$ \\
$\kappa_4$ & $0.1326$ & $0.0752(05)$ & $0.488( 2)$ &  $0.1339$ & $0.0562(2)$  & $0.360(2)$ & $0.1324$ & $0.0960(2)$ & $0.449(1)$ \\
$\kappa_5$ & $0.1333$ & $0.0561(04)$ & $0.416( 2)$ &  $0.1344$ & $0.0419(2)$  & $0.307(2)$ & $0.1334$ & $0.0682(2)$ & $0.364(1)$ \\
$\kappa_6$ & $0.1342$ & $0.0308(04)$ & $0.308( 3)$ &  $0.1348$ & $0.0306(2)$  & $0.261(2)$ & $0.1343$ & $0.0429(1)$ & $0.278(2)$ \\
$\kappa_7$ & $0.1345$ & $0.0236(35)$ & $0.265(12)$ &  $0.1350$ & $0.0248(1)$  & $0.235(2)$ & $0.1348$ & $0.0285(1)$ & $0.223(2)$ \\
$\kappa_c^{(1)}$ &$0.13528(2)$  &         &  & 
                  $0.135854(5)$ &         &  &
		  $0.135786(3)$ &         &    \\
$\kappa_c^{(2)}$ &$0.13530(1)$  &         &  &  
                  $0.135875(4)$ &         &  &
		  $0.135784(3)$ &         &    \\
$\kappa_c^{(3)}$ &$0.13539(3)$  &         &  & 
                  $0.13594(2)$  &         &  &
		  $0.13578(2)$  &         &    \\
\hline    
\end{tabular}
\end{center}
\caption{Values of the hopping parameter used in the various
simulations, and the corresponding pseudoscalar mass $a M_\pi$ and
quark mass $a \tilde m$ defined using the $c_A(\tilde m)$ and
two-point discretization (see
Sec.~\protect\ref{sec:cA}).  
The three estimates of $\kappa_c$,
obtained using quadratic fits, correspond to (1) the zero of $\tilde
m$ with mass dependent $c_A$, (2) the zero of $\tilde m$ with chirally
extrapolated $c_A$, and (3) the zero of $M_\pi^2$.}
\label{tab:qmasses}
\end{table}

In Table~\ref{tab:qmasses} we give our results for the critical
hopping parameter $\kappa_c$, which
is needed to define the vector Ward identity (VWI) quark mass $m$. 
These are obtained from the fits shown in
Fig.~\ref{fig:kappac} (for the {\bf 64NP} dataset).
We fit three quantities to quadratic functions of $1/2\kappa$
(which, up to an additive shift, is the tree-level quark mass).
The first quantity fit is the quark mass $\tilde m$ extracted from
the axial Ward identity, using a mass dependent $c_A (\tilde m)$ and
two-point discretization (see Section~\ref{sec:cA} below for definitions
of these quantities). This is the middle curve in the plot.
The second fit quantity is also
$\tilde m$, but now obtained using the chirally extrapolated value
of $c_A$. This is the lower curve in the plot.
Finally, the third quantity fit is $M_\pi^2$, and gives the upper curve in the plot.
Only the last fit includes results for both
degenerate and non-degenerate quarks,
using the average value of $1/\kappa$ for the latter.
As the curve shows, we find no noticeable dependence on the mass difference. 
The respective fit parameters are
\begin{eqnarray}
a \tilde m (c_A(\tilde m)) &=& -12.24(13) 
+ 5.57(7) \frac{1}{2\kappa} - 0.610(9) \frac{1}{(2\kappa)^2} \,, \nonumber \\
a \tilde m (c_A(0))        &=& -15.12(11) 
+ 7.13(6) \frac{1}{2\kappa} - 0.822(8) \frac{1}{(2\kappa)^2} \,, \\
a^2 M_\pi^2                &=& +48.3(9)   
- 27.9(5) \frac{1}{2\kappa} + 4.03(6)  \frac{1}{(2\kappa)^2} \,.  \nonumber 
\label{eq:kcfits}
\end{eqnarray}
From these we get three estimates of $\kappa_c$ which we find to be
consistent; this was not the case for {\bf 60NP} and {\bf 62NP} data.
The first estimate, $\kappa_c^{(1)}$, is the most direct as $\tilde m$ and
$c_A(\tilde m)$ are extracted together from the same two-point Ward
identity (also see below), so we use it in subsequent analyses and,
henceforth, drop the superscript.

If both $M_\pi$ and $\tilde m$ are extracted from fits that include
large times, where only the ground state survives, then it follows
from eq.~\ref{eq:cA} below that $2 \tilde m \equiv M_\pi^2(2/B_\pi + a
c_A) + O(a^2)$, with $B_\pi \propto \langle 0|P|\pi\rangle/f_\pi$ a
quantity which is non-zero in the chiral limit (and which we will use
in several places below).  Thus $M_\pi$ and $\tilde m$ should vanish
at the same point.  {\color{red} We use this fact to test the adequacy
of our quadratic fits of $M_\pi^2$ versus $\tilde m (c_A(\tilde m))$
or $\tilde m (c_A(0))$. The {\bf 64NP}} {\color{red} data, illustrated for two-point
discretization of derivatives, give significant intercepts:}
\begin{eqnarray}
a^2 M_\pi^2 &=&   0.0090(23) + 2.61(5) a \tilde m + 
5.84(27) (a \tilde m)^2 \quad (c_A(m), \beta=6.0) \,, \nonumber \\
a^2 M_\pi^2 &=&   0.0098(27) + 2.54(7) a \tilde m + 
7.48(40) (a \tilde m)^2 \quad (c_A(0), \beta=6.0) \,,  \nonumber \\
a^2 M_\pi^2 &=&   0.0049(13) + 1.87(3) a \tilde m + 
6.32(15) (a \tilde m)^2 \quad (c_A(m), \beta=6.2) \,, \nonumber \\
a^2 M_\pi^2 &=&   0.0043(13) + 1.86(3) a \tilde m + 
7.24(17) (a \tilde m)^2 \quad (c_A(0), \beta=6.2) \,,  \\
a^2 M_\pi^2 &=&   0.0020(12) + 1.49(2) a \tilde m + 
6.11(09) (a \tilde m)^2 \quad (c_A(m), \beta=6.4) \,, \nonumber \\
a^2 M_\pi^2 &=&   0.0059(12) + 1.34(2) a \tilde m + 
7.73(12) (a \tilde m)^2 \quad (c_A(0), \beta=6.4) \,. \nonumber
\end{eqnarray}
Using $c_A(\tilde m)$ leads to smaller intercepts, and because of this
we use $a \tilde m(c_A(\tilde m))$ (rather than $a\tilde m(c_A(0))$)
when making chiral extrapolations in the subsequent analyses.  We
note, however, that the intercept is not small when converted into
physical units ($\sim (160\ {\rm MeV})^2$), and does not show any
significant decrease with $a$.  In this context it is important to
note that the range of the fits in physical units is different in the
three cases and the lightest ``pions'' are heavy. The range of pion
masses in the three cases are $550 - 1500$, $680 - 1770$ and $850 -
2900$ MeV respectively.  Thus neglected contributions from chiral
logarithms, which become significant only at lower quark masses, and
higher order terms in the chiral expansion, could account for the
intercept.  {\color{red} Since the present data are well fit by a
quadratic, we cannot empirically resolve the issue of what additional
terms need to be included in the fits.}

An important point to keep in mind is that
the extrapolations to extract renormalization and improvement
constants are in $a \tilde m$, and are different from
the usual chiral extrapolations where the control parameter is
$M_\pi^2/\Lambda_{\chi}^2$ with $\Lambda_{\chi} \sim 1$ GeV. 
We do not need to be in the chiral regime for our method to work.
In the ratios of correlators that appear in the Ward identities we use,
the same intermediate states contribute to both numerator and denominator,
and possible non-analytic behavior in the quark mass (including that
from enhanced quenched chiral logarithms) cancels.
What matters is that $a \tilde m \ll 1$, which, as can be seen
from Table~\ref{tab:qmasses}, is reasonably well satisfied for all of our masses.
 Indeed, a striking
feature of our results is that the quadratic fits we use
work very well over our entire range of quark masses.

%
%
\newcommand{\rZVa}{\ref{eq:ZVfitmtilde}}
\newcommand{\rZVb}{\ref{eq:ZVfitm}}
\newcommand{\rZAa}{\ref{ZAZV-1}}
\newcommand{\rcV}{\ref{cV}}

\begin{table}[!ht]
\setlength{\tabcolsep}{4pt}
\renewcommand{\arraystretch}{1.1}
\begin{center}
\begin{tabular}{|c|c|c|c|c|}
\hline
                                                 & {\bf 60NPf}	    & {\bf 60NPb}      & {\bf 62NP}      & {\bf 64NP}  	   \\
\hline
$c_A$                                            & $ -0.039 ( 08) $ & $ -0.037 ( 09) $ & $ -0.034 ( 03) $ & $ -0.032 ( 03) $ \\
\hline						                                      	                                    
$Z^0_V$                                          & $ +0.7689( 08) $ & $ +0.7703( 09) $ & $ +0.7880( 04) $ & $ +0.8033( 05) $ \\
$\tilde b_V$                                     & $ +1.448 ( 20) $ & $ +1.413 ( 23) $ & $ +1.273 ( 10) $ & $ +1.212 ( 11) $ \\
$Z^0_V$                                          & $ +0.7689( 08) $ & $ +0.7697( 09) $ & $ +0.7876( 03) $ & $ +0.8016( 05) $ \\
$b_V$                                            & $ +1.530 ( 12) $ & $ +1.519 ( 13) $ & $ +1.402 ( 08) $ & $ +1.370 ( 09) $ \\
\hline						                                      	                                    
$Z^0_V$                                          & $ +0.773 ( 05)*$ & $ +0.761 ( 06)*$ & $ +0.790 ( 02)*$ & $ +0.801 ( 02) $ \\
$\tilde b_A-\tilde b_V$                          & $ -0.309 ( 76)*$ & $ -0.469 ( 77)*$ & $ -0.096 ( 31)*$ & $ -0.123 ( 54) $ \\
$Z^0_V$                                          & $ +0.774 ( 05)*$ & $ +0.762 ( 06)*$ & $ +0.791 ( 02)*$ & $ +0.800 ( 02) $ \\
$b_A-b_V$                                        & $ -0.288 ( 69)*$ & $ -0.433 ( 71)*$ & $ -0.095 ( 29)*$ & $ -0.130 ( 49) $ \\
$Z^0_V/(Z^0_A)^2$                                & $ +1.203 ( 11)*$ & $ +1.209 ( 13)*$ & $ +1.191 ( 06)*$ & $ +1.173 ( 05) $ \\
$\tilde b_A-\tilde b_V$                          & $ -0.094 ( 89)*$ & $ +0.016 ( 96)*$ & $ -0.075 ( 75)*$ & $ -0.131 ( 65) $ \\
$Z^0_A$                                          & $ +0.799 ( 04)*$ & $ +0.794 ( 05)*$ & $ +0.818 ( 04) $ & $ +0.825 ( 02) $ \\
$Z^0_P/Z^0_AZ^0_S$                               & $ +1.052 ( 10)*$ & $ +1.062 ( 12)*$ & $ +1.084 ( 05)*$ & $ +1.089 ( 04) $ \\
$\tilde b_P-\tilde b_S$                          & $ -0.058 ( 65)*$ & $ -0.178 ( 52)*$ & $ -0.096 ( 25)*$ & $ -0.104 ( 56) $ \\
$c_T$                                            & $ +0.083 ( 12)*$ & $ +0.088 ( 12)*$ & $ +0.063 ( 10)*$ & $ +0.054 ( 05)*$ \\
\hline						                                                                            
$Z^0_P/Z^0_AZ^0_S$  [$c_A(m)$]                   & $ +1.051 ( 10)*$ & $ +1.057 ( 12)*$ & $ +1.084 ( 05)*$ & $ +1.077 ( 02) $ \\
$\tilde b_A-\tilde b_P-\tilde b_m$ [$c_A(m)$]    & $ +0.598 ( 43)*$ & $ +0.629 ( 49)*$ & $ +0.674 ( 21)*$ & $ +0.511 ( 09) $ \\
$-2\tilde b_m$      [$c_A(m)$]                   & $ +1.052 ( 72)*$ & $ +1.251 ( 81)*$ & $ +1.313 ( 27)*$ & $ +1.193 ( 15) $ \\
$Z^0_P/Z^0_AZ^0_S$  [$c_A(0)$]                   & $ +1.055 ( 09)*$ & $ +1.060 ( 10)*$ & $ +1.090 ( 05)*$ & $ +1.077 ( 03) $ \\
$\tilde b_A-\tilde b_P-\tilde b_m$ [$c_A(0)$]    & $ +0.943 (114)*$ & $ +0.932 (137)*$ & $ +0.957 ( 27)*$ & $ +0.646 ( 28) $ \\
$-2\tilde b_m$      [$c_A(0)$]                   & $ +1.406 (107)*$ & $ +1.472 (124)*$ & $ +1.428 ( 28)*$ & $ +1.346 ( 13) $ \\
\hline
\end{tabular}
\vspace{10pt}
\caption{Summary of results for the different combinations of renormalization 
and improvement constants extracted using the two-point derivative.
The horizontal lines separate the extraction of quantities using the
divergence of axial current, the conservation of charge, three-point
axial chiral Ward identities, and the relation between quark masses
given in Eq.~\protect\ref{eq:massVI}.  All unmarked estimates are
based on quadratic fits in both $\tilde m_1 \equiv \tilde m_2$ and
$\tilde m_3$.  Asterisks mark values extracted using linear
extrapolations in both $\tilde m_1 \equiv \tilde m_2$ and $\tilde m_3$. 
All seven masses are used at $\beta=6.2$ and $6.4$ while
at $\beta=6.0$ the lightest quark ($\kappa_7$) is dropped.
Labels $c_A(m)$ and $c_A(0)$ refer, respectively,
to whether the mass dependent or chirally extrapolated value
of $c_A$ is used in the analysis.}
\label{tab:2ptdata}
\end{center}
\end{table}
%


\begin{table}[!ht]
\setlength{\tabcolsep}{4pt}
\renewcommand{\arraystretch}{1.1}
\begin{center}
\begin{tabular}{|c|c|c|c|c|}
\hline
                                                 & {\bf 60NPf}	    & {\bf 60NPb}     & {\bf 62NP}      & {\bf 64NP}  	\\
\hline
$c_A$                                            & $ -0.036 ( 16) $ & $ -0.038 ( 18) $ & $ -0.040 ( 05) $ & $ -0.035 ( 03) $ \\
\hline						                                      	                                    
$Z^0_V$                                          & $ +0.7677( 25) $ & $ +0.7679( 31) $ & $ +0.7877( 04) $ & $ +0.8027( 06) $ \\
$\tilde b_V$                                     & $ +1.492 ( 68) $ & $ +1.499 ( 83) $ & $ +1.296 ( 10) $ & $ +1.233 ( 11) $ \\
$Z^0_V$                                          & $ +0.7672( 21) $ & $ +0.7678( 23) $ & $ +0.7875( 03) $ & $ +0.8016( 05) $ \\
$b_V$                                            & $ +1.533 ( 13) $ & $ +1.522 ( 14) $ & $ +1.402 ( 08) $ & $ +1.370 ( 09) $ \\
\hline						                                      	                                    
$Z^0_V$                                          & $ +0.771(16)@$ & $ +0.760(17)@$ & $ +0.785(03)*$ & $ +0.797(02) $ \\
$\tilde b_A-\tilde b_V$                          & $ -0.234(61)@$ & $ -0.359(57)@$ & $ -0.053(30)*$ & $ -0.174(56) $ \\
$Z^0_V$                                          & $ +0.773(16)@$ & $ +0.762(17)@$ & $ +0.785(03)*$ & $ +0.797(02) $ \\
$b_A-b_V$                                        & $ -0.210(56)@$ & $ -0.324(54)@$ & $ -0.053(29)*$ & $ -0.168(51) $ \\
$Z^0_V/(Z^0_A)^2$                                & $ +1.198(08)@$ & $ +1.196(08)@$ & $ +1.189(05)*$ & $ +1.173(04) $ \\
$\tilde b_A-\tilde b_V$                          & $ -0.079(74)@$ & $ -0.002(83)@$ & $ -0.057(69)*$ & $ -0.009(56) $ \\
$Z^0_A$                                          & $ +0.800(04)@$ & $ +0.797(04)@$ & $ +0.814(02)*$ & $ +0.824(02) $ \\
$Z^0_P/Z^0_AZ^0_S$                               & $ +1.051(08)@$ & $ +1.062(10)@$ & $ +1.084(05)*$ & $ +1.087(04) $ \\
$\tilde b_P-\tilde b_S$                          & $ +0.003(42)@$ & $ -0.142(36)@$ & $ -0.096(25)*$ & $ -0.033(55) $ \\
$c_T$                                            & $ +0.087(11)@$ & $ +0.085(12)@$ & $ +0.071(10)*$ & $ +0.058(05)*$ \\
\hline						                                      	                                    
$Z^0_P/Z^0_AZ^0_S$  [$c_A(m)$]                   & $ +1.011 ( 27)*$ & $ +1.009 ( 26)*$ & $ +1.071 ( 05)*$ & $ +1.074 ( 03) $ \\
$\tilde b_A-\tilde b_P-\tilde b_m$ [$c_A(m)$]    & $ +0.513 (110)*$ & $ +0.470 ( 89)*$ & $ +0.617 ( 27)*$ & $ +0.520 ( 14) $ \\
$-2\tilde b_m$      [$c_A(m)$]                   & $ +1.120 (104)*$ & $ +1.168 ( 92)*$ & $ +1.196 ( 34)*$ & $ +1.183 ( 19) $ \\
$Z^0_P/Z^0_AZ^0_S$  [$c_A(0)$]                   & $ +1.014 ( 23)*$ & $ +1.010 ( 26)*$ & $ +1.068 ( 05)*$ & $ +1.072 ( 03) $ \\
$\tilde b_A-\tilde b_P-\tilde b_m$ [$c_A(0)$]    & $ -0.208 (253)*$ & $ -0.209 (291)*$ & $ +0.187 ( 27)*$ & $ +0.233 ( 31) $ \\
$-2\tilde b_m$      [$c_A(0)$]                   & $ +0.917 (125)*$ & $ +0.953 (135)*$ & $ +1.150 ( 26)*$ & $ +0.918 ( 09) $ \\
\hline
\end{tabular}
\vspace{10pt}
\caption{Summary of results for the different combinations of renormalization 
and improvement constants extracted using the three-point derivative.
All unmarked estimates are based on quadratic fits in both $\tilde m_1
\equiv \tilde m_2$ and $\tilde m_3$.  Asterisks mark values extracted
using linear extrapolations in both $\tilde m_1 \equiv \tilde m_2$ and
$\tilde m_3$. At $\beta=6.2$ and $6.4$, all seven masses are
used. At $\beta=6.0$ the lightest quark $\kappa_7$ is dropped,
and, for estimates marked $@$, only masses
$\kappa_1-\kappa_5$ and linear fits are used.
Labels $c_A(m)$ and $c_A(0)$ refer, respectively,
to whether the mass dependent or chirally extrapolated value
of $c_A$ is used in the analysis.}
\label{tab:3ptdata}
\end{center}
\end{table}
%


\begin{table}[!ht]
\setlength{\tabcolsep}{1pt}
\renewcommand{\arraystretch}{0.9}
\begin{center}
\begin{tabular}{|c||c|c||c|c|}
\hline
\multicolumn{5}{|c|}{\bf 64NP} \\ \hline
	& \multicolumn{2} {c||}{\bf 2pt} & \multicolumn{2} {c|}{\bf 3pt} \\
\cline{2-5}
	& $c_A(m)$ & $c_A(0)$ & $c_A(m)$ & $c_A(0)$ \\ \hline 
 extrapolation& $ -0.167 ( 90 ) $ & $ -0.136 ( 90 ) $& $ -0.042 ( 71 ) $ & $ -0.092 ( 69 ) $ \\ \hline
 $1/m$ fit    & $ -0.079 ( 07 ) $ & $ -0.090 ( 08 ) $& $ -0.099 ( 08 ) $ & $ -0.079 ( 06 ) $ \\ \hline
 slope ratio  & $ -0.088 ( 10 ) $ & $ -0.085 ( 10 ) $& $ -0.066 ( 08 ) $ & $ -0.074 ( 08 ) $ \\ \hline
\hline
\multicolumn{5}{|c|}{\bf 62NP} \\ \hline
	& \multicolumn{2} {c||}{\bf 2pt} & \multicolumn{2} {c|}{\bf 3pt} \\
\cline{2-5}
	& $c_A(m)$ & $c_A(0)$ & $c_A(m)$ & $c_A(0)$ \\ \hline 
 extrapolation& $ -0.143 (166 ) $ & $ -0.128 (164 ) $& $ +0.008 (160 ) $ & $ -0.036 (157 ) $ \\ \hline
 $1/m$ fit    & $ -0.104 ( 17 ) $ & $ -0.124 ( 20 ) $& $ -0.161 ( 20 ) $ & $ -0.120 ( 19 ) $ \\ \hline
 slope ratio  & $ -0.116 ( 23 ) $ & $ -0.117 ( 23 ) $& $ -0.103 ( 21 ) $ & $ -0.105 ( 21 ) $ \\ \hline
\hline
\multicolumn{5}{|c|}{\bf 60NPf} \\ \hline
	& \multicolumn{2} {c||}{\bf 2pt} & \multicolumn{2} {c|}{\bf 3pt} \\
\cline{2-5}
	& $c_A(m)$ & $c_A(0)$ & $c_A(m)$ & $c_A(0)$ \\ \hline
 extrapolation& $ -0.058 (157 ) $ & $ -0.075 (170 ) $& $ +0.158 (115 ) $ & $ -0.015 (188 ) $ \\ \hline
 $1/m$ fit    & $ -0.138 ( 25 ) $ & $ -0.189 ( 47 ) $& $ -0.230 ( 43 ) $ & $ -0.143 ( 58 ) $ \\ \hline
 slope ratio  & $ -0.136 ( 23 ) $ & $ -0.143 ( 28 ) $& $ -0.092 ( 16 ) $ & $ -0.102 ( 16 ) $ \\ \hline
\hline
\multicolumn{5}{|c|}{\bf 60NPb} \\ \hline
	& \multicolumn{2} {c||}{\bf 2pt} & \multicolumn{2} {c|}{\bf 3pt} \\
\cline{2-5}
	& $c_A(m)$ & $c_A(0)$ & $c_A(m)$ & $c_A(0)$ \\ \hline
 extrapolation& $ +0.068 (162 ) $ & $ +0.112 (186 ) $& $ -0.087 (156 ) $ & $ -0.036 (211 ) $ \\ \hline
 $1/m$ fit    & $ -0.132 ( 23 ) $ & $ -0.176 ( 44 ) $& $ -0.221 ( 43 ) $ & $ -0.139 ( 60 ) $ \\ \hline
 slope ratio  & $ -0.123 ( 27 ) $ & $ -0.128 ( 35 ) $& $ -0.089 ( 25 ) $ & $ -0.104 ( 26 ) $ \\ \hline
\hline
\end{tabular}
\vspace{10pt}
\caption{Results for $c_V$ using the two-point discretization
data. See text (sec.~\protect\ref{sec:cV}) and
Ref.~\cite{LANL:Zfac:00} for details. The labels $c_A(m)$ and $c_A(0)$
refer to whether the mass-dependent or chirally extrapolated value of
$c_A$ was used in the analysis.}
\label{tab:cV}
\end{center}
\end{table}

With $a \tilde m$ and $\kappa_c$ in hand we carry out the analysis for
two-point and three-point Ward identities discussed in
Ref.~\inlinecite{LANL:Zfac:00}. Each identity allows us to extract one or
more combinations of on-shell improvement and normalization
constants. Since many of the results for {\bf 60NP} and {\bf 62NP}
data sets given in~\cite{LANL:Zfac:00} have changed as a result
of our reanalysis, estimates from all three lattice spacings
are given in Tables~\ref{tab:2ptdata} and \ref{tab:3ptdata}. 
Similarly, a detailed comparison of the results for $c_V$ obtained using
the methods discussed in Ref.~\inlinecite{LANL:Zfac:00} is
given in Table~\ref{tab:cV} for all three values of $\beta$.

Our final results for the individual constants are collected in
Table~\ref{tab:finalcomp}.  We quote both a statistical error
(given by the single elimination jackknife procedure, in which we
repeat the entire analysis on each jackknife sample), and an estimate
of the residual $O(a)$ uncertainty.  The latter is taken to be the
difference in results obtained using two- and three-point discretizations of
the derivatives except for $b_V$ where there is no three-point estimate. 
A different estimate of the uncertainties can be obtained by comparing
our results to previous estimates by the ALPHA
collaboration~\cite{ALPHA:Zfac:97A,ALPHA:Zfac:97B,ALPHA:Zfac:98}
summarized in Table~\ref{tab:finalcomp}, by the QCDSF collaboration
given in Table~\ref{tab:ZVbV}~\cite{QCDSF:ZVbV:03} and by the SPQcdR
collaboration~\cite{SPQcdR:Z:04}.

\begin{table*}[!ht]
\setlength{\tabcolsep}{1.9pt}
\begin{tabular}{|| >{\scriptsize}c 
	|| >{\scriptsize}l | >{\scriptsize}l | >{\scriptsize}l 
	|| >{\scriptsize}l | >{\scriptsize}l | >{\scriptsize}l 
	|| >{\scriptsize}l | >{\scriptsize}l | >{\scriptsize}l ||}
\hline
\multicolumn{1}{||c||}{}&
\multicolumn{3}{c||}{\(\beta=6.0\)}&
\multicolumn{3}{c||}{\(\beta=6.2\)}&
\multicolumn{3}{c||}{\(\beta=6.4\)}\\
\hline
         & LANL              & ALPHA            & P. Th.
         & LANL              & ALPHA            & P. Th.
         & LANL              & ALPHA            & P. Th.     \\
         &                   &                  &
         &                   &                  &
         &                   &                  &                \\[-12pt]
\hline			     		       
         &                   &                  &
         &                   &                  &
         &                   &                  &                \\[-12pt]
$c_{SW}$ &  1.769            &  1.769           &  1.521
         &  1.614            &  1.614           &  1.481
         &  1.526            &  1.526           &  1.449         \\
         &                   &                  &
         &                   &                  &
         &                   &                  &                \\[-12pt]
\hline			     		       
         &                   &                  &
         &                   &                  &
         &                   &                  &                \\[-12pt]
$Z^0_V$  & $+0.7695(8)(19)$  & $+0.7809(6)$     &  $+0.810$  
         & $+0.7878(4)(2) $  & $+0.7922(4)(9)$  &  $+0.821$
         & $+0.8024(5)(2) $  & $+0.8032(6)(12)$ &  $+0.830$   \\
$Z^0_A$  & $+0.797(4)(2)$    & $+0.7906(94)$    &  $+0.829$   
         & $+0.815(2)(1)$    & $+0.807(8)(2) $  &  $+0.839$   
         & $+0.825(2)(1)$    & $+0.827(8)(1) $  &  $+0.847$     \\
$Z^0_P/Z^0_S$		     		       				    
         & $+0.840(7)(15)$   &  $+0.840(8)$     & $+0.956$  
         & $+0.883(4)(6)$    &  $+0.886(9)$     & $+0.959$
         & $+0.894(3)(4)$    &  $+0.908(9)$     & $+0.962$    \\
         &                   &                  &
         &                   &                  &
         &                   &                  &                \\[-12pt]
\hline			     		       
         &                   &                  &
         &                   &                  &
         &                   &                  &                \\[-12pt]
$c_A$    & $-0.036(8)(3)$    &  $-0.083(5)$     &  $-0.013$ 
         & $-0.034(3)(6)$    &  $-0.038(4)$     &  $-0.012$ 
         & $-0.032(3)(3)$    &  $-0.025(2)$     &  $-0.011$   \\
$c_V$    & $-0.13 (2)(3)$    &  $-0.32 (7)$     &  $-0.028$  
         & $-0.12 (2)(2)$    &  $-0.21(7)$      &  $-0.026$
         & $-0.09 (1)(2)$    &  $-0.13(5)$      &  $-0.024$   \\
$c_T$    & $+0.085(12)(1)$   &  N.A.            &  $+0.020$
         & $+0.063(10)(8)$   &  N.A.            &  $+0.019$
         & $+0.054(5)(4)$    &  N.A.            &  $+0.018$   \\
         &                   &                  &
         &                   &                  &
         &                   &                  &             \\[-12pt]
\hline			     		       
         &                   &                  &
         &                   &                  &
         &                   &                  &             \\[-11pt]
$\tilde b_V$		     		       				    
         & $+1.43 (2)(7)$    &  N.A.            &  $+1.106$ 
         & $+1.27 (1)(2)$    &  N.A.            &  $+1.099$
         & $+1.21 (1)(2)$    &  N.A.            &  $+1.093$   \\
$b_V$    & $+1.52(1)$        & $+1.48(2)$       &  $+1.274$ 
         & $+1.40(1)$        & $+1.41(2)$       &  $+1.255$
         & $+1.37(1)$        & $+1.36(3)$       &  $+1.239$   \\
$\tilde b_A-\tilde b_V$	     		       				    
         & $-0.21(6)(5) $    &  N.A.            &  $-0.002$  
         & $-0.09(4)(2) $    &  N.A.            &  $-0.002$
         & $-0.13(5)(6) $    &  N.A.            &  $-0.002$   \\
$b_A-b_V$	     		       				    
         & $-0.36(7)(13)$    &  N.A.            &  $-0.002$  
         & $-0.10(3)(8) $    &  N.A.            &  $-0.002$
         & $-0.13(5)(3) $    &  N.A.            &  $-0.002$   \\
$\tilde b_P-\tilde b_S$	     		       				    
         & $-0.12 (6)(5)$    &  N.A.            &  $-0.066$  
         & $-0.10 (3)(1)$    &  N.A.            &  $-0.062$
         & $-0.10 (6)(7)$    &  N.A.            &  $-0.059$   \\
$\tilde b_A-\tilde b_P+\tilde b_S/2$	     		       				    
         & $+0.61(5)(12)$    &  N.A.            &  $+0.585$
         & $+0.67(2)(6)$     &  N.A.            &  $+0.579$
         & $+0.51(1)(1)$     &  N.A.            &  $+0.575$   \\
$\tilde b_S$		     		       				    
         & $+1.15(8)(1)$     &  N.A.            &  $+1.172$ 
         & $+1.31(3)(12) $   &  N.A.            &  $+1.161$
         & $+1.19(2)(1) $    &  N.A.            &  $+1.151$   \\[3pt]
         &                   &                  &
         &                   &                  &
         &                   &                  &             \\[-12pt]
\hline			     		       
         &                   &                  &
         &                   &                  &
         &                   &                  &             \\[-11pt]
$\tilde b_A$		     		       				    
         & $+1.22(6)(11) $   &  N.A.            &  $+1.104$ 
         & $+1.19(4)(5) $    &  N.A.            &  $+1.097$
         & $+1.09(5)(6) $    &  N.A.            &  $+1.092$   \\
$b_A$		     		       				    
         & $+1.16(7)(10)$    &  N.A.            &  $+1.271$ 
         & $+1.31(3)(4)$     &  N.A.            &  $+1.252$
         & $+1.24(5)(4)$     &  N.A.            &  $+1.237$   \\
$\tilde b_P$		     		       				    
         & $+1.02(9)(15)$    &  N.A.            &  $+1.105$ 
         & $+1.19(4)(3) $    &  N.A.            &  $+1.099$
         & $+1.13(6)(1) $    &  N.A.            &  $+1.093$   \\
\hline
\end{tabular}
\caption{Final results for improvement and renormalization constants.
The first error in LANL estimates (this work) is statistical, and the second,
where present, corresponds to the difference between using 2-point and
3-point discretization of the derivative. We quote both \({\tilde
b}_V, {\tilde b}_A\) and \(b_V, b_A\) to simplify comparison with
previous results.  Estimates for $Z_P^0/Z_S^0$ by the ALPHA
collaboration are taken from Ref.~\cite{SPQcdR:Z:04} and the rest from 
Refs.~\cite{ALPHA:Zfac:97A,ALPHA:Zfac:97B,ALPHA:Zfac:98}. The final column
gives the results from one-loop tadpole improved perturbation 
theory~Ref.~\cite{LANL:Zfac:00}.}
\label{tab:finalcomp}
\end{table*}

We collect separately, in Table~\ref{tab:c'}, our results for the
improvement constants $c_X^\prime$, the coefficients of the
equation-of-motion operators.  These are discussed in
Sec.~\ref{sec:offshell}.  In sections~\ref{sec:ALPHA} and
\ref{sec:alphaanda} we present an analysis of residual discretization
errors by comparing our estimates with those by the ALPHA
collaboration and with one-loop tadpole improved perturbation theory
estimates summarized in Table~\ref{tab:finalcomp}.

\begin{table}
\setlength{\tabcolsep}{1pt}
\renewcommand{\arraystretch}{1.2}
\begin{center}
\begin{tabular}{|c|c|c|c|c|}
\hline
\multicolumn{1}{|c|}{}&
\multicolumn{1} {c|}{\bf 60NPf}&
\multicolumn{1} {c|}{\bf 60NPb}&
\multicolumn{1} {c|}{\bf 62NP} &
\multicolumn{1} {c|}{\bf 64NP}\\
\hline
 $c'_V+c'_P$  &  $ 2.97(26)  $ &  $ 2.66(36)  $ &  $ +2.68 (   9 )  $ &  $ 2.51(13)  $  \\
 $c'_A+c'_P$  &  $ 2.56(20)  $ &  $ 2.54(24)  $ &  $ +2.44 (  10 )  $ &  $ 2.23(12)  $  \\  
 $2c'_P    $  &  $ 2.98(41)  $ &  $ 2.76(43)  $ &  $ +2.99 (  13 )  $ &  $ 2.12(22)  $  \\   
 $c'_S+c'_P$  &  $ 2.54(14)  $ &  $ 2.54(15)  $ &  $ +2.38 (   7 )  $ &  $ 2.36(11)  $  \\   
 $c'_T+c'_P$  &  $ 2.58(19)  $ &  $ 2.48(23)  $ &  $ +2.45 (  11 )  $ &  $ 2.40(13)  $  \\ \hline
 $c'_V     $  &  $ 1.48(31)  $ &  $ 1.28(32)  $ &  $ +1.19 (  10 )  $ &  $ 1.45(13)  $  \\ 
 $c'_A     $  &  $ 1.07(27)  $ &  $ 1.16(24)  $ &  $ +0.94 (  11 )  $ &  $ 1.17(11)  $  \\ 
 $c'_P     $  &  $ 1.49(21)  $ &  $ 1.38(22)  $ &  $ +1.50 (   7 )  $ &  $ 1.06(11)  $  \\ 
 $c'_S     $  &  $ 1.05(22)  $ &  $ 1.16(19)  $ &  $ +0.89 (   9 )  $ &  $ 1.30(10)  $  \\ 
 $c'_T     $  &  $ 1.09(25)  $ &  $ 1.10(22)  $ &  $ +0.95 (  12 )  $ &  $ 1.34(12)  $  \\   
\hline
\end{tabular}
\vspace{10pt}
\caption{Results for off-shell mixing coefficients using the two-point derivative data
and $c_A(\tilde m)$.}
\label{tab:c'}
\end{center}
\end{table}

\section{Calculation of $\lc c_A$}
\label{sec:cA}
%
%

The calculation of $c_A$ exploits the two-point axial Ward identity 
\begin{equation}
\frac{ \sum_{\vec{x}} \langle 
 \partial_\mu [A_\mu + 
  a c_A \partial_\mu P]^{(ij)}(\vec{x},t) J^{(ji)}(0) \rangle} 
 {\sum_{\vec{x}} \langle P^{(ij)}(\vec{x},t) J^{(ji)}(0) \rangle} 
 = 2 {\tilde m}_{ij}   + O(a^2)\,,
\label{eq:cA}
\end{equation}
which also defines the quark mass ${\tilde m}_{ij}$.  Here the superscript
$(ij)$ refer to the mass (flavor) labels ($\kappa_i\kappa_j$) of the
quark and the antiquark. 
Up to corrections of $O(a^2)$,
this ratio of correlators should be independent
of the source $J$ and the
time \(t\) provided $c_{SW}$ (the coefficient of the
Sheikholeslami-Wohlert term in the action) and $c_A$ are tuned to their
non-perturbative values.  Since this criterion is automatically
satisfied when the correlators are saturated by a single state, the
determination of $c_A$ relies on the contribution of excited states at
small $t$.

\begin{figure}[tbp]   
\begin{center}
\includegraphics[width=0.7\hsize]{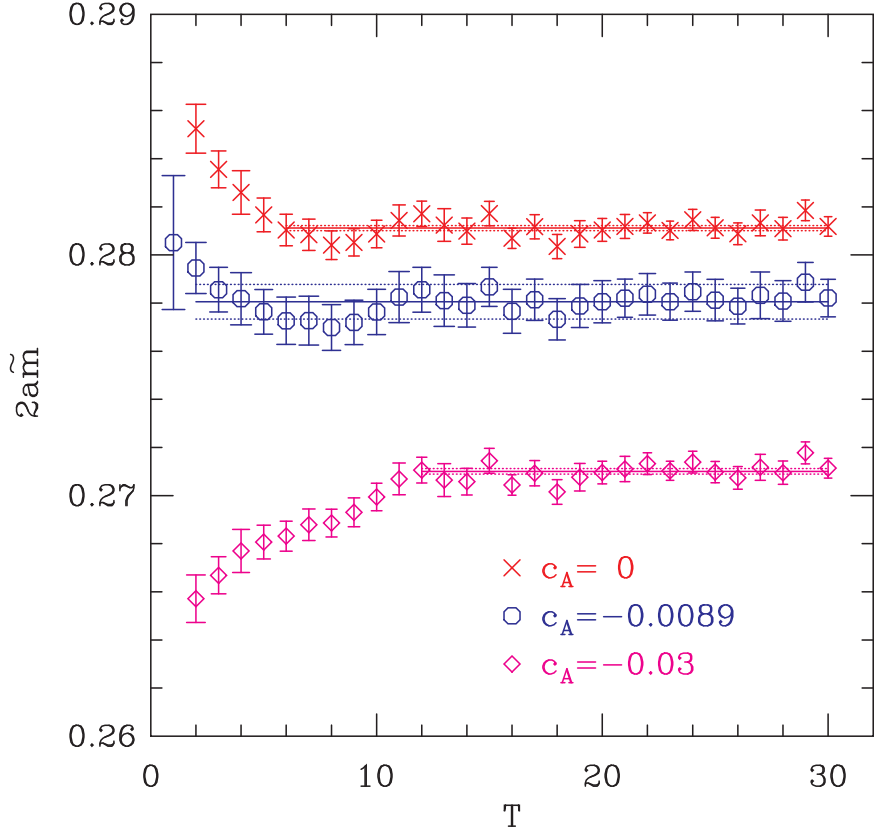}
\end{center}
\caption{Estimates of $2a {\tilde m}_{ij}$ for different values of
$c_A$ illustrated using $i=j=\kappa_3$, $J=P$, and two-point
discretization in the {\bf 64NP} data set.
For this quark mass, $c_A = -0.0089$ extends the plateau to the
earliest allowed time slice $t=2$. To show sensitivity to the tuning
we contrast this best fit with those using $c_A=0$ and $c_A=-0.03$,
the latter being close to the chirally extrapolated value.}
\label{fig:cAtune64}
\end{figure}

The sensitivity of the ratio (\ref{eq:cA}) 
to $c_A$ is illustrated for the {\bf 64NP} data in Fig~\ref{fig:cAtune64}. 
We find that, for $J=P$, the contribution
of higher excited states is significant only at time-slices $t=1-5$
for all three values of $\beta$ (see \cite{LANL:Zfac:00} for data at
$\beta=6.0$ and $6.2$).  For two-point and four-point discretization,
the data at $t=1$ cannot be used to extract
$c_A$ as the discretization of $\partial_4^2 P(t=1)$ in
Eq.~\ref{eq:cA} overlaps with the source at time slice
$t=0$. Consequently, only the range $2 \le t \le 5$ is sensitive to
tuning $c_A$ and we choose $c_A$ to make
$\tilde m_{ij}$ as flat as possible for timeslices $t \ge 2$.
This is done by minimizing the $\chi^2$ for a fit to a constant, as
illustrated in Fig.~\ref{fig:cAtune64}. 
For three-point discretization we can
implement the same choice only for $t \ge 3$.  The $J=A_4$ data are
not presented as they are dominated by the
ground state already at $t \le 4$ and are thus not
sensitive to the choice of $c_A$.
Results for $\tilde m$ with different choices of discretization
are collected in Table~\ref{tab:mtilde}.

\begin{table}
\begin{center}
\advance\tabcolsep by -1pt
\begin{tabular}{|c|c|c|c|c|c|c|c|c|c|}
\hline
\multicolumn{1}{|c|}{}&
\multicolumn{3} {c|}{{\bf 60NP}}&
\multicolumn{3} {c|}{{\bf 62NP}}&
\multicolumn{3} {c|}{{\bf 64NP}}\\
\multicolumn{1}{|c|}{Label}  &
\multicolumn{1} {c|}{2-pt}  &
\multicolumn{1} {c|}{3-pt}  &
\multicolumn{1} {c|}{4-pt}  &
\multicolumn{1} {c|}{2-pt}  &
\multicolumn{1} {c|}{3-pt}  &
\multicolumn{1} {c|}{4-pt}  &
\multicolumn{1} {c|}{2-pt}  &
\multicolumn{1} {c|}{3-pt}  &
\multicolumn{1} {c|}{4-pt}  \\
\hline    
%
%
%
$(\kappa_1,\kappa_1)$ & $0.1442(10)$ & $0.1425(13)$ & $0.1450(10)$ & $0.1345( 6)$ & $0.1340( 7)$ & $0.1349( 6)$ & $0.2106( 6)$ & $0.2108( 9)$ & $0.2110( 6)$ \\
$(\kappa_2,\kappa_2)$ & $0.1182( 8)$ & $0.1167(11)$ & $0.1188( 8)$ & $0.1053( 4)$ & $0.1048( 5)$ & $0.1058( 4)$ & $0.1754( 5)$ & $0.1752( 7)$ & $0.1756( 5)$ \\
$(\kappa_3,\kappa_3)$ & $0.0913( 6)$ & $0.0902( 8)$ & $0.0920( 6)$ & $0.0728( 3)$ & $0.0722( 3)$ & $0.0727( 3)$ & $0.1391( 4)$ & $0.1389( 5)$ & $0.1394( 4)$ \\
$(\kappa_4,\kappa_4)$ & $0.0752( 5)$ & $0.0744( 7)$ & $0.0757( 5)$ & $0.0562( 2)$ & $0.0558( 3)$ & $0.0561( 2)$ & $0.0960( 2)$ & $0.0958( 3)$ & $0.0963( 2)$ \\
$(\kappa_5,\kappa_5)$ & $0.0561( 4)$ & $0.0555( 6)$ & $0.0563( 4)$ & $0.0419( 2)$ & $0.0417( 2)$ & $0.0421( 2)$ & $0.0682( 2)$ & $0.0681( 2)$ & $0.0686( 2)$ \\
$(\kappa_6,\kappa_6)$ & $0.0308( 4)$ & $0.0312(15)$ & $0.0310( 4)$ & $0.0306( 2)$ & $0.0304( 2)$ & $0.0307( 2)$ & $0.0429( 1)$ & $0.0428( 1)$ & $0.0430( 1)$ \\
$(\kappa_7,\kappa_7)$ & $0.0236(35)$ & $0.0251(55)$ & $0.0246(30)$ & $0.0248( 1)$ & $0.0247( 2)$ & $0.0250( 1)$ & $0.0285( 1)$ & $0.0284( 1)$ & $0.0286( 1)$ \\
\hline    
\end{tabular}
\end{center}
\caption{Values of the quark mass $a \tilde m$ defined using the mass-dependent $c_A$ for 
two-point, three-point and four-point discretization schemes at the three couplings. 
}
\label{tab:mtilde}
\end{table}

{\color{red}

As noted in Ref.~\cite{Collins:cA:03}, a possible problem with our
criterion for determining $c_A$ is that it does not involve the same
physical distances at all couplings.  The ``physical'' criterion we
wish to implement is that the same value of $\tilde m_{ij}$ is
obtained in the AWI from both the ground and the first excited
states. This requires that we tune the source to produce the same
mixture of ground and excited states at all values of $a$. While the
lattice size and the \color{red}radius of the smeared source in the generation of quark
propagators were increased with $\beta$,} {\color{red} they were not tuned.  In fact, we find
that our fit is sensitive to the same range, $t=2-5$ ($3-5$ for
three-point discretization), for all three lattice spacings, and thus
is sensitive to significantly shorter Euclidean times at $\beta=6.4$
than at $\beta=6$.

\color{red} The part of our analysis which is, therefore, sensitive to
the extent to which our criterion is physical is the manner in which
the continuum limit is approached. With a physical criterion, the
dominant correction to scaling will be proportional to $a^2$. If one
changes the criterion as $a$ is varied, the simple power dependence
can be distorted. While the data suggests that the contribution from
higher states is small, we cannot rule out some distortion and the
scaling analysis has to be taken with caution. To study the question
in detail, however, would require a more extensive data set than ours.

}

\begin{figure}[tbp]   
\begin{center}
\includegraphics[width=0.7\hsize]{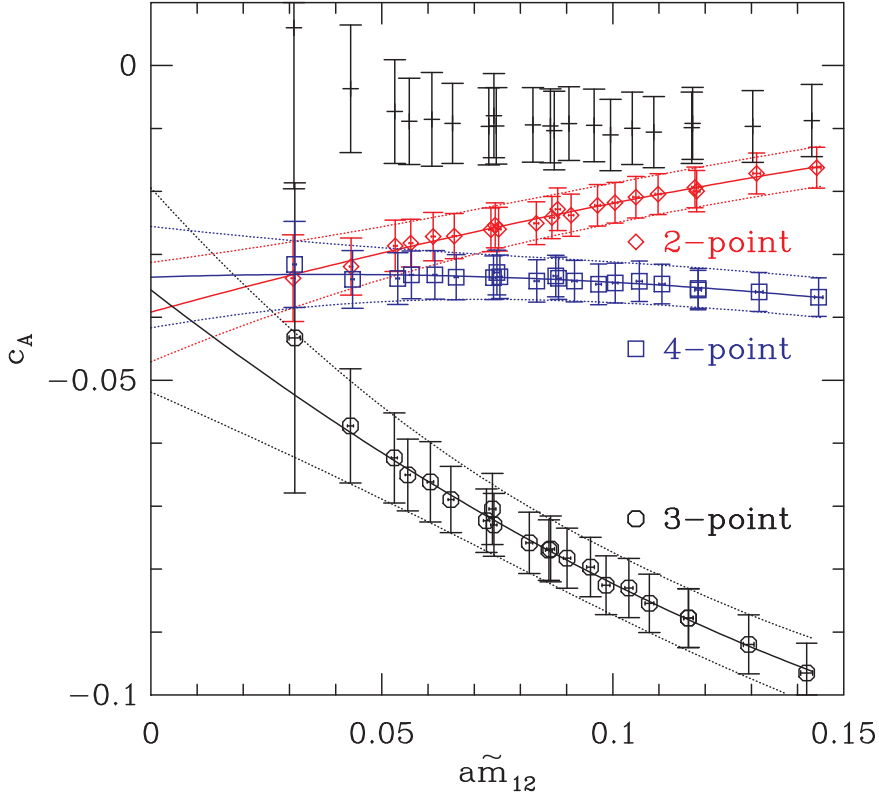}
\end{center}
\caption{Results of chiral extrapolation of $c_A$ data at $\beta=6.0$ ({\bf 60NPf})
for the two-point, three-point and four-point discretization of the
derivative. Quadratic fits are made to all degenerate 
and non-degenerate mass combinations using
$\kappa_1-\kappa_7$.  We also show the quantity $c_A^{3-pt} -
c_A^{2-pt} + a \tilde m/2$ discussed in the text using the symbol
plus.}
\label{fig:cAext60}
\end{figure}

\begin{figure}[tbp]   
\begin{center}
\includegraphics[width=0.7\hsize]{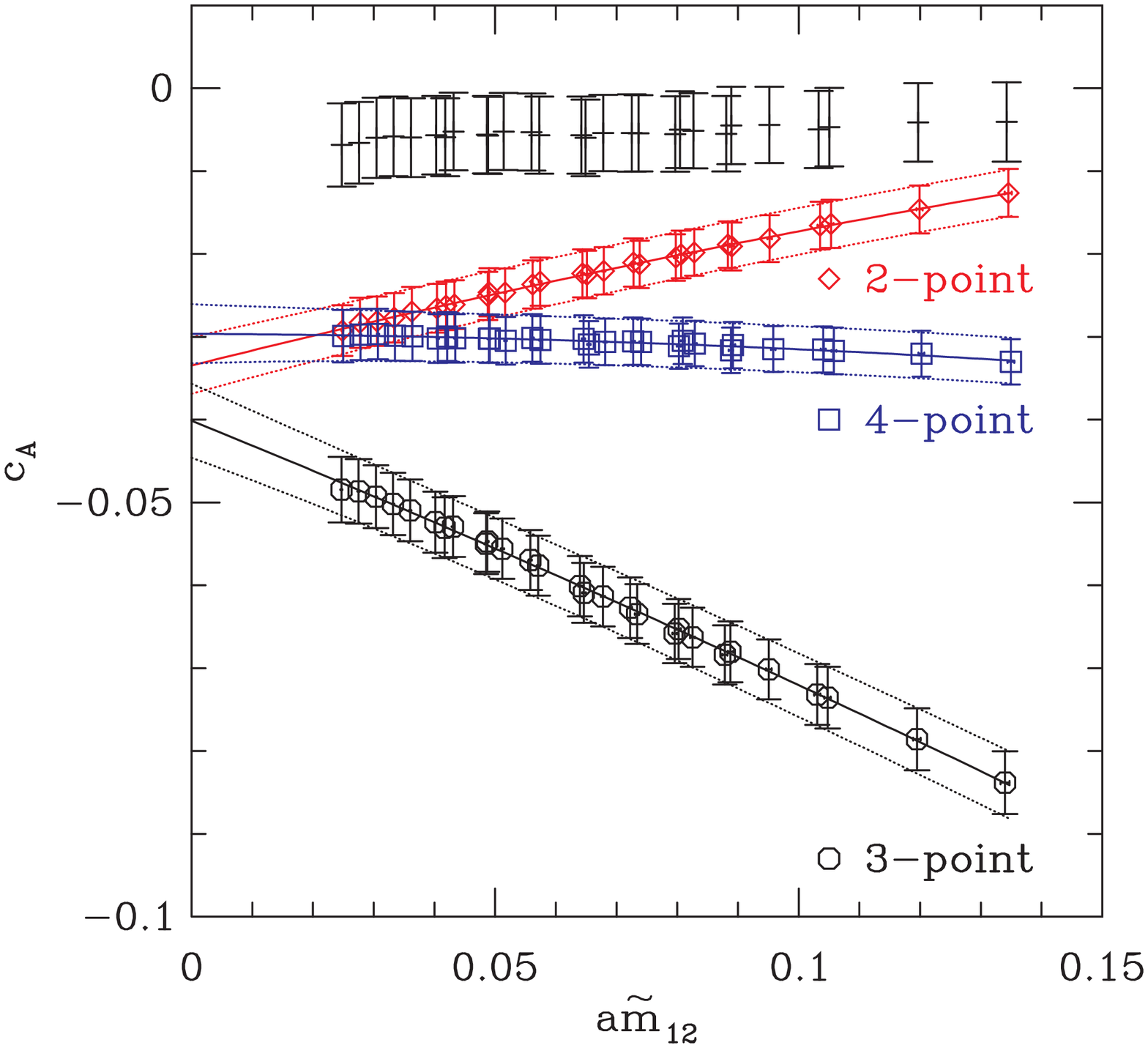}
\end{center}
\caption{Same as Fig.~\protect\ref{fig:cAext60} but at $\beta=6.2$.}
\label{fig:cAext62}
\end{figure}

\begin{figure}[tbp]   
\begin{center}
\includegraphics[width=0.7\hsize]{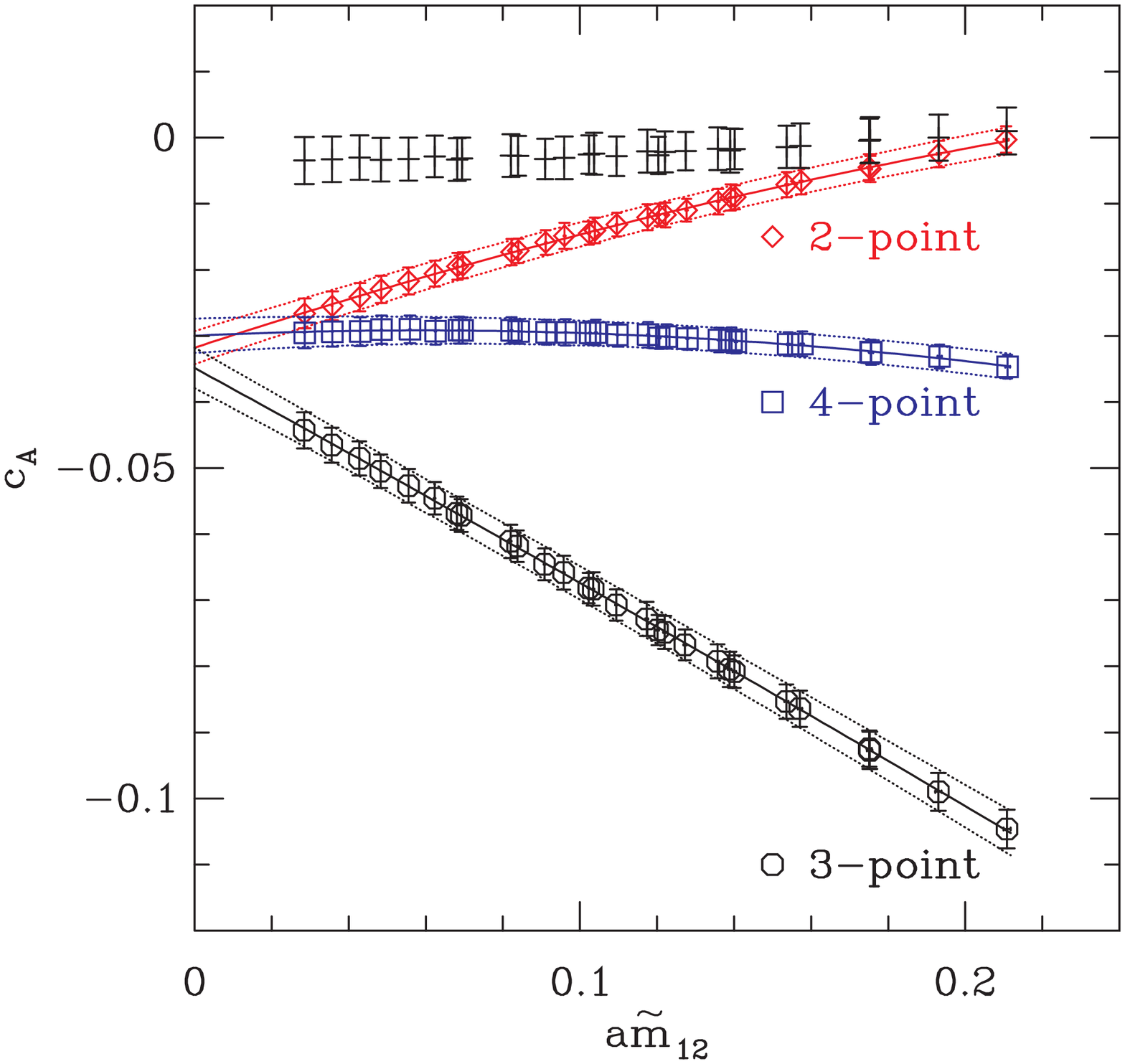}
\end{center}
\caption{Same as Fig.~\protect\ref{fig:cAext60} but at $\beta=6.4$.}
\label{fig:cAext64}
\end{figure}

In Figs~\ref{fig:cAext60}, \ref{fig:cAext62} and \ref{fig:cAext64} we
show quadratic fits to $c_A(\tilde m)$ versus $\tilde m$ for the
two-point, three-point and four-point discretization of the
derivative. The data are for zero momentum correlators at $\beta=6.0$,
$6.2$ and $6.4$, and include degenerate and non-degenerate mass
combinations. The results for two-point and three-point
discretizations are given in Tables~\ref{tab:2ptdata} and
\ref{tab:3ptdata}.  The four-point estimates are $0.034(8)$,
$0.034(9)$, $0.030(4)$ and $0.030(3)$ for {\bf 60NPf}, {\bf 60NPb},
{\bf 62NP} and {\bf 64NP} data sets respectively.  We find that the
three estimates agree within errors in the chiral limit.  By
contrast, the $O(a m)$ contributions are significant, as shown by the
large, roughly linear, dependence of $c_A$ on $\tilde m$.  However, as
we now explain, the bulk of this linear dependence has a simple
kinematic origin and can be understood analytically.

In Ref.~\cite{LANL:Zfac:00} we showed that if, by tuning $c_A$ and
$\tilde m$, eq.~(\ref{eq:cA}) can be satisfied
over a common range of time-slices where two- and three-point
discretizations schemes are implemented then, to $O(a^3)$, $\tilde m$
is the same in both schemes and the $c_A$ at any $a \tilde m$ are related
as $c_A^{3-pt} = c_A^{2-pt} - {\tilde m}a/2 + O(a^2)$. 
It is useful to generalize this argument to provide the relation
between $c_A$ determined in any two discretization schemes.
The equation we wish to satisfy is 
$\partial_4 A_4 + a c_A\partial^2_4 P - 2 {\tilde m} P = 0$. 
Taylor expansion of any lattice version of this relation 
using a symmetric discretization scheme for the derivatives gives 
\begin{equation}
[\partial_4 A_4 + a^2 \alpha \partial^3_4 A_4 + O(a^4)] + 
a c_A [\partial_4^2 P + a^2 \beta \partial^4_4 P + O(a^4)] - 
2 {\tilde m} [P + a^2 \gamma \partial^2_4 P + O(a^4)] = 0 \,.
\label{eq:higherorderdisc}
\end{equation}
The key assumption is that this relation can be satisfied by
two discretization schemes over a common range of timeslices.
If so, then it follows, first, that the two schemes will give the same
$\tilde m$ to $O(a^3)$ and, second, that
\begin{equation}
c_A^{(2)} - c_A^{(1)} = 2a \tilde m(\alpha^{(1)} - \alpha^{(2)} 
                       - \gamma^{(1)} + \gamma^{(2)}) + O(a^2) \,.
\label{eq:cArelation}
\end{equation}
For the two-point and three-point derivatives we have used, this
condition reduces to $c_A^{3-pt} = c_A^{2-pt} - a \tilde m/2 + O(a^2)$
because $\alpha^{2-pt} = 1/24$, $\gamma^{2-pt} = 1/8$, $\alpha^{3-pt}
= 1/6$, and $\gamma^{3-pt} = 0$. Our data confirm these two
predictions to good accuracy: the ratio of correlators ($2\tilde
m_{ij}$) is the same within errors for the three discretization schemes,
as shown in Table~\ref{tab:mtilde}, and Eq.~\ref{eq:cArelation} holds,
as illustrated in Figs.~\ref{fig:cAext60}, \ref{fig:cAext62} 
and~\ref{fig:cAext64} where we also plot the quantity $c_A^{3-pt} -
c_A^{2-pt} + a \tilde m/2$ and show it is consistent with zero
at the 1-$\sigma$ level.

The UKQCD collaboration~\cite{Collins:cA:03} has pointed out that the
mass dependence of $c_A$ can be reduced using higher order
discretization schemes.  This also follows from
Eq.~\ref{eq:cArelation}.  For any $O(a^3)$ improved scheme, $\alpha =
\gamma = 0$, and consequently $c_A^{imp} = c_A^{2-pt} - a{\tilde m}/6
+ O(a^2)$.  In fact we find that the slope of $c_A^{2-pt}$ versus $a
\tilde m_{ij}$ is $\Delta\approx 0.18$, $0.18$ and $0.19$ respectively
for $\beta=6$, $6.2$ and $6.4$, so that the slope obtained using any
$O(a^3)$ improved scheme should indeed be very small, $\Delta-1/6\sim
0.02$.  For our four-point discretization scheme (which is $O(a^3)$
improved, but differs from the five-point scheme used in
Ref.~\inlinecite{Collins:cA:03}) we find that the slope is $\le 0.03$
for all four data sets {\color{red} as shown in
Figs~\ref{fig:cAext60}, \ref{fig:cAext62} and \ref{fig:cAext64}.  We
also find that the contribution of the $(a\tilde m)^2 $ term in the
four-point scheme is comparable to the linear term over the range of
quark masses simulated, and to the errors. Because of the size of
these higher order terms, further improvements in the discretization
of the derivative are not expected to reduce the undertainty.}


We stress that we do not expect a
higher order scheme to completely remove $a\tilde m$ contributions in
$c_A$, for to do so would require complete implementation of an
$O(a^2)$ improvement program.\footnote{%
This means that the
assumption leading to eq.~(\ref{eq:cArelation}), namely that the
relation (\ref{eq:higherorderdisc}) can be satisfied by two schemes
over a range of timeslices, cannot hold precisely.}
Nevertheless, our results indicate that the bulk of the slope 
for two- and three-point discretization is
due to errors associated with discretization of the derivative.

The upshot of this discussion is as follows.  On the one hand, it would
have been advantageous to use a higher-order discretization scheme
with a smaller slope $\Delta$. This would have 
reduced the uncertainty in the results for some of the $b_\CO$ that are
proportional to $\Delta$, as discussed in later sections.  On the
other hand the $O(a \tilde m)^2$ and $O(a \tilde m)$ terms become
comparable in our four-point data, and the error in the extrapolated
value does not decrease compared to the lower-order 
schemes.{\color{red}\footnote{\color{red}Unfortunately, we are not able to determine the efficacy
of the four-point discretization scheme for analyzing the three-point
Ward identities as some of the required raw data has been lost due to
disk corruption.}  }
Having demonstrated that the dominant effect is
kinematic, we can remove it in our two- and three-point schemes simply
by using the mass-dependent $c_A(\tilde m)$ in the improved axial
current, rather than $c_A(0)$.  This is indeed what we do in the axial
rotation $\delta S$ (which, we recall, is always defined with
two-point derivatives and in which we always use the two-point
$c_A(\tilde m)$).

We now return to the numerical results for $c_A(0)$, which are given
in Tables \ref{tab:2ptdata} and \ref{tab:3ptdata}.  Results from two-,
three- and four-point discretizations should differ only by terms of
relative size $a\Lambda_{QCD}$.  In fact, as already noted,
Figs~\ref{fig:cAext60}, \ref{fig:cAext62} and \ref{fig:cAext64} show
that the quadratically extrapolated values for $c_A$ at each of the
three $\beta$ from all three discretization schemes agree.  Estimates
using linear fits, however, differ by combined $1\sigma$ errors, due
to the curvature. 

{\color{red} In Ref.~\inlinecite{LANL:Zfac:00} we chose, for our
central value, $c_A$ from the two-point discretization of the
derivative over that from the three-point derivative for the following
two reasons. First, the \color{red}$O(a^2)$ discretization errors in the
derivatives are smaller, which leads to a smaller slope of $c_A$
versus $\tilde m$; and second, because the statistical errors are
smaller. Now that we understand the relative size of the slope to be
largely a kinematical effect, and the extrapolated values
overlap, and the uncertainty in the estimates are comparable, we take
the weighted mean of the two-, three- and four-point results for our
central value at all three lattice spacings.  In addition to
statistical errors we quote the spread of the results to estimate the
residual $O(a)$ errors.  Note that, as pointed out in the
Introduction, the choice of the discretization scheme used for the
derivative does not affect results for $c_A(0)$ at the leading
order of overall improvement, and the estimates from any scheme can be used 
to define the improved theory. However,
if a calculation requires the axial vector Ward identity be respected, then the
appropriate discretization scheme and the corresponding $c_A(m)$
should be used.  }


\section{Extracting $c_A$ using states at finite momentum}
\label{sec:cAmom}

\begin{figure}[!ht]  
\begin{center}
\includegraphics[width=0.7\hsize]{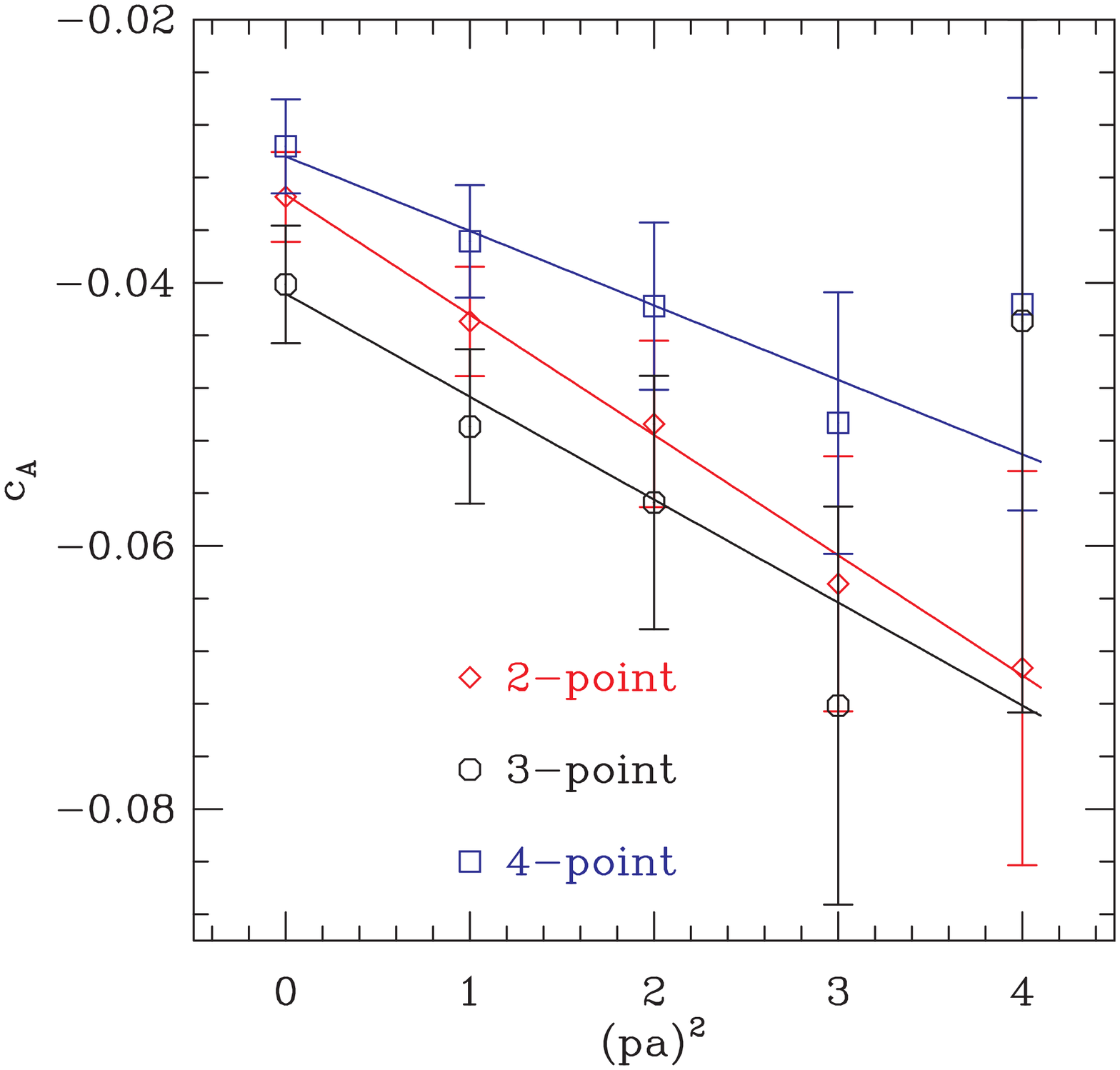}
\end{center}
\caption{$c_A$ using states of non-zero momentum
plotted against $(pa)^2$, along with a linear fit.
We show results for the two-point, three-point and four-point discretization of the derivative. The
data are for $\beta = 6.2$ and $pa$ is in units of $2\pi/24$. }
\label{fig:cA62vspa2pt_3pt}
\end{figure}

\begin{figure}[!ht]  
\begin{center}
\includegraphics[width=0.7\hsize]{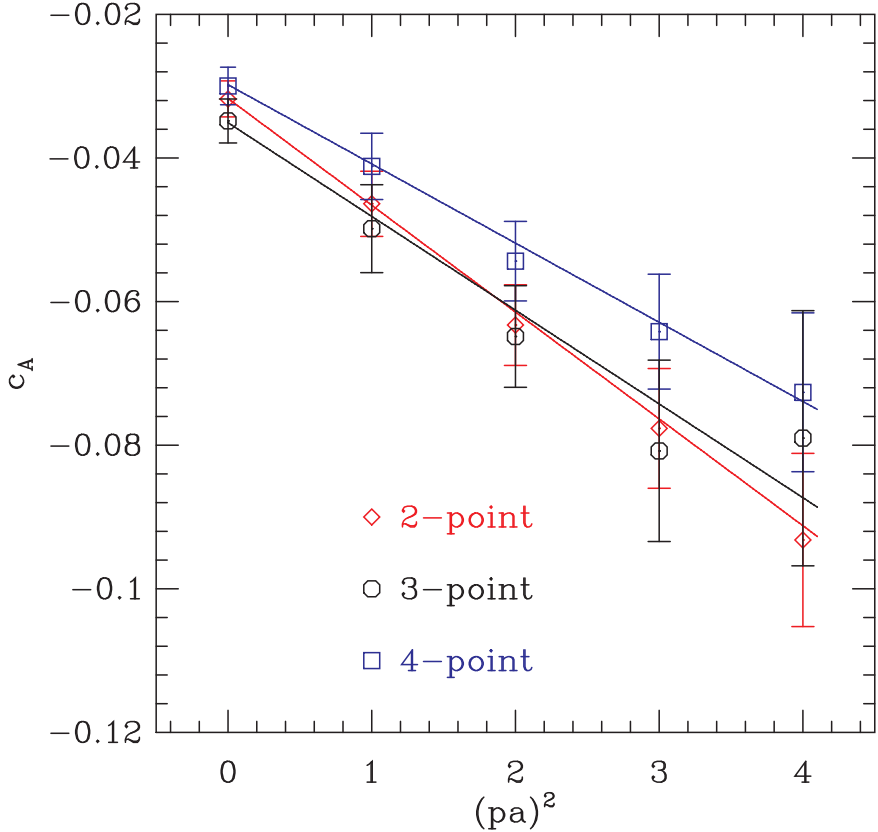}
\end{center}
\caption{$c_A$ using states of non-zero momentum at $\beta=6.4$.
Notation as in Fig.~\protect\ref{fig:cA62vspa2pt_3pt}.}
\label{fig:cA64vspa2pt_3pt}
\end{figure}

The data at $\beta =6.2$ and $6.4$ are precise enough to extract $c_A$
using states having non-zero momenta. In figures~\ref{fig:cA62vspa2pt_3pt}
and \ref{fig:cA64vspa2pt_3pt} we show the results of linear
fits for the chirally extrapolated value of $c_A$ for two-, three-, and 
four-point discretization of the derivative as a function of
$(pa)^2$. For the chiral extrapolation of $c_A(\tilde m)$, quadratic
fits to all mass combinations of $\kappa_1-\kappa_7$ propagators work
very well at all five values of $pa$. The signal in three-point data
at $pa=2$ is noisy and this is reflected in the errors. The data
exhibit the following two features:
\begin{itemize}
\item 
Additional discretization errors of $O(p^2a^2)$ are generated 
when using states with non-zero momenta. 
The coefficients of these corrections are significant,
lying in the range $0.12 - 0.22$. 
\item 
The difference in results between the two-, three- and four-point
discretization of the derivative decreases significantly between
$\beta=6.2$ and $6.4$. 
\end{itemize}
Overall, the consistency of the estimates between the two-, three- and
four-point discretization schemes, the added information from fits versus
$(pa)^2$, and the expected improvement with $\beta$ enhance our
confidence in our quoted estimate of $c_A$.

\section{$Z_V^0$ and $\lc b_V$}
\label{sec:ZV}
\begin{figure}[tbp]  
\begin{center}
\includegraphics[width=0.7\hsize]{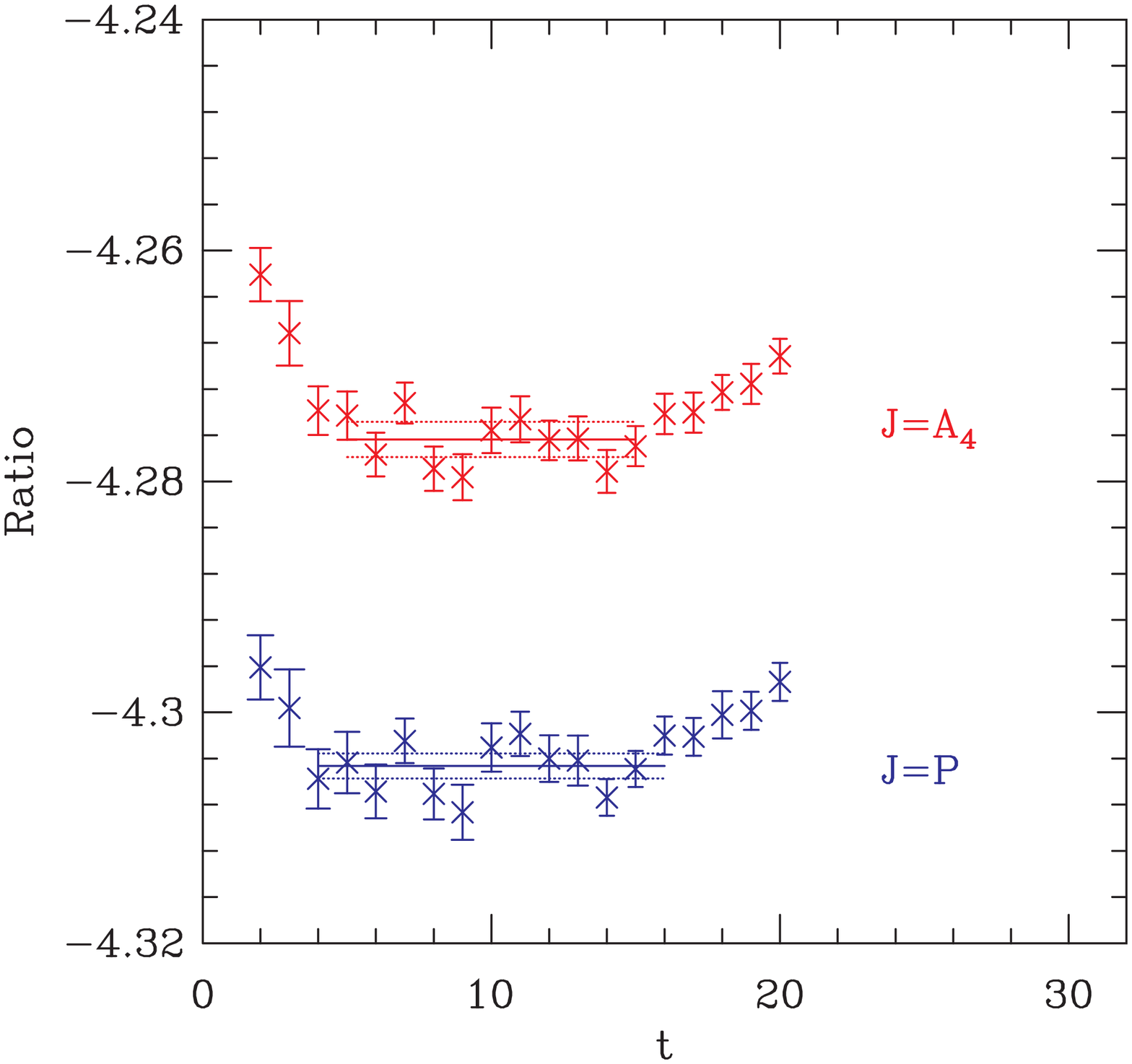}
\end{center}
\caption{The ratio defined in eq.~\ref{eq:ZV} for 
sources $J=P$ and $J=A_4$. Data from the {\bf 64NP} set with
all propagators having mass $\kappa=\kappa_5$. }
\label{fig:ZVsignal}
\end{figure}

Our best estimate of $Z_V^0$ comes from the matrix elements of the
vector charge $\int d^3 x V_4^{(23)}(x)$ between pseudoscalar mesons
\begin{eqnarray}
  \frac{1}{ Z_V^0 (1+\tilde b_V a\tilde m_2) } &=&
  \frac{ \sum_{\vec{x}, \vec{y}}
  \langle P^{(12)}(\vec{x},\tau) 
	(V_I)_4^{(23)}(\vec{y},t) J^{(31)}(0) \rangle }
  { \langle \sum_{ \vec{x}} P^{(12)}(\vec{x},\tau) J^{(21)}(0) \rangle } \,.
\label{eq:ZV}
\end{eqnarray}
with $ \tau > t > 0 $ and the superscript $(23)$ denoting the flavor
of the two fermions in the bilinear (which are taken to be
degenerate). 
Our results for this ratio, illustrated by those in
Fig.~\ref{fig:ZVsignal}, show two features of particular interest:
first, there is a significant dependence on $t$ for time slices
close to the source or sink; and, second, there is a clear difference
between the results using $J=P$ and $J=A_4$. Both features are indicative
of $O(a^2)$ corrections, since, aside from such corrections the ratio
should be independent both of $t$ and the choice of source. The observed
effect is $(1-5)\ a^2\Lambda_{QCD}^2$ using $\Lambda_{QCD}=300\;$MeV. 
Since neither feature would be present if the source and the sink
coupled to a single state (irrespective of improvement), our results
show that the separation between source and sink in our calculation
is insufficient to isolate the lowest state for any value of $t$.
We stress, however, that this is not a problem for implementing the
improvement program (since, after all, we expect ambiguities of
$O(a^2)$).

\begin{figure}[tbp]  
\begin{center}
\includegraphics[width=0.7\hsize]{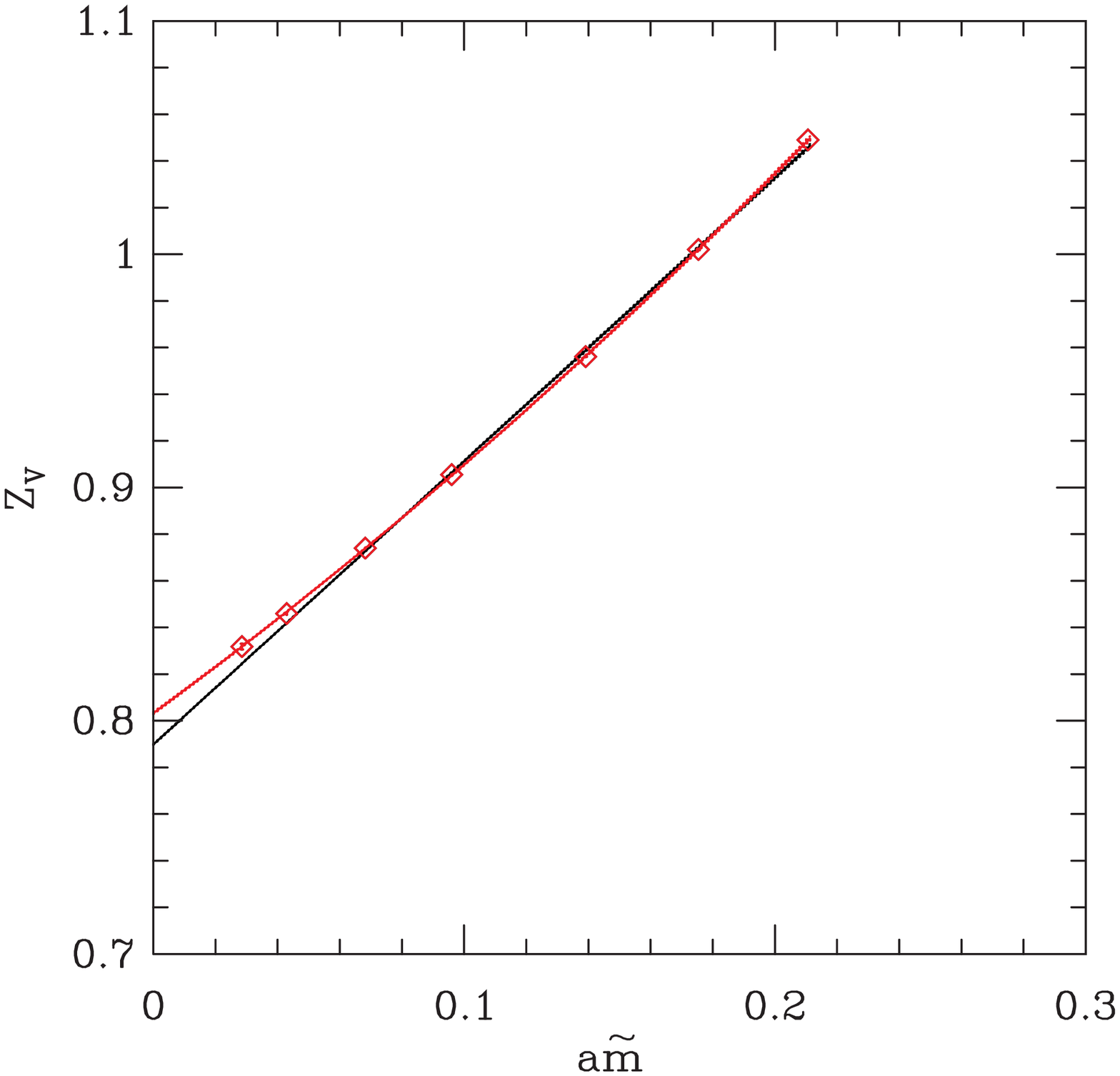}
\end{center}
\caption{Linear and quadratic fit to $Z_V$ versus ${a \tilde m}$ 
for the {\bf 64NP} data set. }
\label{fig:Zvslope}
\end{figure}

To obtain our central values we average the $J = P$ and $J=A_4$ 
data within the jackknife procedure as they are of similar quality.
There is a slight difference in results for $Z_V^0$ using two-
and three-point derivatives, as shown in Tables~\ref{tab:2ptdata} and
\ref{tab:3ptdata}.  Also, at $\beta=6.0$, the errors in the
three-point estimates are almost three times as large. These
differences arise during the chiral extrapolation because the $\tilde
m$ and $\kappa_c$ are slightly different for the two cases and the errors  
in $c_A(\tilde m)$ are $2-3$ times larger for the three-point data.  
The difference between linear and
quadratic chiral extrapolation, as shown in Fig.~\ref{fig:Zvslope}, is
significant. For our central values we use quadratic extrapolations
of the two-point data. 

To extract $Z_V$, ${\tilde b}_V$ and ${b}_V$ we fit 
the ratio in eq.~(\ref{eq:ZV}) to a quadratic function of both $\tilde m$
and the VWI mass. At $\beta=6.4$ the fits yield
\begin{eqnarray}
Z_V &=& 0.8033(5) \left[ 1 + 1.212(11) {\tilde m}a 
			   + 1.134(39) ({\tilde m}a)^2 \right] \,, 
\label{eq:ZVfitmtilde} \\
Z_V &=& 0.8016(5) \left[ 1 + 1.370(9)  ma + 0.033(24) (ma)^2 \right] \,.
\label{eq:ZVfitm}
\end{eqnarray}
The two intercepts, which give $Z_V^0$, differ by 3-$\sigma$ at
$\beta=6.4$ and by $\le 1$-$\sigma$ at $\beta=6.0$ and $6.2$. For the
final estimate of $Z_V^0$ we choose the weighted average as they are
of similar quality. The coefficient of the linear term in the two fits
gives ${\tilde b}_V$ and ${b}_V$ respectively.

In Table~\ref{tab:ZVbV} we compare results with those
from other  non-perturbative
calculations that have been done with the same $O(a)$ improved fermion
action but utilizing different initial and final states to measure the
charge. We find that the results for $b_V$ agree within
the combined $1\sigma$ uncertainties and the expected differences of
$O(a)$. For $Z_V^0$, there is a significant difference between the
LANL and the ALPHA~\cite{ALPHA:Zfac:97A,ALPHA:Zfac:97B,ALPHA:Zfac:98}
collaboration values, which we show, in section~\ref{sec:ALPHA}, can
be explained as residual $O(a^2)$ effects. Estimates by the
QCDSF~\cite{QCDSF:ZVbV:03} and the SPQcdR~\cite{SPQcdR:Z:04}
collaboration lie in the range defined by the LANL and ALPHA data.

\begin{table*}[!ht]
\setlength{\tabcolsep}{1.9pt}
\begin{tabular}{|| >{\scriptsize}c 
	|| >{\scriptsize}l | >{\scriptsize}l | >{\scriptsize}l 
	|| >{\scriptsize}l | >{\scriptsize}l | >{\scriptsize}l 
	|| >{\scriptsize}l | >{\scriptsize}l | >{\scriptsize}l ||}
\hline
\multicolumn{1}{||c||}{}&
\multicolumn{3}{c||}{\(\beta=6.0\)}&
\multicolumn{3}{c||}{\(\beta=6.2\)}&
\multicolumn{3}{c||}{\(\beta=6.4\)}\\
\hline
         & LANL              & ALPHA            & QCDSF
         & LANL              & ALPHA            & QCDSF
         & LANL              & ALPHA            & QCDSF     \\
         &                   &                  &
         &                   &                  &
         &                   &                  &                \\[-12pt]
\hline			     		       
         &                   &                  &
         &                   &                  &
         &                   &                  &                \\[-12pt]
\hline			     		       
$Z^0_V$  & $0.7695(8)  $     & $0.7809(6)$      &  $0.7799(7)$  
         & $0.7878(4)  $     & $0.7922(4)(9)$   &  $0.7907(3)$
         & $0.8024(5)  $     & $0.8032(6)(12)$  &  $0.8027(2)$   \\
\hline			     		        
$b_V$    & $1.52(1)$         & $1.48(2)$        &  $1.497(13)$ 
         & $1.40(1)$         & $1.41(2)$        &  $1.436(8)$
         & $1.37(1)$         & $1.36(3)$        &  $1.391(5)$  \\
\hline
\end{tabular}
\caption{Non-perturbative estimates of $Z_V^0$ and $b_V$ from the
LANL, ALPHA, and QCDSF~\cite{QCDSF:ZVbV:03} collaborations. For
consistency LANL estimates at all three $\beta$ are taken from fits 
versus the VWI mass $m$. }
\label{tab:ZVbV}
\end{table*}

In Ref.\cite{LANL:Zfac:00} it was observed that extrapolations using a
quadratic fit in ${\tilde m}$ give estimates closer to measured values
of $Z_V^0$ near the charm quark mass than did fits versus $m$.  At
$\beta=6.4$ the two fits agree within 1\% up to $\tilde m a\approx
0.3$, whereas the charm quark mass is smaller in lattice units, $i.e.$
$am_c \approx 0.22$. Since, as noted above, $O(a^2)$ errors are $\sim
1\%$, we conclude that either fit can be used for quark masses in the
range $0-m_c$.

\section{$\lc c_V$ and $\lc {\tilde b}_A - \lc {\tilde b}_V$ }
\label{sec:cV}

Up to this stage, we have used only two-point correlation functions 
or three-point correlators involving the vector charge. We now turn
to axial Ward identities involving three-point correlators.
These allow us to determine $c_V$, $\tilde b_A-\tilde b_V$,
$Z_A^0$, $Z_P^0/Z_S^0$, ${\tilde b}_P - {\tilde
b}_S$, and $c_T$, as well as giving an alternate determination
of $Z_V^0$. 
We first consider the improvement coefficient $c_V$ whose precise
determination feeds into the calculation of $Z_A^0$, $Z_P^0/Z_S^0$,
$c_T$, and $c_A^\prime$.  The best signal for $c_V$ is obtained by
enforcing $N_1 = N_2 + c_V D$, with
\begin{eqnarray}
N_1 &=&
 \frac	{ \sum_{\vec{y}}
	\langle \delta {\cal S}^{(12)}_I 
	\ (V_I)_4^{(23)}(\vec{y},y_4) \  P^{(31)}(0) \rangle }
	{ \sum_{\vec{y}} 
	\langle (A_I)_4^{(13)}(\vec{y},y_4) \ P^{(31)}(0) \rangle } \,,
\label{eq:cV1}\\
N_2 &=& 
 \frac{ \sum_{\vec{y}}  
        \langle \delta {\cal S}^{(12)}_I \ 
	V_i^{(23)}(\vec{y},y_4) \  
	A_i^{(31)}(0) \rangle }
        { \sum_{\vec{y}} 
		\langle (A_I)_i^{(13)}(\vec{y},y_4) 
\  A_i^{(31)}(0) \rangle } \,, 
\label{eq:cV2}\\
D &=&  \frac{ \sum_{\vec{y}}  
        \langle \delta {\cal S}^{(12)}_I \ 
	a \partial_\mu T_{i\mu}^{(23)}(\vec{y},y_4) \  
	A_i^{(31)}(0) \rangle }
        { \sum_{\vec{y}} 
		\langle (A_I)_i^{(13)}(\vec{y},y_4) \  A_i^{(31)}(0) \rangle } \,,
\label{eq:cV3} 
\end{eqnarray}
so that
\begin{equation}
c_V = \frac{N}{D} \equiv \frac{N_1-N_2}{D}
\,.
\end{equation}
{\color{red}We recall that $\delta{\cal S}$ uses two-point discretization
(and the corresponding value of $c_A(\tilde m_1)$), 
but that the other improved currents in these expressions are
discretized both with the two- and three-point forms giving two 
sets of estimates. Within each set we provide two estimates using 
$c_A(0)$ and $c_A(m)$ in the expression for $A_I$.}
We also recall that we always use $\tilde m_1=\tilde m_2$.

\begin{figure}[tbp]   
\begin{center}
\includegraphics[width=0.7\hsize]{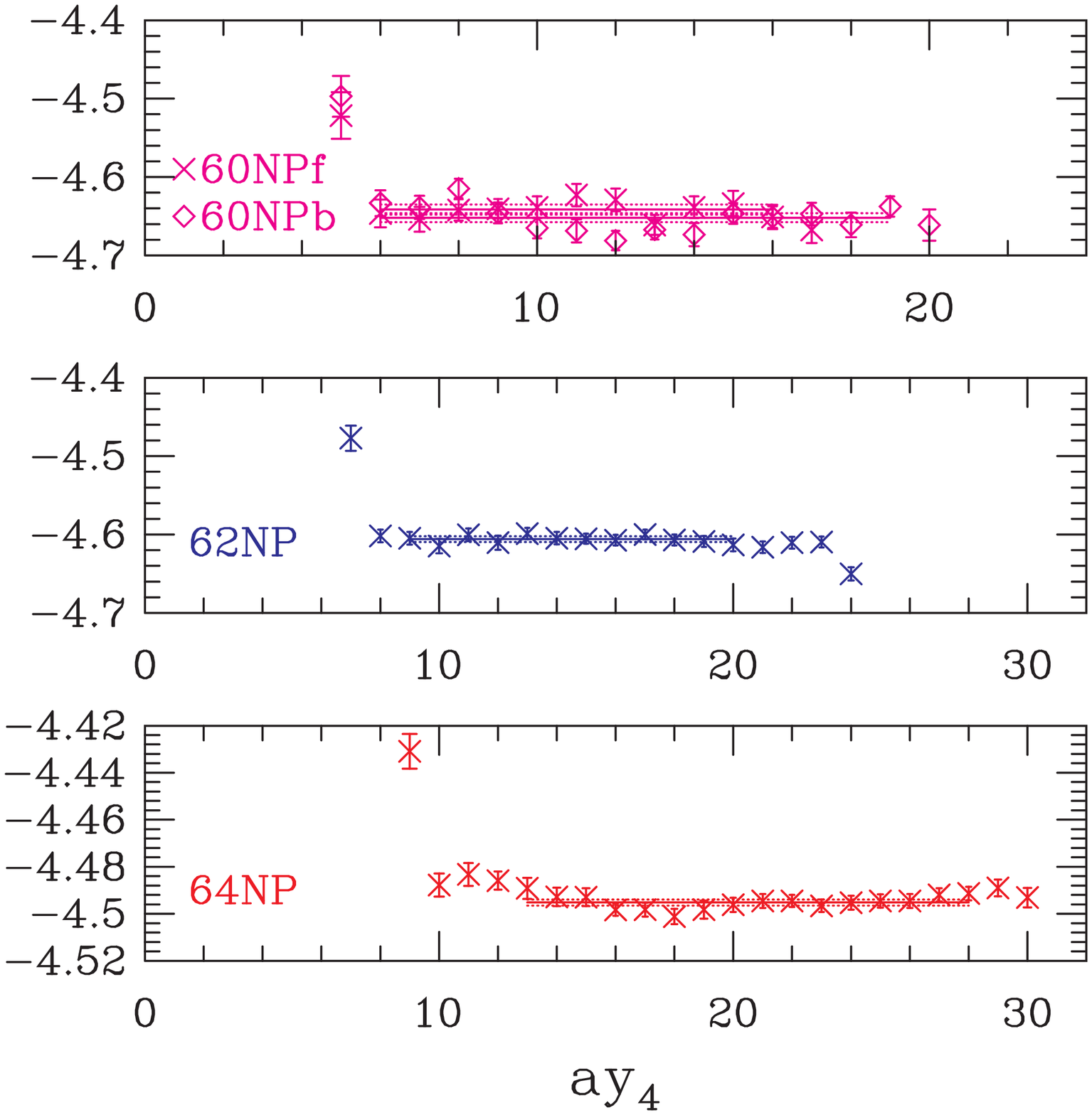}
\end{center}
\caption{Illustration of the quality of the signal for the 
quantity $N_1$ of Eq.~(\protect\ref{eq:cV1}) for all four data sets
with two-point discretization and $c_A(\tilde m)$. In all cases the
data have to be multiplied by the respective values of $2\kappa_3$, 
the lattice normalization of the additional propagator in the numerator.}
\label{fig:cVfit}
\end{figure}

\begin{figure}[tbp]   
\begin{center}
\includegraphics[width=0.7\hsize]{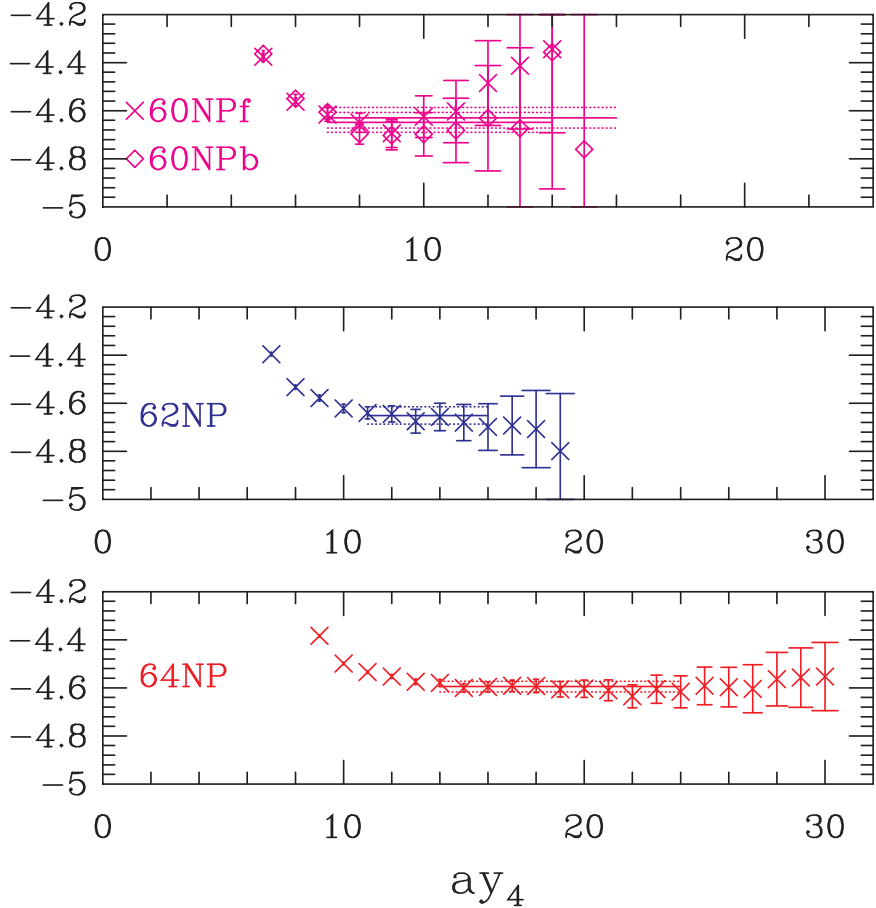}
\end{center}
\caption{Illustration of the quality of the signal for $N_2$
of Eq.~(\protect\ref{eq:cV2}) for all four data sets
with two-point discretization and $c_A(\tilde m)$. In all cases the
data are for $\kappa_3$ and have to be multiplied by the 
respective values of $2\kappa_3$, 
the lattice normalization of the additional propagator in the numerator.}
\label{fig:cVfit1}
\end{figure}

\begin{figure}[tbp]   
\begin{center}
\includegraphics[width=0.7\hsize]{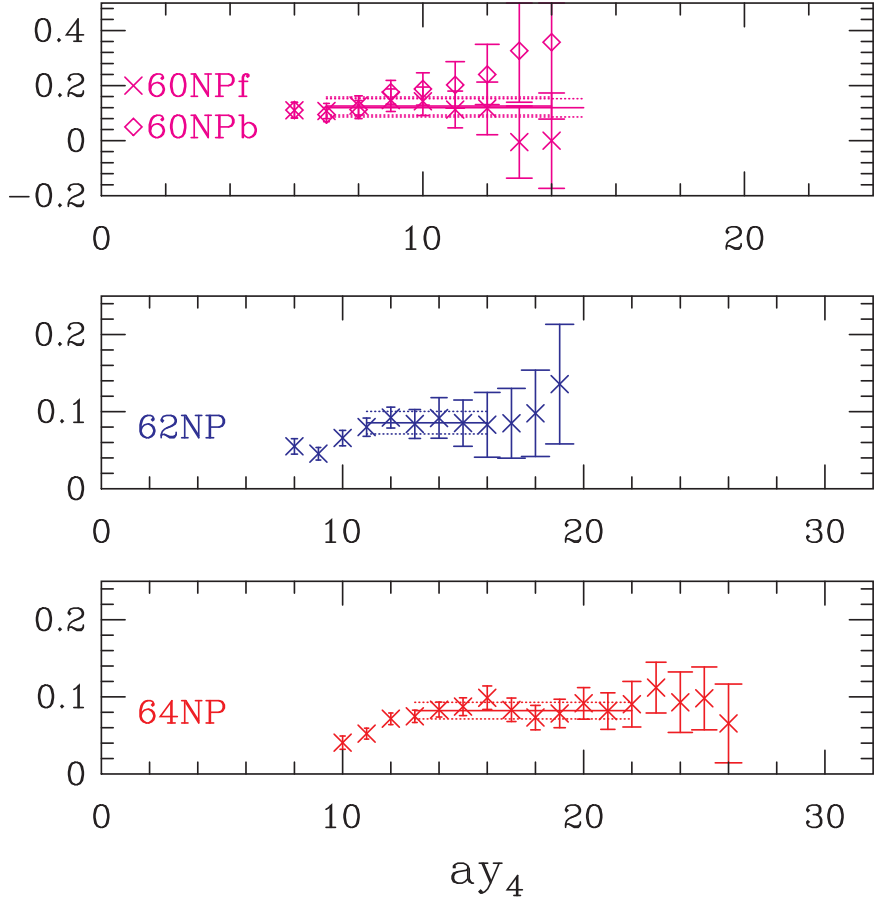}
\end{center}
\caption{Illustration of the quality of the signal for $D$ of
Eq.~(\protect\ref{eq:cV3}) for all four data sets with two-point
discretization and $c_A(\tilde m)$, using $\kappa_3$ propagators in
all cases.}
\label{fig:cVfit2}
\end{figure}

Figures \ref{fig:cVfit}, \ref{fig:cVfit1} and \ref{fig:cVfit2}
illustrate the quality of our data for $N_1$, $N_2$ and
$D$. The improvement in errors and overall quality as $\beta$
increases is evident. Note that $N_2$ and $D$ are expected to have
larger errors than $N_1$ since the lightest state which contributes
is the axial-vector rather than the pion.

\begin{figure}[tbp]   
\begin{center}
\includegraphics[width=0.7\hsize]{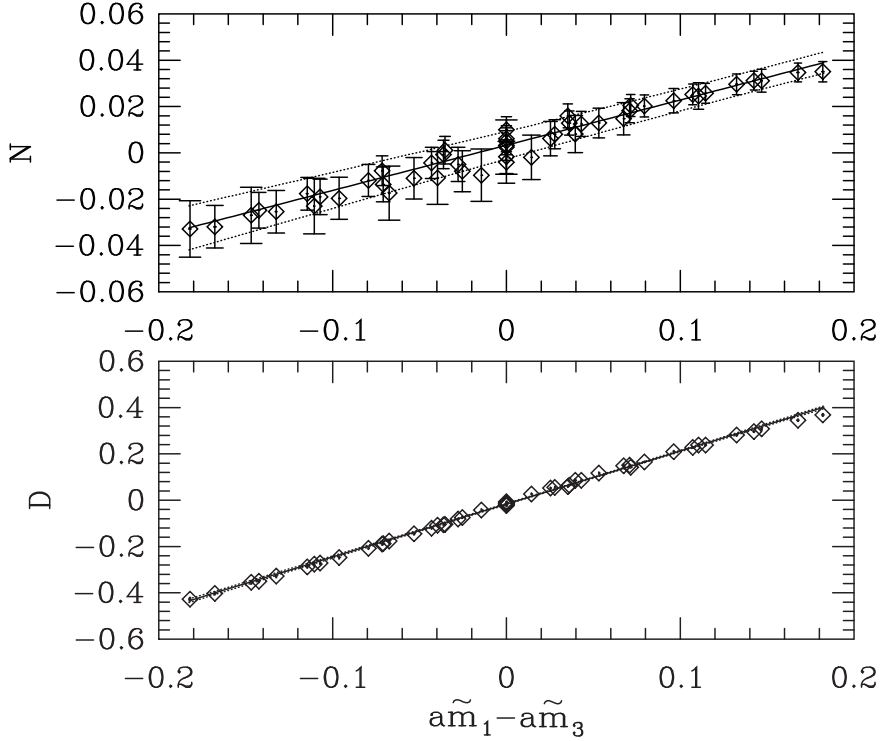}
\end{center}
\caption{Data for $N$ and $D$, defined in the text and used to
extract $c_V$, plotted as a function of
${\tilde m}_1 - {\tilde m}_3 $ for the {\bf 64NP} dataset
with two-point discretization and $c_A(\tilde m)$. }
\label{fig:cVND}
\end{figure}

\begin{figure}[tbp]   
\begin{center}
\includegraphics[width=0.7\hsize]{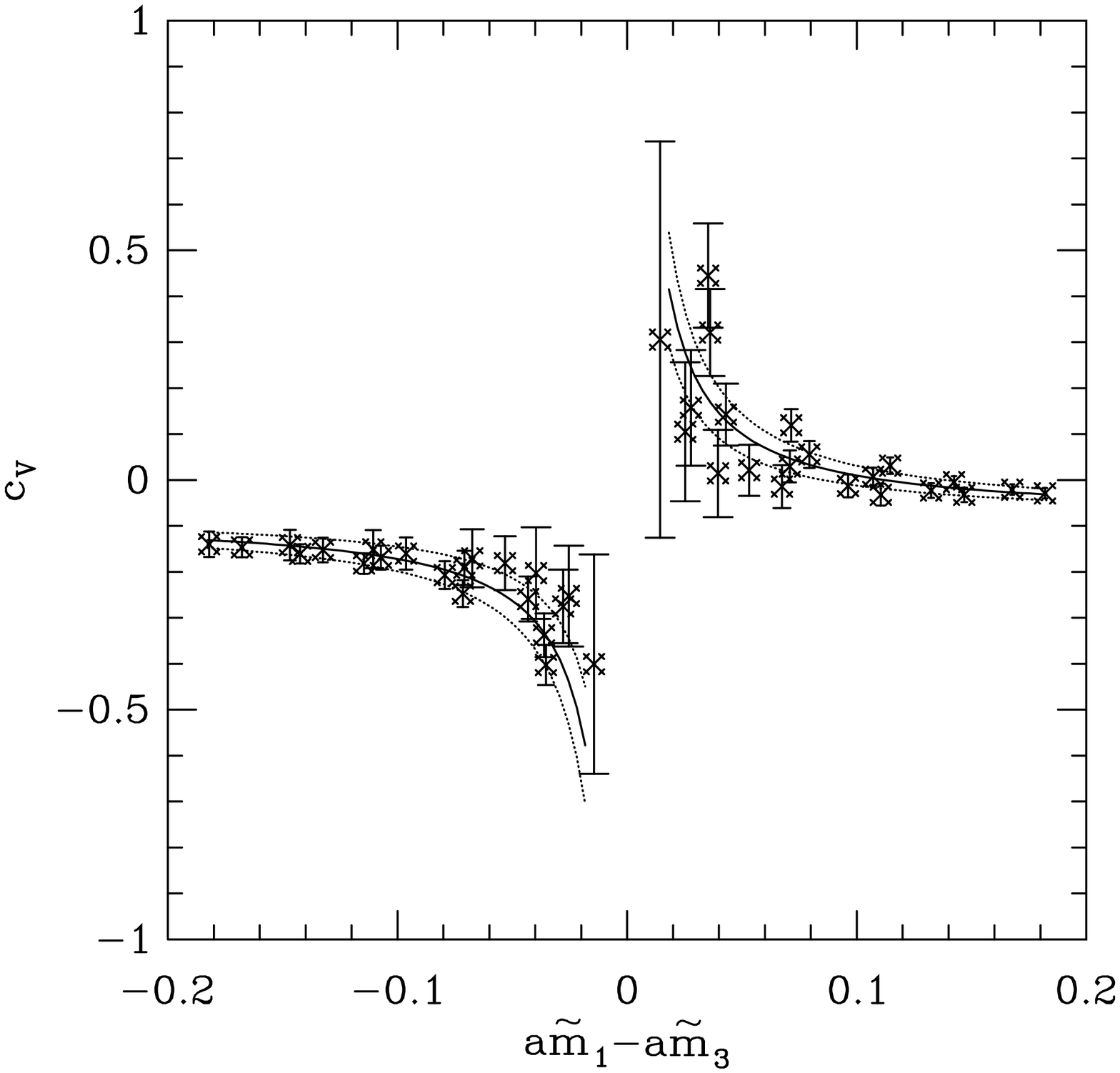}
\end{center}
\caption{A fit of the form 
$c_V = c_V^{(0)} + c_V^{(1)}/({\tilde m}_1 - {\tilde m}_3) $
to the {\bf 64NP} data with two-point discretization and $c_A(\tilde m)$.}
\label{fig:cVpolefit}
\end{figure}

Our procedure is to determine $N_1$, $N_2$ and $D$ from fits to
the plateaus and then combine the first two to form $N\equiv N_1-N_2$.
The results for $N$ and $D$ at $\beta=6.4$ are shown in Fig.~\ref{fig:cVND},
where it is apparent that the errors in $N$ determine the quality
of the result for $c_V=N/D$. As noted in Ref.~\inlinecite{LANL:Zfac:00} 
for the data at $\beta=6$ and $6.2$,
both $N$ and $D$ are to good approximation
functions of $\tilde m_1-\tilde m_3$ that vanish when 
$\tilde m_1\approx\tilde m_3$. 
Since they do not, however, vanish at exactly the same point
(presumably due to statistical and residual discretization errors),
their ratio diverges, as shown in Fig.~\ref{fig:cVpolefit}.

In Ref.~\inlinecite{LANL:Zfac:00}, we used three methods to extract
$c_V$ that try to minimize the effect of this spurious
singularity, and we follow the same strategy here. Details of
the methods will not, however, be repeated.
Our estimates at $\beta=6.0$ and $6.2$ 
have changed after redoing the chiral fits to $N_1$, $N_2$ and $D$,
and so we quote, in Table~\ref{tab:cV}, results for all $\beta$.
For each method we have an additional four choices:
we can use two-point or three-point discretization of the currents,
and for each of these we can use either mass-dependent or
chirally extrapolated values 
of $c_A$ in the operator $(A_I)_4^{(13)}$
appearing in the denominator of $N_1$.

The extrapolation method has the largest uncertainty 
so we discard it.
The consistency between the result using the ``$1/m$ fit'',
shown in Fig.~\ref{fig:cVpolefit}, and the ``slope-ratio'' method,
improves with $\beta$, but the ``slope-ratio'' method is more
stable with respect to the range of quark masses used in the fits
at all three $\beta$ values, and has the smallest dependence
on the choice of $c_A$. We
therefore take our final estimates from the ``slope-ratio'' method and
average the $c_A(\tilde m)$ and $c_A(\tilde m=0)$ values to
get our final estimates. As usual, we take the central value from
the two-point scheme and use the three-point scheme to estimate
the discretization error.

%

%

\begin{figure}[tbp]    
\begin{center}
\includegraphics[width=0.7\hsize]{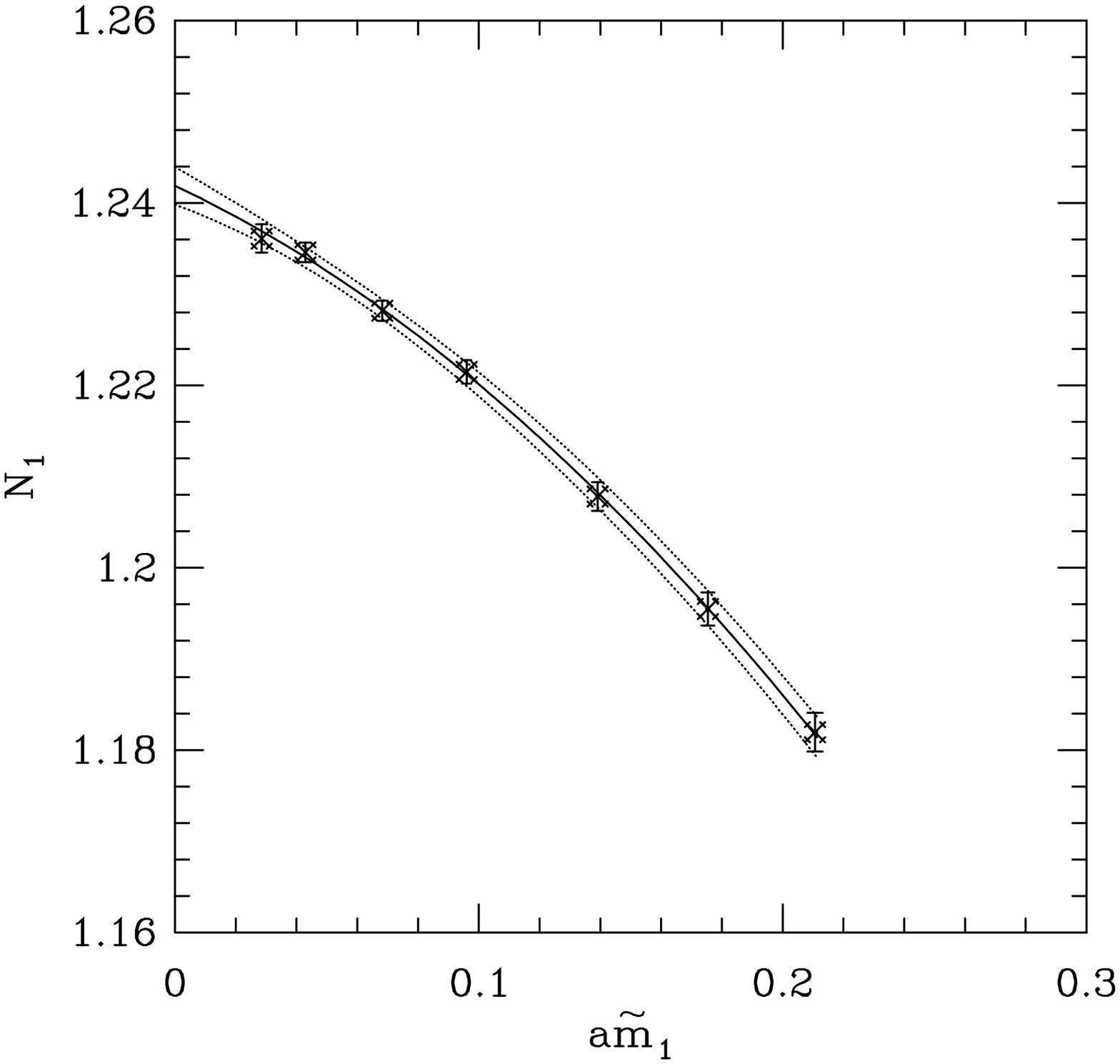}
\end{center}
\caption{Quadratic extrapolation of the ratio in
Eq.~(\protect\ref{eq:cV1}) in $\tilde m_1 = \tilde m_2$ for fixed
$\tilde m_3 = \kappa_4$ for {\bf 64NP} data set 
with two-point discretization and $c_A(\tilde m)$.}
\label{fig:Zvbvba_m1}
\end{figure}
\begin{figure}[tbp]    
\begin{center}
\includegraphics[width=0.7\hsize]{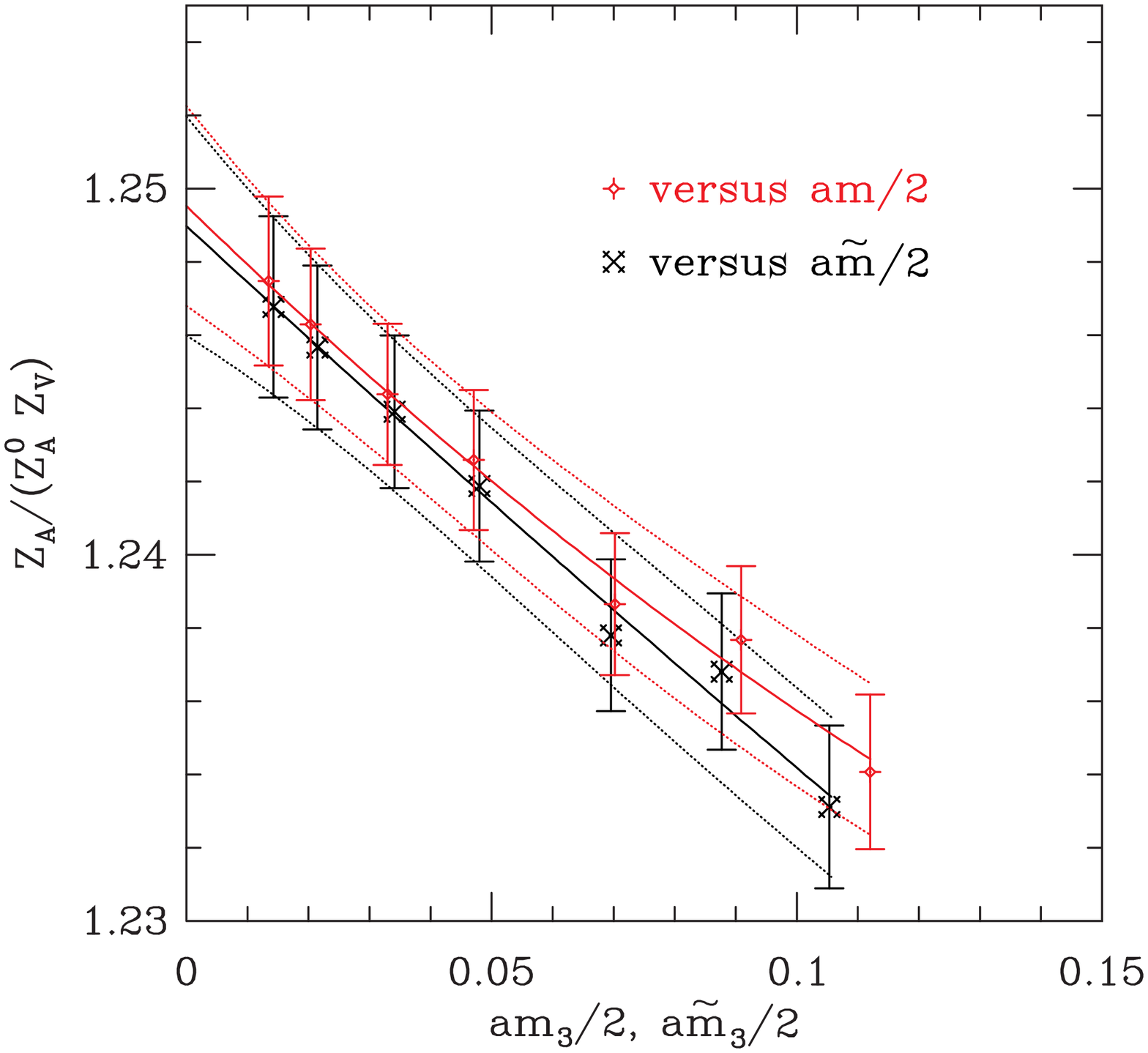}
\end{center}
\caption{Quadratic fits to Eq.~(\protect\ref{eq:cV1}) to
extract ${\tilde b}_A - {\tilde b}_V$ and $b_A - b_V$ for {\bf 64NP}
data set with two-point discretization and $c_A(\tilde m)$. 
The crosses show data and fits versus the AWI quark mass
$\tilde m$, whereas the diamonds show results obtained using 
the VWI quark mass $m$. }
\label{fig:Zvbvba}
\end{figure}

\medskip
We can also use the quantity $N_1$, defined in Eq.~\ref{eq:cV1}, to
determine $ {\tilde b}_A - {\tilde b}_V$, $b_V-b_A$ and to give an
alternate determination of $Z_V^0$. We must first extrapolate to
$\tilde m_1=\tilde m_2=0$ to remove the contribution of
equations-of-motion operators.  In Fig.~\ref{fig:Zvbvba_m1} we
illustrate the quadratic fits used to do this for the {\bf 64NP}
data. We then fit to a quadratic function of $\tilde m_3$ or $m_3$.
These fits, shown in Fig.~\ref{fig:Zvbvba} for two-point
discretization and $c_A(0)$, have parameters

\begin{eqnarray}
\frac{1}{Z_V^0} \big(1 + ({\tilde b}_A - {\tilde b}_V) \frac{a {\tilde m}_3}{2} + O(a^2)\big) &=& 
       1.249(3) \big(1 - 0.123(54) \frac{a {\tilde m}_3}{2}
                       + 0.06(38) (\frac{a {\tilde m}_2}{2})^2  \big) \nonumber \\
\frac{1}{Z_V^0} \big(1 + ({       b}_A - {       b}_V) \frac{a         m_3 }{2} + O(a^2)\big) &=& 
       1.250(3) \big(1 - 0.130(49) \frac{a m_3}{2}
                       - 0.25(33) (\frac{a         m_3 }{2})^2 \big) \,.
\end{eqnarray}
The estimates for $Z_V^0$ are consistent with those obtained using the
conserved vector charge, Eq.~\ref{eq:ZVfitmtilde}, but have larger
errors, so our preferred value is from the analysis presented in
Section~\ref{sec:ZV}. 

The coefficient of the term linear in $\tilde m$ ($m$) gives ${\tilde
b}_A - {\tilde b}_V$ ($b_A - b_V$).  We find that the errors in both
${\tilde b}_A - {\tilde b}_V$ and $b_A - b_V$ are large and
comparable. In addition, there can be large $O(a)$ errors feeding in
from the dependence of $c_A$ on $\tilde m$ as discussed below.  

It is easy to see that when using Eq.~(\ref{eq:cV1})
to extract $\tilde b_A - \tilde b_V$ the result will depend on the
choice whether $c_A(m)$ or the chirally extrapolated $c_A(0)$ is
used. As explained in \cite{LANL:Zfac:00}, a shift $c_A \to c_A + \xi$
in the definition of $(A_I)_4$ in the denominator produces a change
in $Z_V^0$ of the form $\xi a B_\pi$.  If, instead, we use 
$c_A(\tilde m) = c_A + \Delta \tilde m a$ 
in the calculation then the slope, not
the intercept, changes, i.e., one gets 
$ \tilde b_A - \tilde b_V - \Delta a B_\pi/2$ 
instead of $\tilde b_A - \tilde b_V$. For the
two-point data at $\beta=6.4$ the two estimates are $-0.32(5)$ and
$-0.12(5)$ for $c_A(\tilde m)$ and $c_A(0)$ respectively. The difference,
$\sim 0.20$, even though formally of higher order in $a$, is large
because $a B_\pi \sim 1.5$ and $\Delta = 0.19$ as discussed in the
extraction of $c_A$. We do not have an a priori argument that, to this
order, favors one choice over another. Anticipating that calculations
of physical quantities will use improvement constants defined in the
chiral limit and understanding that the slope $\Delta$ is almost
entirely an artifact of the discretization scheme used to calculate
$c_A$, we take results obtained using $c_A(m=0)$ for the two-point
discretization as our estimates. We stress that we do not include the
difference between the results using
$c_A(m)$ and $c_A(0)$ as part of the error. 
These new results supercede those given in~\cite{LANL:Zfac:00}.

Overall, ${\tilde b}_A - {\tilde b}_V$ is small and the uncertainty is
comparable to the signal.  The expected relation $({\tilde b}_A -
{\tilde b}_V) = (Z_A^0Z_S^0/Z_P^0) (b_A - b_V) + O(a)$ holds at the $1
\sigma$ level.

\section{$Z_A^0$ } 
\label{sec:ZA}

\begin{figure}[tbp]    
\begin{center}
\includegraphics[width=0.7\hsize]{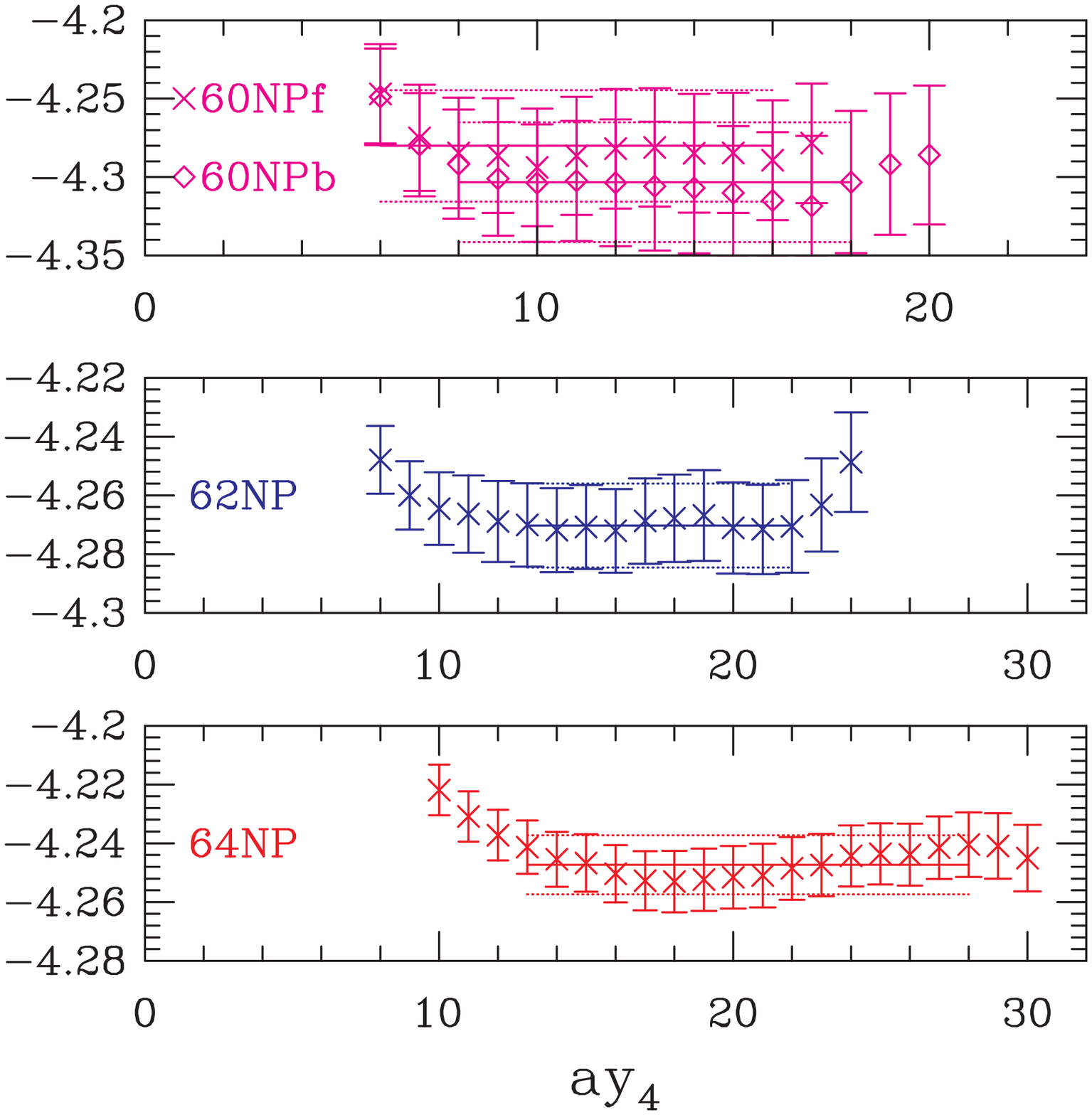}
\end{center}
\caption{Illustration of the signal for the ratio defined in
Eq.~(\ref{ZAZV-1}) for the four data sets using two-point
discretization. In all four cases the data have to be multiplied by
the respective values of $2\kappa_3$, the lattice normalization of the
additional propagator in the numerator.}
\label{fig:ZAcomp}
\end{figure}

\begin{figure}[tbp]    
\begin{center}
\includegraphics[width=0.7\hsize]{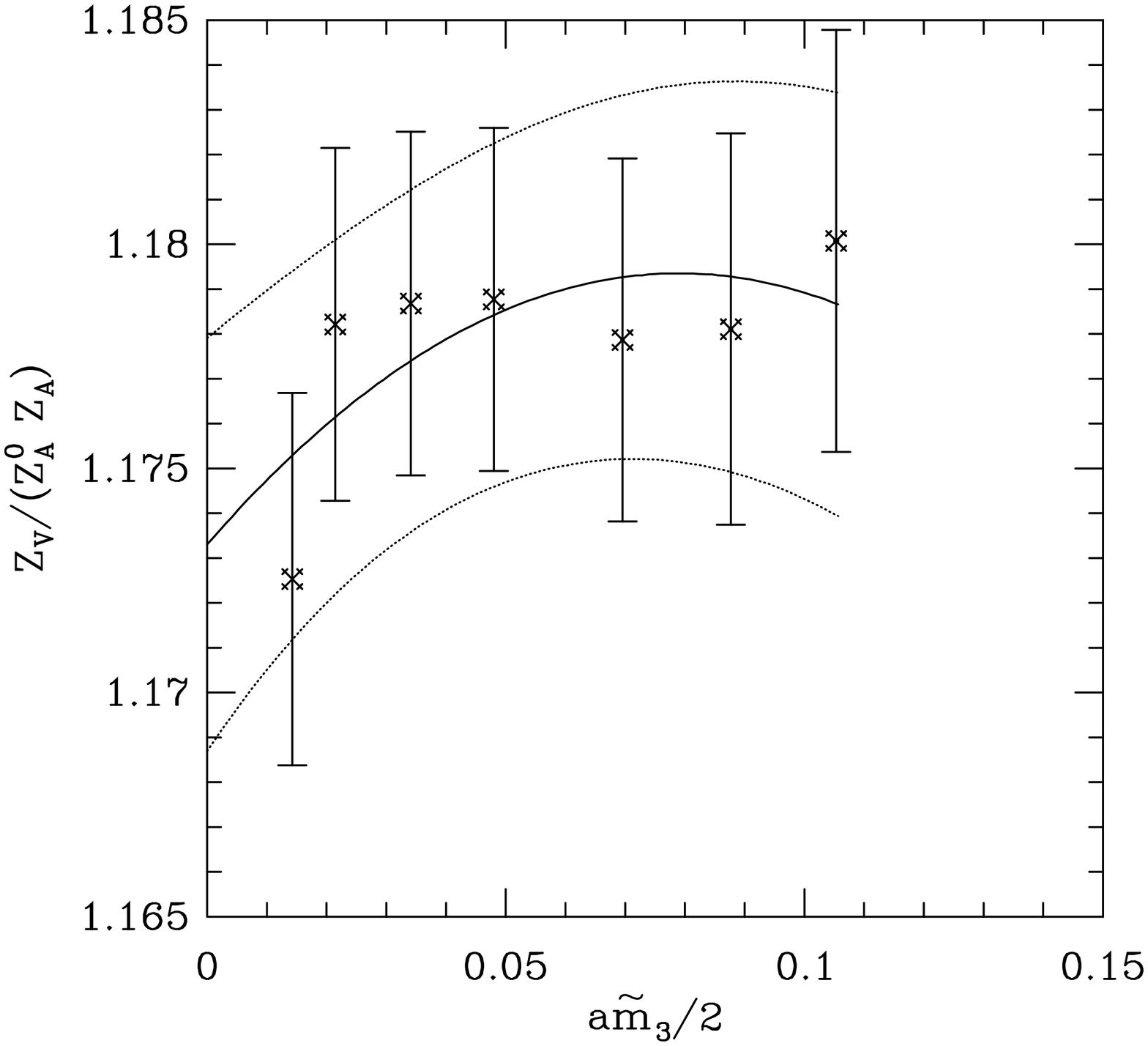}
\end{center}
\caption{The {\bf 64NP} two-point data for $Z_V/Z_A^0 Z_A
(\tilde m_3)$ are obtained by extrapolating the ratio defined in
Eq.~\protect\ref{ZAZV-1} to $\tilde m_1 = \tilde m_2 = 0$ using a
quadratic fit. The intercept of the quadratic fit in $\tilde m_3$,
gives $Z_V^0/Z_A^0 Z_A^0$.}
\label{fig:ZVZAZA}
\end{figure}

The Ward identity 
\begin{eqnarray}
    \frac       { \sum_{\vec{y}} \langle \delta {\cal S}^{(12)}_I
            \ (A_I)_i^{(23)}(\vec{y},y_4) \  V_i^{(31)}(0) \rangle }
        { \sum_{\vec{y}}
        \langle (V_I)_i^{(13)}(\vec{y},y_4) \ V_i^{(31)}(0) \rangle }
 &=&  \frac{ Z_V^0 (1+\tilde b_V a \tilde m_3/2) }
     { Z_A^0 \cdot Z_A^0 (1+ \tilde b_A a \tilde m_3/2) }\,,
\label{ZAZV-1}
\end{eqnarray}
gives $Z_V^0/(Z_A^0)^2$ and a second estimate of ${\tilde b}_A -
{\tilde b}_V$. The quality of the signal for the ratio of correlation
functions, as illustrated in Fig.~\ref{fig:ZAcomp}, is good as the
intermediate state is the vector meson. Data in Fig.~\ref{fig:ZVZAZA}
show that quadratic fits in both $\tilde m_1 \equiv \tilde m_2$ and
$\tilde m_3$ are preferred at $\beta=6.4$.  Linear fits are sufficient
at $\beta=6.0$ and $6.2$. The resulting values are given in
Tables~\ref{tab:2ptdata} and \ref{tab:3ptdata}.

Including the results in Section~\ref{sec:ZV} we have two estimates
for ${\tilde b}_A - {\tilde b}_V$ with similar errors. These estimates
come from Ward identities that involve different, pseudoscalar versus
vector, intermediate states. Also, in Eq.~\ref{ZAZV-1} the term
proportional to $c_A$ in $A_I$ does not contribute at zero momentum so
there is no associated uncertainty. Thus, the $O(a)$ errors can be
different in the two cases. As shown in Tables~~\ref{tab:2ptdata} and
\ref{tab:3ptdata}, we find that the two estimates show considerable
$O(a)$ variation, but this is not unexpected given the size of the errors and
the possibility of additional $O(a \Lambda_{QCD} \sim 0.2-0.1)$
uncertainty in previous estimates as discussed in
Section~\ref{sec:ZV}. Had we chosen to use $c_A(\tilde m)$ to extract
${\tilde b}_A - {\tilde b}_V$ in Section~\ref{sec:ZV} the variation
would have been larger by a factor of two or more. Thus, for our final
estimate we average the two two-point estimates and quote the
difference between two-point and three-point discretization schemes as
an estimate of residual $O(a)$ errors. The upshot of the analysis is
that ${\tilde b}_A - {\tilde b}_V$ is small and the systematic errors
are of the same size as the signal.

\begin{figure}[tbp]   
\begin{center}
\includegraphics[width=0.7\hsize]{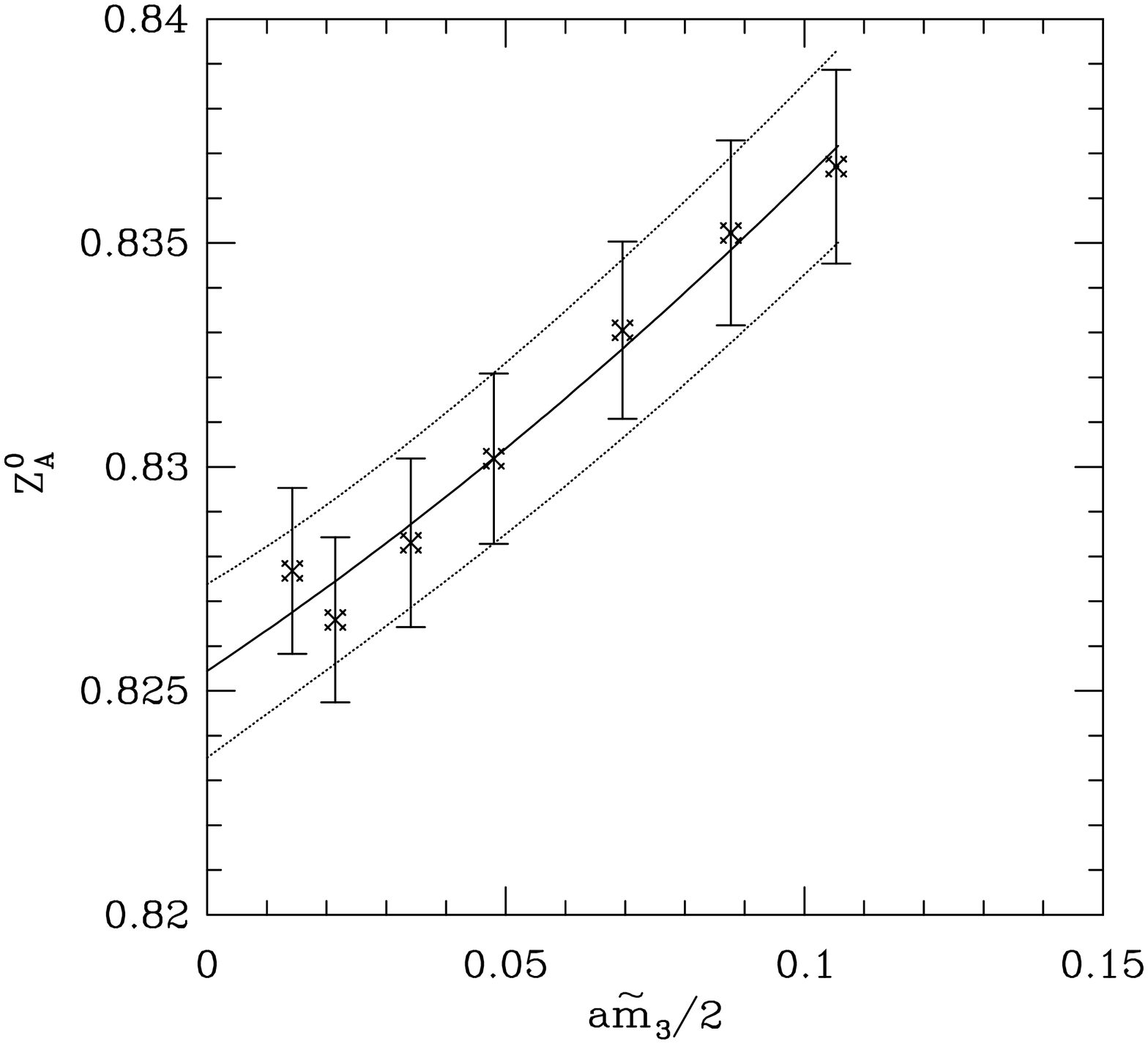}
\end{center}
\caption{$Z_A^0$ is obtained from the product of ratios of correlators
defined in Eqs.~(\protect\ref{ZAZV-1}) and (\protect\ref{eq:cV1}). The
figure shows a quadric extrapolation in ${\tilde m}_3/2$ for the {\bf
64NP} data set with two-point discretization and $c_A(\tilde m)$. }
\label{fig:ZAfitm}
\end{figure}

{\color{red}

To estimate $Z_A^0$ we use the product of Eqs.~(\ref{ZAZV-1})
and~(\ref{eq:cV1}) as it yields $1/(Z_A^0)^2$ directly. The final
chiral extrapolation in $a \tilde m_3$ for the product is shown in
Fig.~\ref{fig:ZAfitm}.  In this product \color{red}the terms proportional to
$a\tilde m_3$ cancel, but nevertheless the data show a clear $a \tilde m_3$
dependence. This we interpret as due to $O(a^2)$ terms of the generic
form $Z_A^0(\tilde m_3) = Z_A^0(0) (1 + a^2 \tilde m_3 \Lambda)$.  The
slopes at $\beta=6.0$, $6.2$ and $6.4$ are $0.22(6)$, $0.12(11)$ and
$0.11(4)$ respectively.  To match the observed slope $a \Lambda
\approx 0.11$ at $\beta=6.4$ requires $\Lambda \sim\Lambda_{QCD}
\approx 0.4\;$GeV, which is a reasonable value. Also, the change 
between $\beta=6.0$ and $6.4$ is consistent with the expected scaling in $a$.
In Ref.~\cite{LANL:Zfac:00} we had ignored this dependence and fit the data to a
constant to extract $Z_A^0$.  In light of our results at $\beta=6.4$
and a better understanding of possible $\tilde m_3$ dependence, we
\color{red}have refit the data at $\beta=6.0$ and $6.2$ also. We now use
quadratic extrapolation in $\tilde m_1 \equiv \tilde m_2$ at $\beta=6.4$ and
linear at $\beta=6.0$ and $6.2$.  Linear extrapolation in $\tilde m_3$
works well at all three couplings, however at $\beta=6.2$
and $6.4$ we use quadratic fits to maintain consistency with the rest of the 
analysis. At these weaker couplings linear and quadratic estimates are consistent.

}


A comparison of Figs.~\ref{fig:ZVZAZA} and \ref{fig:ZAfitm} raises the
following concern.  The slope in Fig.~\ref{fig:ZVZAZA} with respect to
$a \tilde m_3$ relative to the intercept is an $O(a)$ effect,
proportional to $\tilde b_V-\tilde b_A$, while that in
Fig.~\ref{fig:ZAfitm} is, as just discussed, of one higher order.\footnote{Even though the data 
in Fig.~\ref{fig:ZVZAZA} is consistent with no $\tilde m_3 $ dependence, this is not true at 
$\beta=6.0$ and $6.2$. We make a quadratic fit as indicated by all other data at $\beta=6.4$.} 
The two slopes are, however, numerically very similar. This once again
suggests that there can be substantial uncertainty of $O(a)$,
comparable to the value itself, in any result for $\tilde b_V-\tilde
b_A$. In fact, our analysis illustrates a problem common to the
extraction of all measurements of the differences $b_{{\cal O}}-b_{\delta \cal O}$.
The signal, the errors, and the $O(a^2m\Lambda)$ uncertainties are all
comparable.


\section{$Z_P^0/Z_S^0$, $\lc {\tilde b}_P - \lc {\tilde b}_S $}
\label{sec:ZPZS}

\begin{figure}[tbp]   
\begin{center}
\includegraphics[width=0.7\hsize]{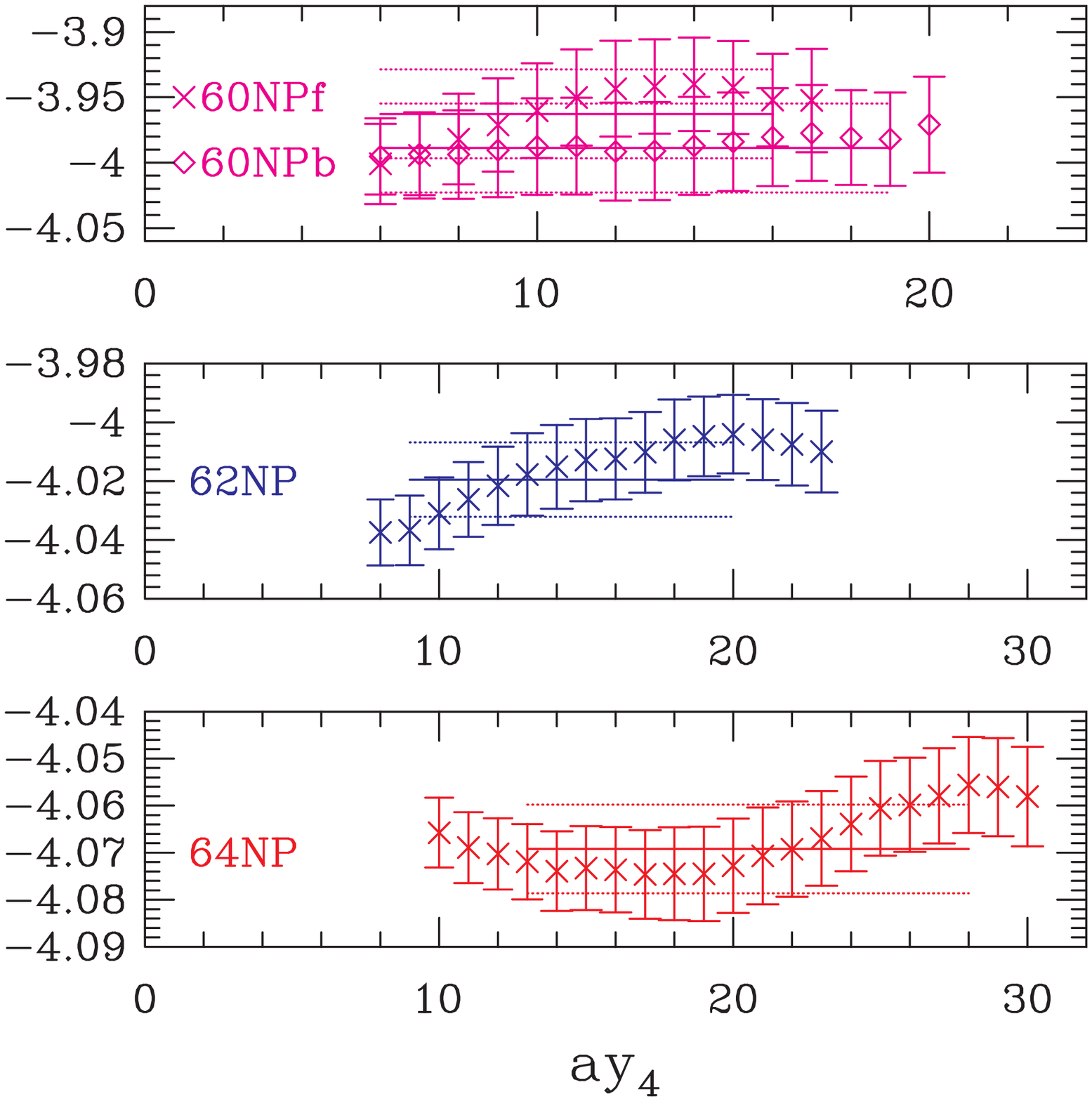}
\end{center}
\caption{Comparison of the signal in the ratio of correlators on the 
{\lhs} of Eq.~(\protect\ref{eq:ZPZS1}) used to extract $Z_P^0/Z_S^0$.
In all four cases the
data have to be multiplied by the respective values of $2\kappa_3$, 
the lattice normalization of the additional propagator in the numerator.}
\label{fig:ZPZS}
\end{figure}

To obtain $Z_P^0/(Z_S^0 Z_A^0)$ and 
$ {\tilde b}_P - {\tilde b}_S$ we use the identity
\begin{eqnarray}
 \frac	{ \sum_{\vec{y}} 
	\langle \delta {\cal S}^{(12)}_I 
		\ S^{(23)}(\vec{y},y_4) \ J^{(31)}(0) \rangle }
	{ \sum_{\vec{y}} 
	\langle P^{(13)}(\vec{y},y_4) \ J^{(31)}(0) \rangle }
 &=&  
\frac{ Z_P^0 (1+\tilde b_P a\tilde m_3/2) } 
	{ Z_A^0 \cdot Z_S^0 (1+ \tilde b_S a\tilde m_3/2) }\,,
	\label{ZPZS-1}
\label{eq:ZPZS1}
\end{eqnarray}
evaluated in the limit $\tilde m_1 \equiv \tilde m_2 \to 0$ with $J =
P$ or $A_4$.  The intermediate state in both the numerator and the
denominator has the quantum numbers of a pion, and the ratio has a
very good signal, whose quality, as a function of $\beta$, is shown in
Fig.~\ref{fig:ZPZS}.  Data at $\beta=6.4$ for the ratio on the left
hand side of Eq.~(\ref{eq:ZPZS1}) favor quadratic fits for $\tilde m_1
\equiv m_2 \to 0$ and $\tilde m_3 \to 0$ extrapolations as illustrated
in Figure~\ref{fig:ZPZSfit}.  The intercept and the slope give
$Z_P^0/(Z_S^0 Z_A^0)$ and ${\tilde b}_P - {\tilde b}_S$ respectively,
and these estimates are quoted in Tables~\ref{tab:2ptdata} and
\ref{tab:3ptdata}. To get $Z_P^0/Z_S^0$ we eliminate $Z_A^0$ by
combining the ratio in Eq.~(\ref{eq:ZPZS1}) with the product of Ward
identities discussed in section~\ref{sec:ZA}.

{\color{red} The value of ${\tilde b}_P - {\tilde b}_S$ is numerically
small, comparable to the errors and of the same order as $O(a^2 m_3
\Lambda)$ effects discussed in Section~\ref{sec:ZA}. In this case we
take the average of the two-point and three-point values as our best
estimate. The reason is that the operators in Eq.~(\ref{eq:ZPZS1}) do
not contain any derivatives (no improvement terms) so the difference between two-point and
three-point estimates arises solely from the chiral extrapolations due
to the tiny differences in $\tilde m$ for the two cases as shown in
Table~\ref{tab:mtilde}.}

Our results for $Z_P^0/Z_S^0$, obtained using Eq.~(\ref{eq:ZPZS1}) and
$Z_A$ from Section~\ref{sec:ZA}, are presented in
Table~\ref{tab:finalcomp}. These are consistent with the recent
estimates by the ALPHA and SPQcdR collaborations~\cite{SPQcdR:Z:04}.
In Section~\ref{sec:alphaanda} we compare our results with predictions
of one-loop perturbation theory and discuss the size of $O(a^2)$ and
$O(\alpha_s^2)$ corrections needed to explain the large difference.

\section{$Z_P^0/(Z_S^0 Z_A^0)$, $\lc {\tilde b}_A - \lc {\tilde b}_P - \lc {\tilde b}_m $, and $\lc {\tilde b}_m$ }
\label{sec:bPbA}

One can derive a relation between the two definitions of quark mass~\cite{ROMETV:Imp:98},
\begin{equation}
\frac{ \tilde m }{ m }  = \frac{Z_P^0 Z_m^0}{ Z_A^0}
[1 - (\tilde b_A - \tilde b_P - \tilde b_m) a{\tilde m}_{av} + 
\tilde b_m a \frac{({\tilde m}^2)_{av} - ({\tilde m}_{av})^2}
                  {{\tilde m}_{av}} + O(a \tilde m)^2 ] \,,
\label{eq:massVI}
\end{equation} 
where ${X}_{av} = ( X_1 + X_2)/2$.  This relation is useful because
$Z_m^0=1/Z_S^0$ and $b_S = - 2 b_m$~\cite{Luscher:bSbm:97,LANL:Zfull:06}.

\begin{figure}[tbp]   
\begin{center}
\includegraphics[width=0.7\hsize]{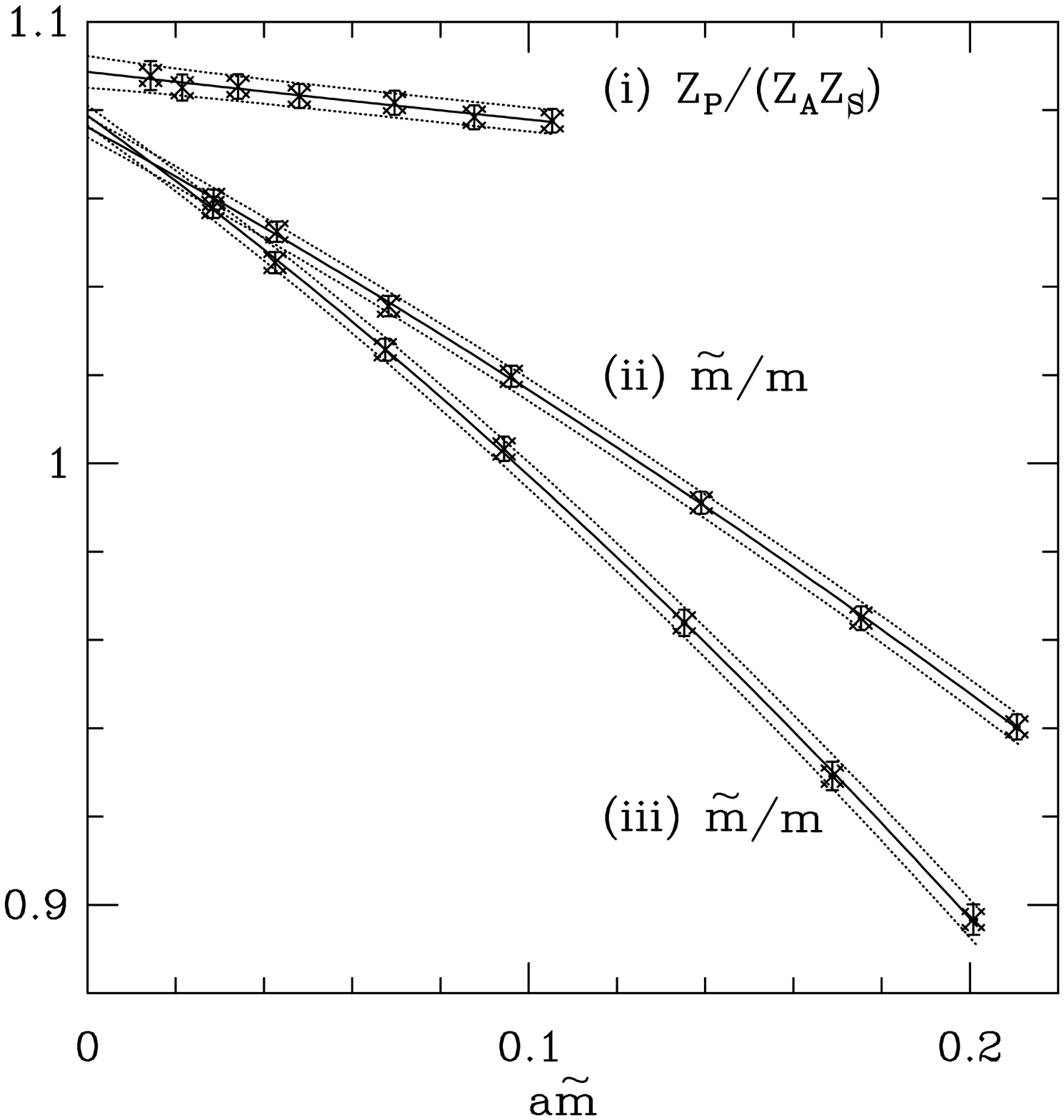}
\end{center}
\caption{Quadratic fits used to extract $Z_A^0 Z_S^0 / Z_P^0$. The
three fits correspond to (i) Eq.~(\ref{eq:ZPZS1}) plotted versus $a
\tilde m = a \tilde m_3/2$, (ii) Eq.~(\ref{eq:massVI}) with $\tilde m$
defined using the mass dependent $c_A$, and (iii)
Eq.~(\ref{eq:massVI}) with $\tilde m$ defined using the chirally
extrapolated $c_A$.  The data are from the {\bf 64NP} set with
two-point discretization.  Note that the
intercepts from all three fits should agree up to errors of $O(a^2)$,
but the slope of (i) is
$b_P-b_S$ whereas those of (ii)
and (iii) give ${\tilde b}_A-{\tilde b}_P-{\tilde b}_m$.}
\label{fig:ZPZSfit}
\end{figure}

In Fig.~\ref{fig:ZPZSfit} we illustrate fits to Eq.~(\ref{eq:massVI})
for the simpler case of degenerate quarks for $\tilde m$ calculated
using both $c_A(\tilde m)$ and $c_A(0)$.  In this case the term
proportional to $\tilde b_m$ does not contribute.  The data show that
for $\beta=6.4$ including a term quadratic in ${\tilde m}_{av}$ gives
a much better fit whereas for $6.0$ and $6.2$ a linear fit
suffices. The term proportional to $b_m$ contributes only to
non-degenerate combinations.  Fits using all combinations of six
($\beta=6.0)$ and seven ($\beta=6.2$ and $6.4$) values of quark masses
allow both $\lc {\tilde b}_P - \lc {\tilde b}_A - {\tilde b}_{m}$ and
${\tilde b}_{m}$ to be extracted reliably. These two sets of results
of the fits, using $c_A(\tilde m)$ and $c_A(0)$, are quoted in
Tables~\ref{tab:2ptdata} and \ref{tab:3ptdata}.

The intercept, which gives ${Z_P^0 }/{ Z_A^0 Z_S^0}$, should be
same for $c_A(m)$ and $c_A(0)$, to the extent that the fits are
good.  Furthermore, 
the difference between two- and three-point
results (which have different results for $c_A$) should be small. 
These features are borne out by the results. The only notable difference 
is that the errors in the two-point data are smaller. 
The results are also consistent with those obtained using
Eq.~(\ref{eq:ZPZS1}), and have similar errors, as illustrated in
Fig.~\ref{fig:ZPZSfit}. In Ref.~\cite{LANL:Zfac:00} we preferred
the results from Eq.~(\ref{eq:ZPZS1}) since the method of this
section has a greater sensitivity to uncertainties in $c_A$
(which are enhanced by the presence of the factor
$B_\pi = M_\pi^2 / \tilde m \approx 4\;$GeV). With better understanding 
of the errors we now choose to take, for our final value of ${Z_P^0 }/{ Z_A^0 Z_S^0}$, 
the weighted mean of the two-point results from Eq.~(\ref{eq:ZPZS1}) 
and those using $\tilde m/m$,
with the latter determined using $c_A(\tilde m)$.

The extraction of $\tilde b_A - \tilde b_P + \tilde b_S/2$ and $\tilde
b_S$ is effected by the choice of $c_A$.  Using Eqs.~\ref{eq:cA} and
\ref{eq:massVI} one can show that, to leading order, $(\tilde b_A -
\tilde b_P + \tilde b_S/2)|_{c_A(m)} = (\tilde b_A - \tilde b_P +
\tilde b_S/2)|_{c_A(0)} - \Delta a B_\pi/2$ and similarly for $\tilde
b_m$.  Our data are roughly consistent with this relation for both the
two-point and three-point discretization methods. For example, in case
of the two-point discretization method, the values of the slope
$\Delta$, illustrated in Fig.~\ref{fig:cAext64} for $\beta=6.4$ data,
are approximately $0.18$, $0.18$ and $0.19$ and $a B_\pi \approx 2.6$,
$1.9$ and $1.5$ at the three couplings respectively. This $O(a)$
effect, enhanced by the large value of $B_\pi$, gives rise to the difference in slopes as
illustrated in Fig.~\ref{fig:ZPZSfit}.

For the two combination of $b$'s the two-point and three-points
results are consistent for $c_A(\tilde m)$.  This is expected because,
as discussed in section~\ref{sec:cA}, at each quark mass the 
$\tilde m$ extracted from the two discretization schemes are, up to $O(a^3)$,
the same, provided the mass dependent $c_A(m)$ are used. 
We also find that the fits to two-point data with $c_A(m)$ are
marginally better. So we use estimates obtained
from the two-point data with $c_A(\tilde m)$ for the central values.

Note that the considerations regarding choice of $c_A(\tilde m)$
versus $c_A(0)$ here are different from those applied in
section~\ref{sec:cV} when determining $\tilde b_A-\tilde b_V$.  To
avoid confusion it is worthwhile summarizing our choices. The quark
mass $\tilde m$ and $c_A(\tilde m)$ are extracted simultaneously from
the two-point AWI.  We then use these $\tilde m$ and $c_A(\tilde m)$
in all calculations of $\delta S$.  For improving the external
current, $A_I$, in the three-point AWI we use $c_A(0)$.  Lastly, the
``slope-ratio'' method, where we use the average of data with
$c_A(\tilde m)$ and $c_A(0)$, gives $c_V(0)$ needed to improve the
vector current.

The estimate of $\tilde b_A - \tilde b_P + \tilde b_S/2$ from fits to
Eq.~\ref{eq:massVI} using the full set of masses (degenerate and
non-degenerate) is very similar to that obtained using only the
degenerate set.  Including non-degenerate combinations we find that
$\tilde b_m$ can also be extracted reliably. With $\tilde b_A - \tilde
b_P + \tilde b_S/2$ and $\tilde b_S$ in hand we can finally extract
$\tilde b_P$ in two ways. The first is obtained by combining $\tilde
b_P - \tilde b_S$ and $\tilde b_S$ and the second combines $\tilde
b_V$, $\tilde b_A - \tilde b_V$, $\tilde b_A - \tilde b_P + \tilde
b_S/2$ and $\tilde b_S$. Both estimates use one combination of $b$'s
from the three-point axial Ward identity. These are of similar quality
and dominate the errors. We find that these two estimates of $\tilde
b_P$, which provide a consistency check, differ at the level of the
uncertainties present in all combinations of $b$'s extracted using
three-point AWI. For our final estimates of $\tilde b_P$ given in
table~\ref{tab:finalcomp} we take the weighted average.

\begin{figure}[tbp]  
\begin{center}
\includegraphics[width=0.7\hsize]{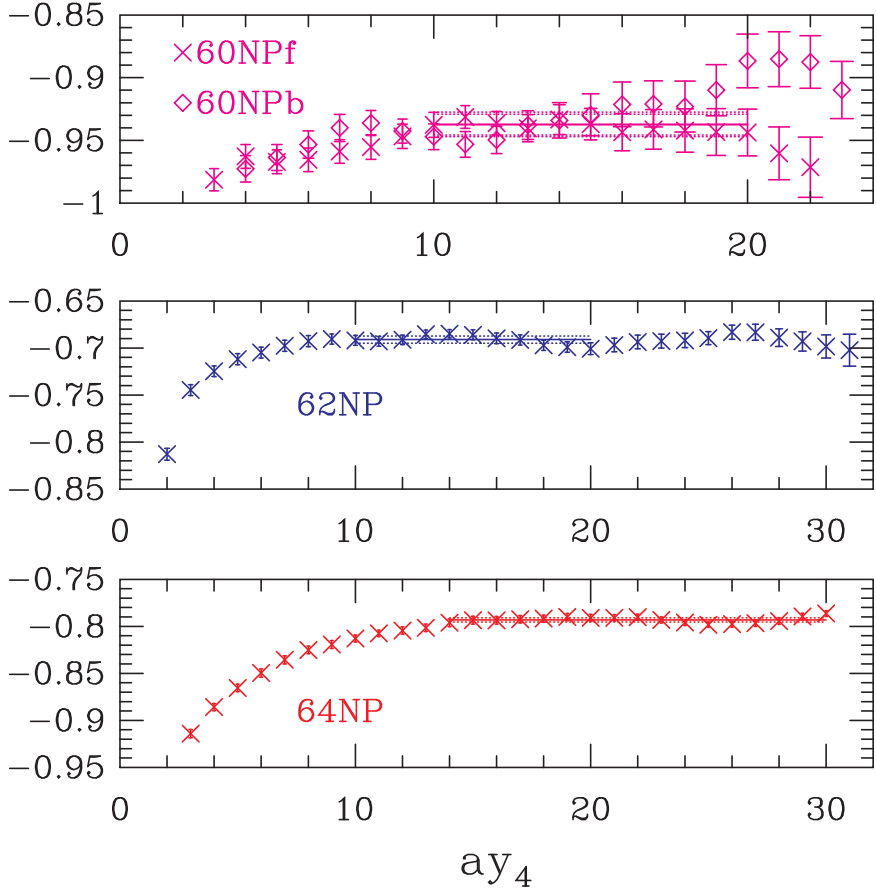}
\end{center}
\caption{The signal in the ratio of correlators used to extract $c_T$
and defined on the left hand side of Eq.~(\ref{eq:cT2}).  The points 
are from {\bf 64NP} two-point data with $\kappa_3$ quark propagators.}
\label{fig:cT2}
\end{figure}

\section{$\lc c_T$}
\label{sec:cT}

\begin{figure}[tbp]   
\begin{center}
\includegraphics[width=0.7\hsize]{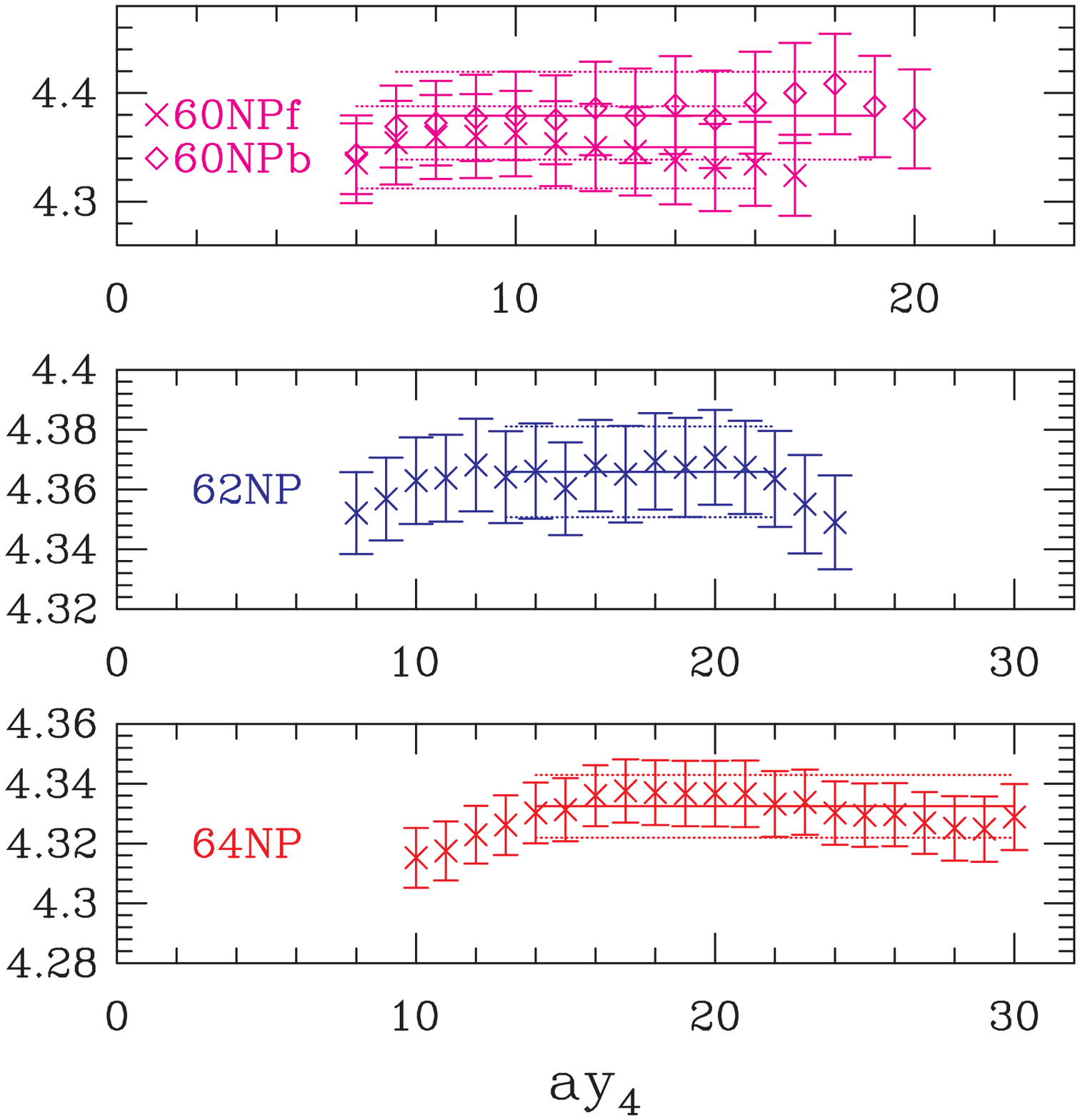}
\end{center}
\caption{The signal in the ratio of correlators defined on the right
hand side of Eq.~(\ref{eq:cT2}).  This {\bf 64NP} data
are used to extract $c_T$. In all four cases the data have to be
multiplied by the respective values of $2\kappa_3$, the lattice
normalization of the additional propagator in the numerator.}
\label{fig:cT3}
\end{figure}

$c_T$ is extracted by solving, for each ${\tilde m}_3$, the Ward
identity 
\begin{eqnarray}
 1 + a c_T \frac{ \sum_{\vec{y}} \langle
    [- \partial_4 V_k ]^{(13)}(\vec{y},y_4) T_{k4}^{(31)}(0) \rangle }
    { \sum_{\vec{y}} \langle T_{k4}^{(13)} (\vec{y},y_4) T_{k4}^{(31)}(0) \rangle }
&= & Z_A^0 \frac { \sum_{\vec{y}} 
        \langle \delta {\cal S}^{(12)}_I \ T_{ij}^{(23)}(\vec{y},y_4) 
	\ T_{k4}^{(31)}(0) \rangle }
        { \sum_{\vec{y}}  
        \langle T_{k4}^{(13)} (\vec{y},y_4) \ T_{k4}^{(31)} (0) \rangle } \,,
        \label{cT-1}
\label{eq:cT2}
\end{eqnarray}
and extrapolating these estimates to $\tilde m_3 = 0$ as discussed in
Ref.~\inlinecite{LANL:Zfac:00}.  The quality of the data for the
ratios on the left and right hand sides of this equation is very good
as illustrated in Figs.~\ref{fig:cT2} and \ref{fig:cT3}. We find that
the two-point and three-point methods give consistent estimates after
the chiral extrapolations. We take the two-point value as our final
estimate and the difference from the three-point result as a
systematic uncertainty.

\begin{figure}[tbp]   
\begin{center}
\includegraphics[width=0.7\hsize]{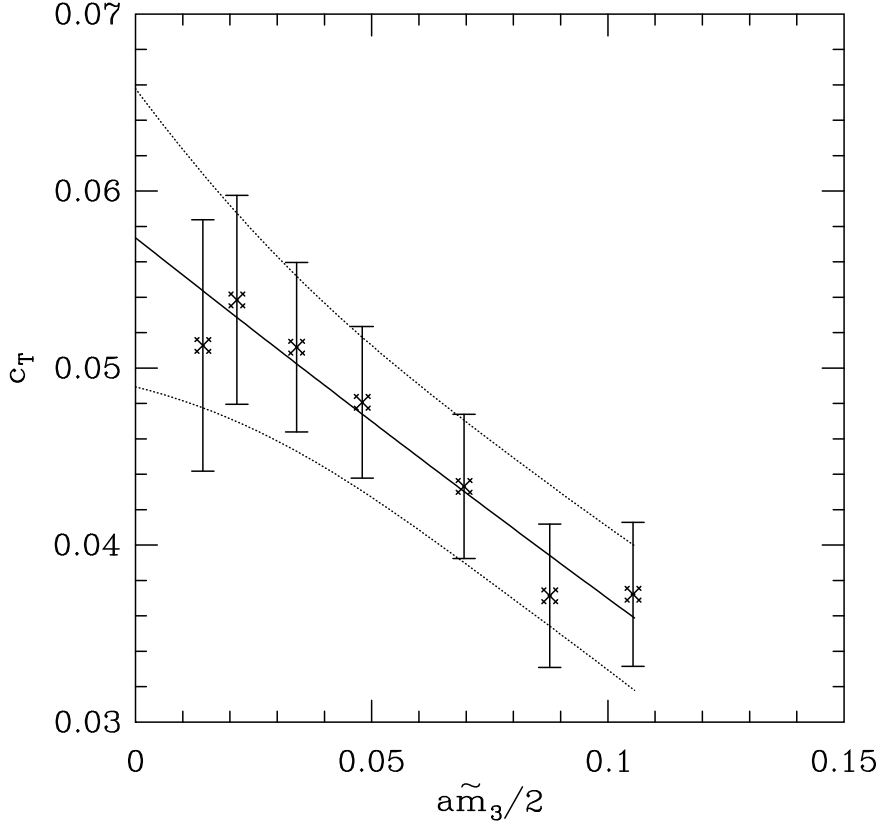}
\end{center}
\caption{$c_T$ is extracted using a quadratic fit to all seven
quark masses ($\kappa_1 - \kappa_7$) for the {\bf 64NP} data
set with two-point discretization and $c_A(\tilde m)$.}
\label{fig:cTvsm}
\end{figure}

{\color{red} The data, illustrated in Fig.~\ref{fig:cTvsm}, exhibit a
behavior linear in $\tilde m_3$.  This can arise due to corrections of
the form $O(a \Lambda a \tilde m_3)$.  We had erroneously neglected
this $O(\tilde m_3 a)$ dependence in $c_T$ in previous analyses.  The
slopes, $-0.33(11)$, $-0.21(10)$ and $-0.17(3)$ at $\beta = 6.0$,
$6.2$ and $6.4$ respectively, are consistent with an $a \Lambda$
behavior.  The change in $c_T$ between a constant and a linear fit are
significant at the $1\sigma$ level, $i.e.$ they change from $0.063(7)
\to 0.085(12)$, $0.051(7) \to 0.063(10)$, $0.041(3) \to 0.054(5)$ for
the three $\beta$ values respectively. Thus, our new estimates, based
on linear fits, differ from those quoted in
Ref.~\inlinecite{LANL:Zfac:00}.  
}

\section{Equation-of-motion operators}
\label{sec:offshell}

We extract the coefficients, $ c'_\CO $, of the equation-of-motion
operators from the $\tilde m_{12}$ dependence of the three-point
AWI~\cite{LANL:Zfac:00}: 
\begin{equation}
\frac{
	\langle \int_V d^4x\, \delta S_I^{(12)}
	\ \CO_I^{(23)}(y_4, \vec y) \ J^{(31)}(0)
	\rangle }
	{  \langle 
	\delta \CO_I^{(13)}(y_4, \vec y) \ J^{(31)}(0)
	\rangle }
=  \frac{ Z^{(13)}_{\delta\CO} }
	{ Z^{(12)}_A \ Z^{(23)}_\CO }
	+ a \frac{c'_P + c'_\CO}{2} {\tilde m}_{12} + O(a^2) \,.
\label{eq:WI-c'}
\end{equation}
This can be rewritten as
\begin{eqnarray}
c'_P + c'_\CO = 
	2 s_\CO - X_\CO 
	\big(\tilde b_{\delta \CO} - \tilde b_\CO - 2 \tilde b_A \big) \,,
\label{eq:c'+c'}
\end{eqnarray}
where $X_\CO = { Z^0_{\delta\CO} } / { Z^0_A \ Z^0_\CO }$ and $s_\CO$
is the slope, in the limit $\tilde m_3 \to 0$, of the left hand side
of Eq.~(\ref{eq:WI-c'}) with respect to $\tilde m_1$ for fixed $\tilde
m_3$. The results for $c'_P + c'_\CO $ are shown in
Table~\ref{tab:c'}, and for the three individual pieces $s_\CO$,
$X_\CO(b_{\delta \CO} - b_{\CO})/2$, and $X_\CO b_{A}$ in
Table~\ref{tab:c'breakup-64np}.

\begin{table}
\begin{center}
\begin{tabular}{|l|c|c|c|}
\hline
\multicolumn{4}{|c|}{$62NP$} \\
\hline
\multicolumn{1}{|c|}{$c'_\CO+c'_P$}&
\multicolumn{1} {c|}{$s_\CO$}&
\multicolumn{1} {c|}{$X_\CO(b_{\delta \CO} - b_{\CO})/2$}&
\multicolumn{1} {c|}{$X_\CO b_{A}$ } \\
\hline
$c^\prime_V + c^\prime_P $   & $-0.22(05)$ & $-0.06(2)$     &  $1.50(5) $ \\
$c^\prime_A + c^\prime_P $   & $-0.15(07)$ & $+0.05(5)$     &  $1.42(5) $ \\
$c^\prime_P + c^\prime_P $   & $-0.17(10)$ & $-0.11(6)$     &  $1.56(6) $ \\
$c^\prime_S + c^\prime_P $   & $-0.15(03)$ & $-0.05(1)$     &  $1.29(5) $ \\
$c^\prime_T + c^\prime_P $   & $-0.21(07)$ & $+0.02(5)$     &  $1.45(5) $ \\
\hline    
\hline
\multicolumn{4}{|c|}{$64NP$} \\
\hline
\multicolumn{1}{|c|}{$c'_\CO+c'_P$}&
\multicolumn{1} {c|}{$s_\CO$}&
\multicolumn{1} {c|}{$X_\CO(b_{\delta \CO} - b_{\CO})/2$}&
\multicolumn{1} {c|}{$X_\CO b_{A}$ } \\
\hline
$c^\prime_V + c^\prime_P $   & $-0.18( 5)$ & $-0.08( 3)$   &  $1.36(6) $ \\
$c^\prime_A + c^\prime_P $   & $-0.08( 6)$ & $+0.08( 4)$   &  $1.27(6) $ \\
$c^\prime_P + c^\prime_P $   & $-0.84(36)$ & $-0.41(32)$   &  $1.49(8) $ \\
$c^\prime_S + c^\prime_P $   & $-0.06( 4)$ & $-0.06( 3)$   &  $1.18(6) $ \\
$c^\prime_T + c^\prime_P $   & $-0.17( 7)$ & $-0.05( 5)$   &  $1.31(6) $ \\
\hline    
\end{tabular}
\end{center}
\caption{The three contributions to the coefficient of the equation of
motion operators $c'_\CO+c'_P$ for the {\bf 62NP} and {\bf 64NP} data
sets using the two-point derivative data and $c_A(\tilde m)$ in the calculation of 
$\delta S$ and $c_A(0)$, $c_V(0)$, $c_T(0)$ in the discretization of the operators.}
\label{tab:c'breakup-64np}
\end{table}

The quality of all the results is dominated by how well we can measure
$c'_P$. Unfortunately, the intermediate state in the relevant
correlation functions is a scalar which has a poor signal.  To obtain
a flat region with respect to the time slice of the operator insertion
we fit the ratio on the $l.h.s.$ of Eq.~\ref{eq:WI-c'} allowing
$\tilde m$ in $\delta S$ to be a free parameter. The resulting $\tilde
m$ differ from those obtained using Eq.~\ref{eq:cA} by about $7\%$,
$4\%$, and $2\%$ at $\beta=6.0$, $6.2$ and $6.4$ respectively.

There is an additional systematic uncertainty of $O(a) \sim 0.1$ in
the determination of any slope from the chiral fits as discussed
previously.  This impacts the determination of all three terms 
$s_\CO$, $X_\CO(b_{\delta \CO} - b_{\CO})/2$ and $X_\CO b_{A}$. 

\begin{figure}[!ht]  
\begin{center}
\includegraphics[width=0.7\hsize]{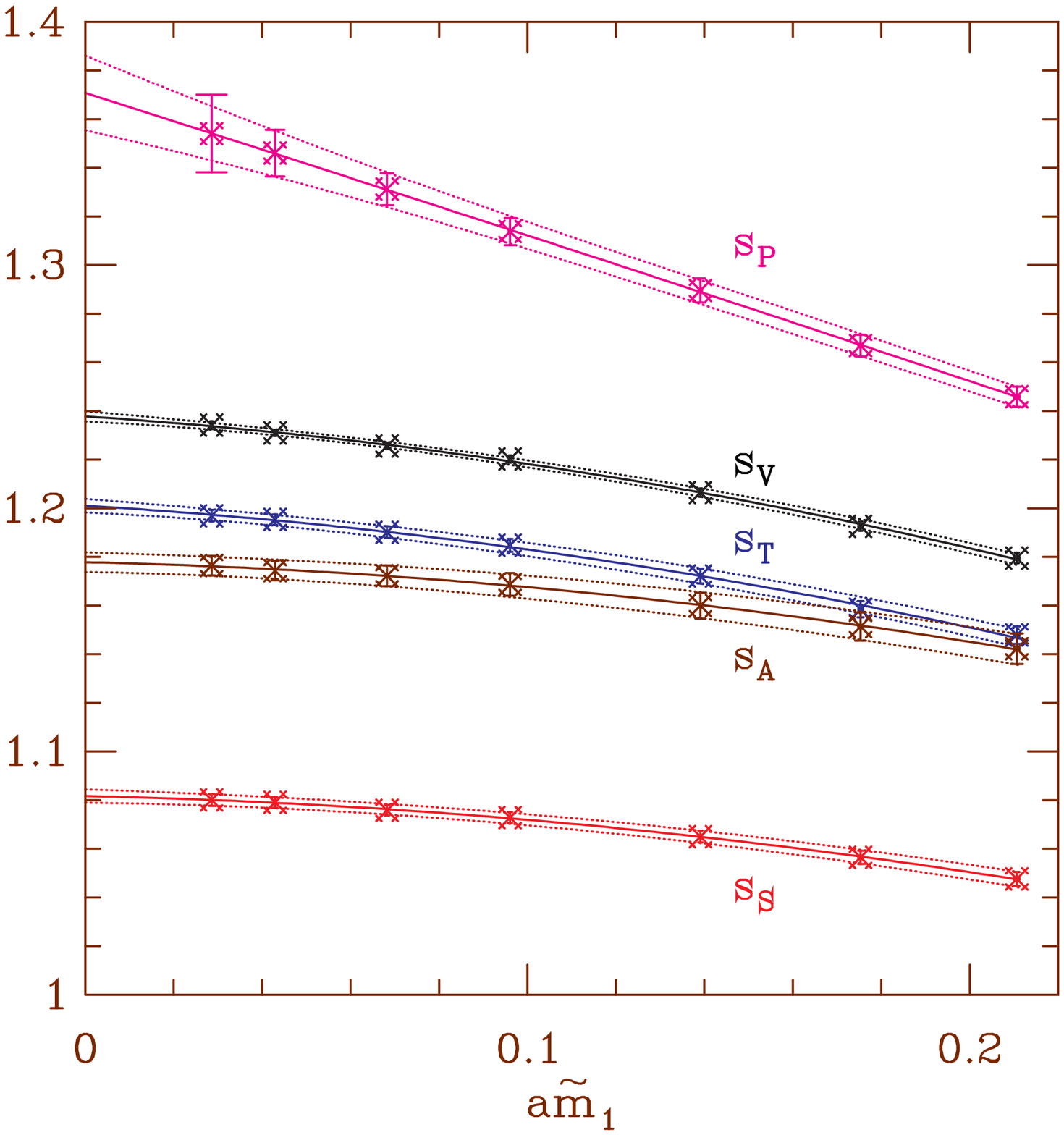}
\end{center}
\caption{Quadratic fits to the {\lhs} of Eq.~(\ref{eq:WI-c'}) versus $a \tilde m$.  The
slopes $s_\CO$ contribute to the coefficient of the equation of
motion operators through Eq.~(\ref{eq:c'+c'}).  The data are for
{\bf 64NP} with two-point discretization, $c_A(\tilde m)$ and 
${\tilde m}_3$ chosen to be $\kappa_3$. }
\label{fig:c'slope}
\end{figure}

Examples of fits to the left hand side of Eq.~\ref{eq:WI-c'} are shown
in Fig.~\ref{fig:c'slope} for the {\bf 64NP} data set. The estimates,
at leading order, should not depend on ${\tilde m}_3$, however, the
data show higher order effects. We, therefore, use a quadratic
extrapolation in ${\tilde m}_3$ at $\beta=6.4$ and linear at
$\beta=6.0$ and $6.2$ to get $s_\CO$. This changes the estimates from
those presented in \cite{LANL:Zfac:00} and \cite{LANL:Zfac:lat01}.

There is a very significant improvement in the signal for both the
individual terms and the final $c'_P + c'_\CO$ as $\beta$ increases.
Nevertheless, due to the uncertainties discussed above, all the $c'$
could have additional systematic uncertainties similar to the errors
quoted in Table~\ref{tab:c'}, whose resolution is beyond the scope of
this work. Thus, we consider our estimates as qualitative and warn the
reader that the difference from the tree level value $c'_{\CO}=1$
should be used with caution.

\section{Comparison with results by the ALPHA collaboration}
\label{sec:ALPHA}

The ALPHA collaboration has used a very different method $i.e.$, the
Schrodinger Functional method, and their estimates have the largest
differences from ours, so it is worthwhile comparing the two sets of
values for $Z_A^0$, $Z_V^0$, $Z_P^0/Z_S^0$, $c_A$, $c_V$, and
$b_V$. We expect the difference to vanish as $O(a^2)$ for $Z_A^0$,
$Z_V^0$ and $Z_P^0/Z_S^0$, and as $O(a)$ for $c_A$, $c_V$ and $b_V$.
We find that, within combined errors, the estimates for $b_V$ by the
LANL, ALPHA and QCDSF collaboration are already consistent at all
three $\beta$ values as shown in Table~\ref{tab:ZVbV}.  Similarly,
estimates for $Z_P^0/Z_S^0$ by the LANL and ALPHA collaborations
agree. For the other four quantities, there is a statistically
significant difference, and we have attempted to see whether the
lattice spacing dependence is consistent with theoretical
expectations.  To do this, we have fit the difference $\Delta X =
X_{LANL} - X_{ALPHA}$ to an appropriate function of $a$, with the
results:
\begin{eqnarray}
\label{eq:diffALPHA}
\Delta Z_V^0 &=& \phantom{-}   0.004(1)   - [261(16) a]^2  \quad \chi^2/{ndf}= 0.03  \\
\label{eq:compALPHAzv}
\Delta Z_A^0 &=&              -0.002(12)  + [222(190) a]^2 \quad \chi^2/{ndf}= 0.5  \\
\label{eq:compALPHAza}
\Delta c_A   &=&              -188(39) a  + [769(74) a]^2  \quad \chi^2/{ndf}= 0.4 \\
\label{eq:compALPHAca}
\Delta c_V   &=&              -0.15(14)   + 703(431) a     \quad \chi^2/{ndf}= 0.01
\label{eq:compALPHAcv}
\end{eqnarray}
where $a$ is in units of MeV${}^{-1}$ so that the coefficients are in
units of MeV. Error estimates in $\Delta X$ were determined by adding
the two independent statistical errors in quadratures. A number of
comments are in order:

\begin{itemize}
\item
These fits are very sensitive to the errors assigned to $\Delta X$ and 
should only be used to draw qualitative conclusions. 
\item
As noted earlier, our condition ($m_{ij}$ independent of $t$ for $t
\geq 2$) for fixing $c_A$, and the variation in the physical size of
our sources with $a$ may lead to a more complicated dependence on $a$
than simply the leading order expectation. Similarly, we have chosen
different forms for the chiral extrapolation at the various
$\beta$'s. These issues have been ignored here given the small number
of values of $\beta$.
\item
For $Z_V^0$ and $Z_A^0$ we expect a vanishing intercept and a
difference proportional to $a^2$. This expectation is borne out
reasonably well.  The non-zero value for the intercept in $\Delta
Z_V^0$ could be a manifestation of higher order terms that are ignored
in our fit. The size of the $a^2$ term is consistent with being
$\sim(a\Lambda_{QCD})^2$.
\item
Fits to $\Delta c_A$ without a quadratic term have large $\chi^2$. The
linear plus quadratic fit given in Eq.~\ref{eq:compALPHAca} does
slightly better than constant plus quadratic.  The fits are dominated
by the difference at $\beta=6.0$ where our estimate agrees with that
given in Ref.~\cite{Collins:cA:03}.
\item
Estimates of $c_V$ by the ALPHA collaboration are systematically much
larger.  The fit in Eq.~\ref{eq:compALPHAcv} has large coefficients,
however the errors are equally large. The calculation of $c_V$
warrants further study since the differences are large.

\end{itemize}

\section{Comparison with perturbation theory}
\label{sec:alphaanda}

The data at three values of the coupling allow us to also fit the
difference between the non-perturbative and tadpole improved one-loop
estimates as a function of $a$ and $\alpha_s^2$, $i.e.$, including
both the leading order discretization and perturbative
corrections. The results of these fits are
\begin{eqnarray}
\label{eq:diffPERT}
\Delta Z_V^0          &=&   -(192 a)^2 -   (1.3 \alpha_s)^2 \quad \chi^2/{ndf}= 0.9 \\
\Delta Z_A^0          &=&   -(159 a)^2 -   (1.2 \alpha_s)^2 \quad \chi^2/{ndf}= 0.5 \\
\Delta (Z_P^0/Z_S^0)  &=&   -(439 a)^2 -   (1.9 \alpha_s)^2 \quad \chi^2/{ndf}= 3.5 \\
\Delta c_V            &=&   -(138  a)  -   (1.5 \alpha_s)^2 \quad \chi^2/{ndf}= 0.4 \\
\Delta c_A            &=&    ( 30  a)  -   (1.4 \alpha_s)^2 \quad \chi^2/{ndf}= 0.01\\
\Delta c_T            &=&    (130  a)  +   (0.4 \alpha_s)^2 \quad \chi^2/{ndf}= 0.1 \\
\Delta {\tilde b_V}   &=&    (1197 a)  -   (3.8 \alpha_s)^2 \quad \chi^2/{ndf}= 3.2 \\
\Delta b_V            &=&    (630  a)  -   (1.8 \alpha_s)^2 \quad \chi^2/{ndf}= 6.3 \\
\Delta {\tilde b_A}   &=&   -(770  a)  -   (3.5 \alpha_s)^2 \quad \chi^2/{ndf}= 0.8 \\
\Delta {\tilde b_P}   &=&   -(857  a)  -   (1.9 \alpha_s)^2 \quad \chi^2/{ndf}= 2.3 \\
\Delta {\tilde b_S}   &=&    (507  a)  -   (2.3 \alpha_s)^2 \quad \chi^2/{ndf}= 8.9 
\end{eqnarray}
where $a$ is expressed in MeV${}^{-1}$, $\Delta X = X_{LANL} -
X_{1-loop}$, and $\alpha_s = g^2 /(4\pi u_0^4)$ is the tadpole
improved coupling with values $0.1340$, $0.1255$ and $0.1183$ at the
three $\beta$. The tadpole factor $u_0$ is chosen to be the fourth
root of the expectation value of the plaquette. 

One conclusion from these fits
is that one-loop tadpole improved perturbation
theory estimates of the $Z$'s and $c$'s underestimate the
corrections. The deviations can, however, be explained by coefficients
of reasonable size, $i.e.$ the coefficient of $O(a)$ is $\approx
\Lambda_{QCD}$ and the perturbative corrections are $ (1- 2)
\alpha_s^2$. The case of $ Z_P^0/Z_S^0$ is marginal, and we point to
non-perturbative calculations using external quark and gluon states
(the RI/MOM method) that show that the majority of the difference
comes from 1-loop perturbation theory significantly underestimating
($1-Z_P^0$)~\cite{SPQcdR:Z:04}.

The most striking differences from perturbation theory are for the $b$'s.
We stress, however, that the fits are very poor as evident from
the $\chi^2/{ndf}$. There are two useful statements we can make. In
the case of ${b_V}$ (and similarly ${\tilde b}_V$), the agreement
between our results and those by the ALPHA, QCDSF, and SPQcdR
collaborations~\cite{ALPHA:Zfac:97A,QCDSF:ZVbV:03,SPQcdR:Z:04},
suggests that 1-loop perturbation theory underestimates the
correction. Second, at $\beta=6.4$ ${\tilde b_A}$, ${b_A}$, ${\tilde
b_P}$ and ${\tilde b_S}$ are in good agreement with perturbation
theory.

\section{Conclusion}
\label{sec:conc}
We have presented new results for renormalization and improvement
constants of bilinear operators at $\beta=6.4$. Combining these with
our previous estimates at $\beta=6.0$, and $6.2$, and with the results
from the ALPHA collaboration we are able to quantify residual
discretization errors. Overall, we find that the efficacy of the
method improves very noticeably with the coupling $\beta$. Using data
at $\beta=6.4$ we are able to resolve higher order mass dependent
corrections in the chiral extrapolation for all the renormalization
and improvement constants presented here. Our final results are
summarized in Table~\ref{tab:finalcomp}.

Determination of $c_A$ is central to $O(a)$ improved calculations. By
comparing results from three different discretization schemes we
improve the reliability of our error estimate. We also show that reliable
estimates from correlators at finite momenta can be extracted and find
that these give consistent results with those from zero-momentum
correlators once additional $O(p^2a^2)$ errors are taken into account.

We find that both $c_A$ and $c_V$ are small, and the most significant
differences from estimates by the ALHPA collaboration are at the
strongest coupling $\beta=6.0$.

We also compare our non-perturbative estimates with one-loop tadpole
improved perturbation theory. Overall, we find estimates based on
1-loop tadpole improved perturbation theory underestimate the
corrections in the $Z$'s and $c$'s. The differences can, however, be
explained by terms of $O(\Lambda_{QCD} a)$ and $(1-2) \alpha_s^2$.

The most significant differences are in $b_V$ and $\tilde b_V$ which
are hard to explain by a combination of $O(a)$ and $\alpha_s^2$ errors
with coefficients of reasonable size. All the other $b$'s show
agreement with perturbative estimates by $\beta=6.4$.

\begin{acknowledgments}
These calculations were done at the Advanced Computing Laboratory at
Los Alamos and at the National Energy Research Scientific Computing
Center (NERSC) under a DOE Grand Challenges grant.  The work of T.B.,
R.G., and W.L. was, in part, supported by DOE grant KA-04-01010-E161
and of S.R.S by DE-FG03-96ER40956/A006.
\end{acknowledgments}


\bibliography{paper}

\printtables
\printfigures

\end{document}